\documentclass{emulateapj-rtx4}

\usepackage{lscape}
\usepackage{appendix}

\slugcomment{Received 2011 August 25; accepted by ApJ 2011 November 29}

\shorttitle{Close Companions to Young Stars. I.}
\shortauthors{Nguyen et al.}

\begin{document}


\title{Close Companions to Young Stars. I. A Large Spectroscopic Survey in Chamaeleon~I and Taurus-Auriga}


\author{Duy Cuong Nguyen\altaffilmark{1,2,3}, Alexis Brandeker\altaffilmark{1,2}, Marten H. van Kerkwijk\altaffilmark{1}, Ray Jayawardhana\altaffilmark{1}}

\email{dcnguyen@pas.rochester.edu}

\altaffiltext{1}{Department of Astronomy \& Astrophysics, University of Toronto, 50 St. George Street, Toronto, ON M5S 3H4, Canada; nguyen, mhvk, rayjay@astro.utoronto.ca}
\altaffiltext{2}{Department of Astronomy, Stockholm University, SE-106 91 Stockholm, Sweden; duy.nguyen, alexis@astro.su.se}
\altaffiltext{3}{Current address: Department of Physics and Astronomy, University of Rochester, Rochester, NY 14627-0171, USA; dcnguyen@pas.rochester.edu}

\begin{abstract}
We present the results of a multiplicity survey of $212$ T~Tauri stars in the Chamaeleon~I and Taurus-Auriga star-forming regions, based on high-resolution spectra from the Magellan Clay $6.5$\,m telescope. From these data, we achieved a typical radial velocity precision of $\sim\!80$\,m\,s$^{-1}$ with slower rotators yielding better precision, in general. For $174$ of these stars, we obtained multi-epoch data with sufficient time baselines to identify binaries based on radial velocity variations. We identified eight close binaries and four close triples, of which three and two, respectively, are new discoveries. The spectroscopic multiplicity fractions we find for Cha~I ($7\%$) and Tau-Aur ($6\%$) are similar to each other, and to the results of field star surveys in the same mass and period regime. However, unlike the results from imaging surveys, the frequency of systems with close companions in our sample is not seen to depend on primary mass. Additionally, we do not find a strong correlation between accretion and close multiplicity. This implies that close companions are not likely the main source of the accretion shut down observed in weak-lined T~Tauri stars. Our results also suggest that sufficient radial velocity precision can be achieved for at least a subset of slowly rotating young stars to search for hot Jupiter planets.
\end{abstract}


\section{Introduction}
\label{sec:Introduction}

Most stars, both in the solar neighborhood and in young clusters are members of binary or multiple systems. Yet, the formation and early evolution of binary and multiple stars is poorly constrained observationally, and not well understood theoretically. For instance, the fraction of wide binaries in dense star-forming regions such as the Orion Nebula Cluster and IC~348 are comparable to that of field stars \citep{1999AA...343..831D,2006AA...458..461K} whereas the frequency of binaries is much higher among young stars in dispersed T~associations like Taurus-Auriga \citep[e.g.,][]{1995ApJ...443..625S,1997ApJ...481..378G,2003AJ....126.2009B}; for reviews see \citet{2000prpl.conf..703M} and \citet{2007prpl.conf..379D}. Furthermore, high-order multiples are more common in nearby star-forming regions than for solar-type main-sequence stars in the solar neighborhood \citep{2006AA...459..909C}.

Some simulations suggest that stars usually form in triples and higher-order multiple systems only to be dispersed later, with the fraction of stars in multiple systems decreasing from $80\%$ down to $40\%$ by about $10$\,Myrs \citep{2004MNRAS.351..617D}. Predictions are that the binary fraction is higher among higher mass stars, and that brown dwarfs are never close companions to stars \citep{2004AA...414..633G}.

While past multiplicity surveys, using speckle imaging and adaptive optics on $4$-m class telescopes, have drawn attention to the ubiquity of binaries in star-forming regions \citep[e.g.,][]{1993AJ....106.2005G,1993AA...278..129L}, their limited contrast and angular resolution have left many key questions unanswered or only partially answered. For instance, the frequency of higher-order multiples is uncertain, and so is the frequency of very low-mass stellar and sub-stellar companions \citep{2007ApJ...671.2074A}. Multiple systems are probably common.

With adaptive optics on $8$-m class telescopes, it has become straightforward to detect all stellar and even all brown dwarf companions down to the deuterium-burning limit with separations of tens of AU for nearby young stars \citep[e.g.,][]{2008ApJ...683..844L}. Like \citet{2008MNRAS.385.2210M}, our radial velocity study is complementary and covers the close separations.

We present the results of a high-resolution spectroscopic survey of $212$ stars spanning $\sim\!0.2$--$3\,M_{\sun}$ in the nearby $\sim\!2$\,Myr old star-forming regions Chamaeleon~I (hereafter Cha~I) and Taurus-Auriga (hereafter Tau-Aur). Cha~I and Tau-Aur are at distances of $\sim\!160$\,pc and $\sim\!140$\,pc, respectively \citep{1997AA...327.1194W, 1994AJ....108.1872K}. Previously, we presented a study of rotation, disk, and accretion signatures for a subsample of $144$ stars showing no evidence of spectroscopic companions and broadening functions showing only a single source \citep{2009ApJ...695.1648N}; we use the derived projected rotational velocities ($v\,\sin\,i$) and accretion signatures (H$\alpha\,10\%$\,width) for this work. Furthermore, we also analyzed the variability in accretion-related emission lines for a subsample of $40$ classical T~Tauri stars \citep{2009ApJ...694L.153N}. \citet{2008ApJ...683..844L} presented a census of wide binaries in Cha~I that encompasses our sample while \citet{2007ApJ...670.1337D} investigated circumstellar disks, including the effect of companions on disks, for a subsample of Cha~I targets with available near-IR data.

Among the issues we address in this work is the dependence of multiplicity on primary mass, i.e., whether higher mass stars are more likely to be in binaries and multiples than their lower mass counterparts. An increase in wide binaries with increasing mass has been observed for both young stars and field dwarfs \citep[e.g.,][]{2007ApJ...662..413K,2008ApJ...683..844L}. Furthermore, we look at how close-in multiplicity varies between different star-forming regions and the field; the multiplicity of $\sim\!0.1$--$2\,M_{\sun}$ field dwarfs in the solar neighborhood has been studied extensively \citep[e.g.,][]{1991AA...248..485D,1992ApJ...396..178F,2010ApJS..190....1R}. We also explore whether close companions contribute to the observed difference between classical T~Tauri stars (CTTS), which are accreting, and weak-lined T~Tauri star (WTTS), which are not accreting based on weak H$\alpha$ emission. Although the source of dichotomy between CTTS and WTTS is currently unknown, the presence of non-accreting $2$\,Myr old stars is surprising. It has been suggested that the inner disks around weak-lined objects may have been truncated by close binary companions \citep[e.g.,][and references therein]{2000prpl.conf..703M}.

\section{Observations}
\label{sec:Observations}

Our star sample consists of a magnitude-limited subset (R $\le 17.6$ for Cha~I; R $\le 13.4$ for Tau-Aur) of targets from \citet{2004ApJ...602..816L} for Cha~I, and from \citet{1993AA...278..129L}, \citet{1993AJ....106.2005G}, \citet{1995ApJ...443..625S}, \citet{1998AA...331..977K}, \citet{2002ApJ...580..317B} and \citet{2004ApJ...617.1216L} for Tau-Aur. Our targets span the spectral type range from F2 to M5 based on published classifications. In addition, we observed a sample of 25 slowly rotating velocity standard stars selected from the list of \citet{2002ApJS..141..503N}; these are listed in Table~\ref{tbl:RVStandards} and cover a similar spectral range to our targets. We determined the spectral type for 13 targets without prior classification by fitting their spectra against those of the standard stars, and identifying the best fits.

\begin{deluxetable}{lcr}
\tablecolumns{3}
\tabletypesize{\scriptsize}
\tablewidth{0pt}
\tablecaption{Slowly Rotating Radial Velocity Standard Stars Used as Templates \label{tbl:RVStandards}}
\tablehead{
\colhead{Name} & \colhead{Spectral Type} & \colhead{$RV$\tablenotemark{a}}\\
& & \colhead{(km~s$^{-1}$)}
}
\startdata

GJ\,156 & K7 & 62.597\\
GJ\,729 & M3.5 & -10.499\\
Gl\,205 & M1.5 & 8.665\\
Gl\,349 & K3 & 29.836\\
Gl\,382 & M1.5 & 7.932\\
Gl\,876 & M4 & -1.591\\
Gl\,880 & M1.5 & -27.317\\
HD\,103932 & K4 & 48.499\\
HD\,111631 & K7 & 5.041\\
HD\,120467 & K5.5 & -37.806\\
HD\,153458 & G0 & 0.641\\
HD\,172051 & G6 & 37.103\\
HD\,193901 & F7 & -171.455\\
HD\,217987 & M0.5 & 8.809\\
HD\,83443 & K0 & 28.994\\
HD\,87359 & G5 & -0.26\\
HD\,88218 & G0 & 36.652\\
HD\,90156 & G5 & 26.934\\
HD\,91638 & F8 & -4.751\\
HD\,92945 & K1.5 & 22.856\\
HD\,96700 & G0 & 12.769\\
HR\,5447 & F2 & 0.141\\
LHS\,1763 & K5 & -55.527\\
NSV\,2863 & M1 & 4.724\\
NSV\,6431 & M2 & 15.809\\

\enddata
\tablenotetext{a}{Radial velocities are adopted from \citet{2002ApJS..141..503N}.}
\end{deluxetable}

Our data were taken using the echelle spectrograph MIKE \citep{2003SPIE.4841.1694B} on the Magellan Clay $6.5$-m telescope at the Las Campanas Observatory, Chile. The MIKE instrument is a slit-fed double echelle spectrograph with blue and red arms. For this study, we used only the red region spanning $4\,800$--$9\,400$\,\AA\, in 32 spectral orders. The $0.35''$ slit was used with no binning to obtain the highest possible spectral resolution, R $\sim\!60\,000$. The pixel scale was $0.14''$ pixel$^{-1}$ in the spatial direction, and approximately $0.024$\,\AA\,pixel$^{-1}$ at $6\,500$\,\AA\, in the spectral direction. The raw data were bias-subtracted and flat-fielded, and before extraction, the scattered background in the spectrograph was subtracted by fitting splines to inter-order pixels. Note that no sky subtraction is done as part of our extraction. For the bright stars studied by Brandeker et al. (in preparation), the sky background is generally insignificant compared to the stellar spectrum, but for the fainter among our targets it is not. As we will discuss below, this generally does not pose a problem for our purpose of inferring binarity. In MIKE, the spatial direction of the projected slit is wavelength dependent, and not aligned with the CCD columns. To extract these slanted spectra, we used customized routines running in the ESO-MIDAS environment (described in detail in Brandeker et al., in preparation). For wavelength calibration, we used exposures of a thorium-argon lamp as well as observed telluric absorption lines. Integration times were chosen such that we obtained signal-to-noise ratios (S/N)\,$>\!30$ per spectral resolution element at $6\,500$\,\AA; they typically ranged from $60$ to $1\,200$ seconds depending on seeing. We also applied barycentric correction to all spectra.

We obtained $813$ high-resolution optical spectra of $212$ members of the Cha~I and Tau-Aur star-forming regions. The data were collected on $15$ nights during four observing runs between 2006 February and 2006 December.

\section{Analysis Techniques}
\label{sec:Analysis}

To find close binaries and higher-order multiples in our sample, we used both broadening function and radial velocity analysis. The former was used to identify double-lined spectroscopic binaries (hereafter SB2s) and triple-lined spectroscopic binaries (hereafter SB3s); these systems are characterized by multiple prominent peaks in their broadening functions. The latter was used to detect single-lined spectroscopic binaries (hereafter SB1s); acceleration would be suggested by significant radial velocity scatter in our data. Both analyses involved fitting the target spectra with comparison spectra to minimize $\chi^2$,
\\
\begin{equation}
\label{equ:chi2}
\chi^2 \equiv \sum_{i} \left( \frac{Y (\lambda_i) - P (\lambda_i) }{\sigma (\lambda_i)} \right) ^2
\end{equation}
\\
where $Y$ and $\sigma$ are the fluxes and uncertainties of the target spectra, respectively, and $P$ are the fluxes of the comparison spectra. The comparison spectra are produced using template spectra from the observed slowly rotating velocity standard stars. To match a standard star with each target star, we examined the spectra from the three standard stars closest in spectral type to the target, and selected the standard star spectrum that best fits that of the target. For the fits, we masked telluric absorption lines, Li-$\lambda$6708, which is present in young stars but absent in our standard stars, and activity-related emission lines H$\alpha$, H$\beta$, Pa6 through 9, Pa11, Pa14, \ion{He}{1} ($\lambda 5876$, $\lambda 6678$, $\lambda 7065$), \ion{O}{1} ($\lambda 7774$, $\lambda 8446$, $\lambda 8456$), [\ion{O}{1}] ($\lambda 5577$, $\lambda 6300$, $\lambda 6364$), \ion{Na}{1} D doublet, \ion{Ca}{2} IR triplet, and \ion{Fe}{2} 42 multiplet from the spectra. We used $\chi^2$ analysis instead of Fourier techniques for SB1s, or TODCOR techniques \citep{1994ApJ...420..806Z} for SB2s and SB3s to take advantage of uncertainties at each individual wavelength. We processed data for each echelle order separately to avoid discontinuities resulting from poorly corrected blaze, and combined the output to get our results.

\subsection{Broadening Functions}
\label{sec:BroadeningFunctions}

The broadening functions of each target were derived based on the technique described in \citet{1999ASPC..185...82R}. Our implementation involves computing the discrete broadening function $B (v_j)$ that minimizes $\chi^2$ in Eq.~\ref{equ:chi2} for,
\\
\begin{equation}
\label{equ:BroadeningFunctionSpectraLambda}
P(\lambda_i) = \left[ \sum_{j} B (v_j) \cdot S ((1 + v_j / c )\lambda_i) \right] \cdot C (\lambda_i)
\end{equation}
\\
where $B$ is the broadening function, $S$ is the template spectrum from a slowly rotating standard star, and $C$ is the best-fit polynomial to the continuum. Here, the broadening function is essentially an estimate of the average line profile for a given spectrum.

Our routine to derive the broadening function of a target at a given epoch consists of two steps. First, for each echelle order, we fit the target spectra with comparison spectra to produce a broadening function spanning $-120$ to $+120$\,km\,s$^{-1}$ with a sampling interval of $6$\,km\,s$^{-1}$. This interval is close to the velocity resolution of the spectrograph, and reduces possible covariance between points from oversampling. Second, we compute the weighted mean of the broadening function across the echelle orders.

For targets that were later identified as SB2 or SB3 candidates, we fit the multiple peaks of their broadening functions with an equal number of analytic rotational broadening functions from \citet{2005oasp.book.....G} in order to measure their component radial velocities, flux ratios and $v\,\sin\,i$. Note, due to our sampling resolution, this technique has a $v\,\sin\,i$ lower limit of $\sim\!9{\rm\,km\,s}^{-1}$. Furthermore, for well-blended or wide broadening functions, the radial velocity precision is poor using this technique, typically a few $100$\,m\,s$^{-1}$. Thus, for single-peak broadening functions, we rely mostly on radial velocities computed directly from spectra as described in \S\ref{sec:RadialVelocities}.

\subsection{Radial Velocities}
\label{sec:RadialVelocities}

The radial velocities of each target were computed by fitting the target spectra with series of velocity shifted comparison spectra. This technique computes the velocity shift $v_{\star}$ that minimizes $\chi^2$ in Eq.~\ref{equ:chi2} for,
\\
\begin{equation}
\label{equ:RadialVelocitySpectraLambda}
P(\lambda_i, v_{\star}) = \left( G (v\,\sin\,i) \ast S \right) [(1 + v_{\star}/c)\lambda_i] \cdot C (\lambda_i)
\end{equation}
\\
where $G$ is the analytic rotational broadening function from \citet{2005oasp.book.....G} assuming a limb darkening coefficient of 0.65, $v\,\sin\,i$ is the projected rotational velocity of the target star, $S$ is the template spectrum from a slowly rotating standard star, and $C$ is the best-fit polynomial to the continuum. The projected rotational velocity for each target star was computed by fitting the target spectra with a series of template spectra rotationally broadened from $0$ to $200{\rm\,km\,s}^{-1}$, and calculating the best fit value \citep[see][]{2009ApJ...695.1648N}.

Our routine to estimate the radial velocity of a target at a given epoch consists of four steps. First, for each echelle order, we fitted the target spectra with comparison spectra velocity shifted from $-120$ to $+120$\,km\,s$^{-1}$ in steps of 10\,km\,s$^{-1}$, and recorded the velocity value of the best fit for each echelle order. Second, we refined our search to radial velocities within 10\,km\,s$^{-1}$ of the first-pass results in steps of 0.01\,km\,s$^{-1}$, and revised our estimates accordingly. Third, comparing the results for different orders, we removed outliers using a standard Tukey filter, i.e., values lying $1.5$ times the interquartile range below the first quartile and above the third quartile were discarded \citep[see][]{ZIIIAAAACAAJ}. For a Gaussian distribution, this filter corresponds to removing data points beyond $2.7\,\sigma$. Fourth, we computed the weighted mean and the associated error in the mean across the echelle orders, and used those values as the radial velocity and measurement uncertainty of the target at that epoch, respectively.

\begin{figure}
\begin{center}
\includegraphics[width=8cm]{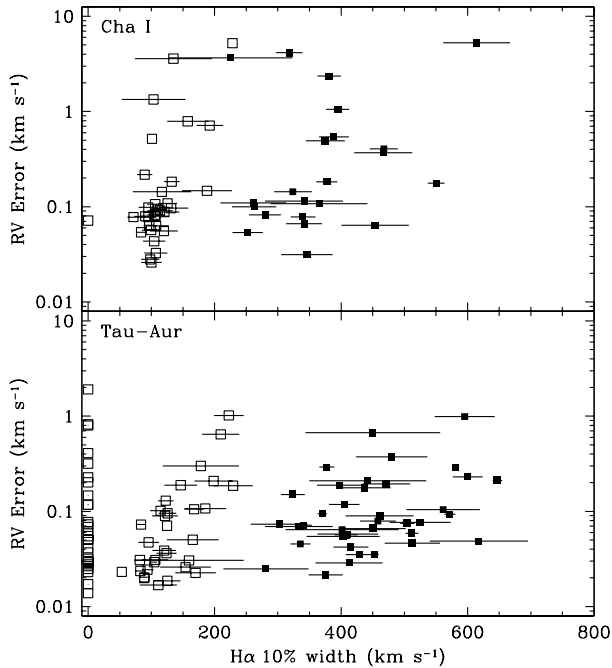}
\caption[Cha~I \& Tau-Aur: Radial velocity error as a function of H$\alpha\,10\%$\,width]{The weighted mean radial velocity measurement uncertainty for members of Cha~I and Tau-Aur with single-peak broadening functions as a function of the accretion signature H$\alpha\,10\%$\,width. Accretors and non-accretors are denoted by solid and hollow symbols, respectively. There appears to be no significant difference in the radial velocity measurement uncertainties between accretors and non-accretors.}
\label{fig:RVevsHa10w}
\end{center}
\end{figure}

\begin{figure}
\begin{center}
\includegraphics[width=8cm]{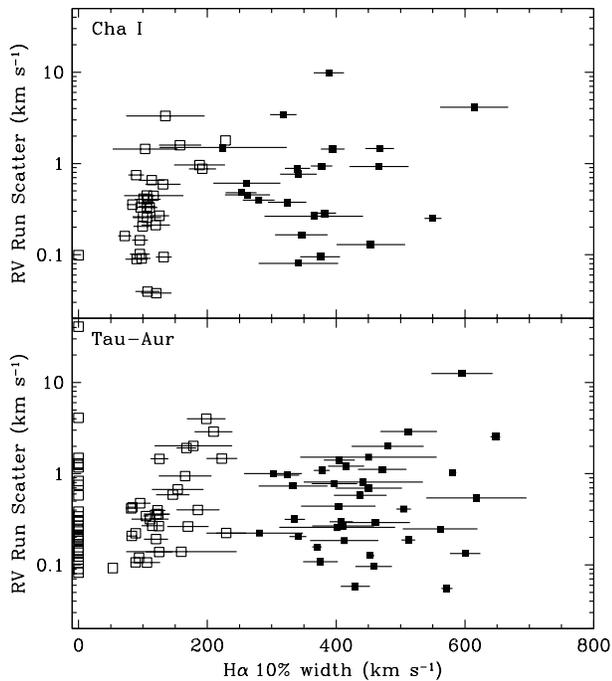}
\caption[Cha~I \& Tau-Aur: Radial velocity scatter within observing runs as a function of H$\alpha\,10\%$\,width]{Radial velocity scatter within the observing runs for members of Cha~I and Tau-Aur with single-peak broadening functions as a function of the accretion signature H$\alpha\,10\%$\,width. Accretors and non-accretors are denoted by solid and hollow symbols, respectively. There appears to be no significant difference in the radial velocity scatter within the observing runs between accretors and non-accretors.}
\label{fig:RVRunWSDShortvsHa10w}
\end{center}
\end{figure}

For some targets with heavily veiled spectra (e.g., DR\,Tau), this technique did not produce consistent radial velocity results. In these cases, we derived the radial velocities by fitting analytic rotational broadening functions to the computed broadening functions (see \S\ref{sec:BroadeningFunctions}), but corrected for veiling. This correction involved adding to Eq.~\ref{equ:BroadeningFunctionSpectraLambda} a linear component alongside the rotational broadening function to estimate and account for the contribution from veiling. Furthermore, we considered the overall potential influence of accretion and veiling on our radial velocity estimates by comparing both the estimated radial velocity measurement uncertainties and scatter within the observing runs against the accretion signature H$\alpha\,10\%$\,width (see \S\ref{sec:BinaryPopulations}). This comparison is illustrated in Fig.~\ref{fig:RVevsHa10w} \& \ref{fig:RVRunWSDShortvsHa10w}. There appears to be no significant difference in the radial velocity measurement uncertainties nor the radial velocity scatter within the observing runs between accretors and non-accretors in Cha~I and Tau-Aur.

The aggregated radial velocity results along with stellar characteristics such as spectral type, H$\alpha\,10\%$\,width, and $v\,\sin\,i$ for our sample are listed in Tables~\ref{tbl:SingleStars} \& \ref{tbl:Binaries}. The reported radial velocities for our targets incorporate the radial velocities and artificial rotational broadening of the templates. Additional information including brief comments and plots of several spectral lines for each object are provided in Fig.~\ref{fig:Cha__T55} through \ref{fig:Tau__Hubble4}. For a more detailed description of these figures, see \S\ref{sec:NotesIndividualSources}.

\section{Identifying Close Companions}
\label{sec:IdentifyingCompanions}

Our approach to identifying close companions consists of two basic parts. First, we checked the broadening functions of each star for multiple peaks suggestive of SB2s. Second, we examined the radial velocity measurements for variations consistent with SB1s.

\subsection{SB2 Candidates}
\label{sec:SB2}

We identified SB2 candidates as targets that have broadening functions with prominent multiple peaks that cannot be accounted for by non-companion sources (e.g., see V826\,Tau; Fig.~\ref{fig:Tau__V826TauA+B}). Plausible non-companion sources include sky or moon contamination, and contribution from a known resolved companion. We discuss these influences below.

An extra peak near the observer's rest frame could be attributed to possible contamination from moonlight or twilight sky. This occurs commonly in our data, since we did not correct for sky in our reduction (see \S\ref{sec:Observations}). To assess this contamination, we checked the angular distance of the target to the moon and the altitude of the sun below the horizon at the time the data were taken, and we inspected the raw data frames.

Additional peaks in the broadening function may be caused by light from nearby resolved companions. We collected information on known visual binaries by cross-referencing our target list with those of speckle and direct imaging surveys: for Cha~I, we used \citet{2008ApJ...683..844L}, and for Tau-Aur we checked \citet{1993AA...278..129L}, \citet{1993AJ....106.2005G}, \citet{1995ApJ...443..625S}, \citet{1997ApJ...490..353G}, and \citet{1998AA...331..977K}. To gauge the significance of the visual binaries on the broadening functions, we used the $\Delta K$ magnitudes and $K$-band flux ratios reported in the surveys to calculate corresponding $R$-band flux ratios. This calculation consists of the following eight steps:

\begin{enumerate}
\item Use the spectral type of the targets to estimate effective temperatures $T_{\rm eff}$ \citep{2004AJ....128.2316S}.
\item With $T_{\rm eff}$, compute luminosities $L$ from the models of \citet{1997MmSAI..68..807D}.
\item With $L$, compute bolometric magnitudes $M_{\rm bol} = 4.75 - 2.5 \log (L / L_{\sun})$.
\item Use the spectral type of the targets and the color-temperature relation of \citet{2004AJ....128.2316S} to compute the bolometric correction $BC$, and the $V$--$R$ and $V$--$K$ colors of the primary stars.
\item Combine the above to calculate the absolute $R$ and $K$ magnitudes of the primary stars.
\item Apply the reported $\Delta K$ magnitudes and $K$-band flux ratios of the systems to derive the absolute $K$ magnitudes of the companions.
\item Use these absolute $K$ magnitudes to estimate the corresponding absolute $R$ magnitudes by interpolating the color-temperature relation of \citet{2004AJ....128.2316S}.
\item Combine the derived absolute $R$ magnitudes of the primary stars and the companions to find the $R$-band flux ratios of the systems.
\end{enumerate}

Obviously, the above calculation relies on numerous assumptions, and the resulting estimated $R$-band flux ratios are thus only approximate. However, they suffice for our purpose of gauging whether a significant contribution to the broadening function from a resolved companion is likely.

Our ability to identify SB2 candidates from the broadening functions is limited by several factors. First, the contribution of companions to the broadening function must be sufficiently large to discriminate them from their primary stars. The SB2 candidates we could identify had a flux ratio lower bound of $\sim\!0.1$. Second, the radial velocity separation of component stars with similar $v~\sin~i$ must be large enough to distinguish each star in the broadening function. The minimum measurable velocity separation is equal to the sampling interval of $6$\,km\,s$^{-1}$; however, we find empirically the lower limit of the velocity separation for our identified SB2 candidates is $\sim\!10$\,km\,s$^{-1}$. Note, resolved companions are generally in long-period orbits, and thus, the radial velocity separation between the host and the companion is usually small, and the peaks in the broadening function from the pair would likely overlap. Moreover, overlapping contributions from two sources would produce an additional sharp central peak in the broadening function, thus a lack of this feature excludes some targets with only two wide peaks, e.g., T20 and HD\,283572, from being identified as SB2s due to bad model fitting. Third, the $v\,\sin\,i$ of the component stars with similar radial velocities must be large enough to distinguish each star in the broadening function. For example, see the broadening function of DF\,Tau in Fig.~\ref{fig:Tau__DFTauA+B}, the two component stars have similar radial velocities and could only be distinguished because of the significant difference in $v\,\sin\,i$ between them. Among applicable SB2 candidates, the smallest difference between component star $v\,\sin\,i$ is $\sim\!24$\,km\,s$^{-1}$.

We identified thirteen SB2s, four suspected SB2s, and four SB3s in our sample. Of the thirteen SB2s, five show significant acceleration based on their broadening functions whereas the other eight display no remarkable radial velocity changes. We re-labeled these latter objects as long-period SB2s. For seven out of the eight long-period SB2s, resolved companions are likely the source of the secondary peaks in their broadening functions, and thus, they are also visual binaries. The one remaining long-period SB2, RX\,J0443.4+1546, could have a binary separation of a few AU, which would be smaller than can be currently resolved by imaging techniques but large enough not to show a significant change in radial velocity over the time baseline of our observations. The sets of radial velocity measurements for all candidate and suspected SB2s, and SB3s are listed in Tables~\ref{tbl:RVSB2} \& \ref{tbl:RVSB3}.

\subsection{SB1 Candidates}
\label{sec:SB1}

We identified SB1 candidates as targets that have radial velocity scatter that cannot be accounted for by noise (e.g., see RX\,J0415.8+3100; Fig.~\ref{fig:Tau__RXJ0415.8+3100}). We consider not only measurement noise but also possible intrinsic noise from the star. Of particular relevance are apparent radial velocity changes due to spots, which effectively attenuate light from part of the star. Combined with rotation, this leads to apparent velocity changes on cycles connected to the rotation period of the star. To test the significance of the radial velocity changes of each target, we calculate several relevant statistics.

First, to evaluate the overall radial velocity scatter of each target, we compute the $\chi^2$ statistic of the radial velocity measurements with the null hypothesis of constant velocity using the relation
\\
\begin{equation}
\label{equ:chi2SB1}
\chi_{\rm RV}^2 = \sum_{i} \left( \frac{v_i - \langle v \rangle}{\sigma_i} \right) ^2
\end{equation}
\\
where $v_i$ and $\sigma_i$ are the radial velocities and measurement uncertainties of the target at epochs $t_i$, respectively, and $\langle v \rangle$ is the weighted mean of the radial velocity of the target. We would like to achieve a confidence level of $>\!95\%$ that no single star is mistakenly identified as an SB1 in our sample. Given that we have $\sim\!200$ targets, this condition requires a confidence level of $>\!99.97\%$ for correct identification of individual targets. For a typical target observed at four epochs, this implies that we require $\chi_{\rm RV}^2\!>\!19.1$ to identify a target as an SB1.

Second, to estimate systematic noise for each target, we consider the measurements for each observing run separately. Since each observing run is only a few days long, velocity changes of a few km\,s$^{-1}$ on this timescale are much more likely to be due to rotation than to orbits. Specifically, for each target, we calculate the systematic noise during each observing run,
\\
\begin{equation}
\label{equ:SystematicNoiseRun}
\sigma_{{\rm N},j} = \sqrt{\sigma_{{\rm S},j}^2 - \sigma_{{\rm E},j}^2}
\end{equation}
\\
where $\sigma_{{\rm S},j}$ is the weighted standard deviation of the radial velocities for observing run $j$, and $\sigma_{{\rm E},j}$ is the weighted mean of the radial velocity errors for observing run $j$. The weighted standard deviation is given by
\\
\begin{equation}
\label{equ:WeightedSigma}
\sigma_{{\rm S},j}^2 = \left( \frac{N'_j}{N'_j - 1} \right) \frac{ \sum_k w_k ( v_k - \langle v \rangle_j)^2}{\sum_k w_k}
\end{equation}
\\
where $v_k$ are the velocities at epochs $t_k$ from observing run $j$, $\langle v \rangle_j$ is the corresponding weighted mean of the velocities, $w_k \equiv 1/\sigma_k^2$ are the statistical weights based on measurement uncertainties at epochs $t_k$, and $N'_j$ is the normalized number of frames defined as
\\
\begin{equation}
\label{equ:NPrime}
N'_j \equiv N_j \frac{\langle w \rangle_j^2}{\langle w^2 \rangle_j}
\end{equation}
\\
where $N_j$ is the number of frames taken of the given target during observing run $j$. Note, in the case of all equal weights, Eq.~\ref{equ:WeightedSigma} reduces to the equation for simple standard deviation. Finally, we adopt as the systematic noise for a given target the weighted mean of the systematic noise for all the observing runs of the target, i.e., $\sigma_{\rm N} \equiv \sqrt{\langle \sigma_{{\rm N},j}^2 \rangle}$.

\begin{figure}
\begin{center}
\includegraphics[width=8cm]{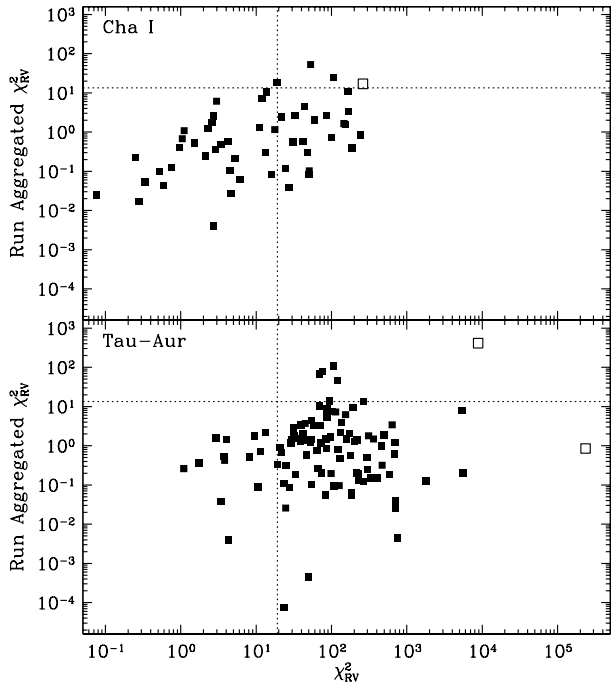}
\caption[Cha~I \& Tau-Aur: Run aggregated $\chi^2_{\rm RV}$ as a function of $\chi^2_{\rm RV}$]{Run aggregated $\chi^2_{\rm RV}$ for members of Cha~I and Tau-Aur with single-peak broadening functions as a function of $\chi^2_{\rm RV}$. Single stars and SB1 candidates are denoted by solid and hollow symbols, respectively. The vertical and horizontal dotted-lines are drawn at $\chi^2_{\rm RV} = 19.1$ and $13.4$, respectively, and represent the typical lower bound criteria of $\chi^2_{\rm RV}$ to identify a target as an SB1 in our survey. The single stars in the upper right quadrant have radial velocity variations likely due to star spots. The exceptional SB1 in Tau-Aur, RX~J0415.8+3100, has a run aggregated $\chi^2_{\rm RV}$ that is well below the threshold of $13.4$ because of the large daily RV variations from its short period companion.}
\label{fig:RVRunChi2vsRVChi2}
\end{center}
\end{figure}

Third, to re-evaluate the radial velocity scatter of each target with compensation for systematic noise, we add in quadrature the systematic noise $\sigma_{\rm N}$ of a given target to the radial velocity uncertainties $\sigma_i$ of the target, and re-evaluate the $\chi^2$ statistic (Eq.~\ref{equ:chi2SB1}) using the radial velocities and uncertainties aggregated by observing run, i.e., using $v_j$ and $\sigma_j$ the weighted means and standard errors of the targets during observing runs $j$. Analogous to the assessment for overall radial velocity scatter, for a typical target observed during two observing runs, we look for a run aggregated $\chi^2_{\rm RV}\!>\!13.4$ to identify a target as an SB1. An exception to this guideline applies to SB1s with periods shorter than a few days because their radial velocity oscillations are on the same timescales as the observing runs themselves. To illustrate our $\chi^2_{\rm RV}$ criteria visually, we plot in Fig.~\ref{fig:RVRunChi2vsRVChi2} the run aggregated $\chi^2_{\rm RV}$ for our targets as a function of $\chi^2_{\rm RV}$. Note, the single stars in the upper right quadrant of Fig.~\ref{fig:RVRunChi2vsRVChi2} have radial velocity variations likely due to star spots.

\begin{figure}
\begin{center}
\includegraphics[width=8cm]{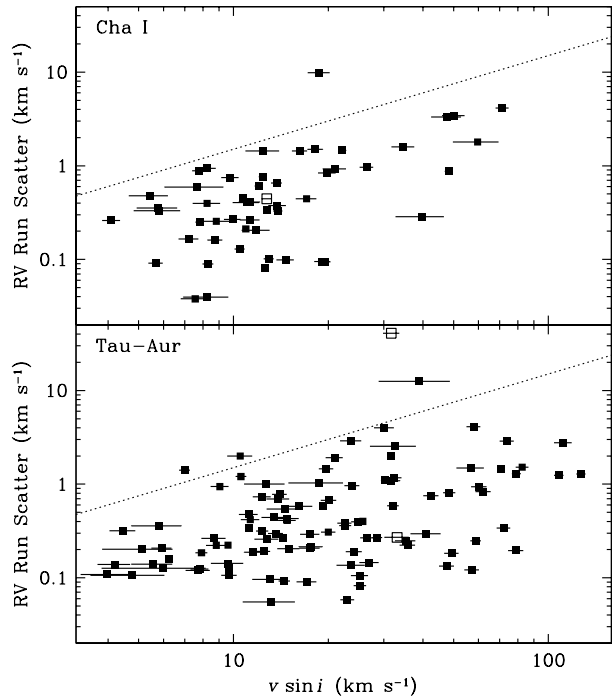}
\caption[Cha~I \& Tau-Aur: Radial velocity scatter within observing runs as a function of $v\,\sin\,i$]{Radial velocity scatter within the observing runs for members of Cha~I and Tau-Aur with single-peak broadening functions as a function of projected rotational velocity $v\,\sin\,i$. Single stars and SB1 candidates are denoted by solid and hollow symbols, respectively. The dotted-line is drawn at $15\%$ of $v\,\sin\,i$ and represents the approximate upper bound of the radial velocity scatter. The single stars above this limit have large measurement uncertainties due to poor spectra, and the SB1s below the limit still show individual epoch radial velocity variability that exceeds the threshold.}
\label{fig:RVRunWSDShortvsVsini}
\end{center}
\end{figure}

Fourth, we estimate the extent of the radial velocity fluctuations induced by star spots (see also \S\ref{sec:RadialVelocityScatter}). Since the influence of star spots on observed radial velocity is related to the rotational velocity of the star, we show in Fig.~\ref{fig:RVRunWSDShortvsVsini} the radial velocity scatter within the observing runs for targets with single-peak broadening functions in our sample as a function of projected rotational velocity $v\,\sin\,i$. For both Cha~I and Tau-Aur, the upper bound of the short-period velocity scatter is $\sim\!15\%$ of $v\,\sin\,i$. Single-line targets with radial velocity scatter larger than this value are likely SB1s.

We identified three SB1s and four suspected SB1s in Cha~I and Tau-Aur. The sets of radial velocity measurements for these objects are listed in Table~\ref{tbl:RVSB1}.

\subsection{Detection Limits}
\label{sec:DetectionLimits}

Both companion mass $M_2$ and orbital period $P$ contribute to the companion detection limits of radial velocity searches. Furthermore, for each mass and period combination, detection is also affected by orbital elements such as inclination angle $i$ and orbital phase $\phi$. Although eccentricity might facilitate or hinder the detection of a companion in an individual case, the effect is canceled out statistically \citep{1992ApJ...396..178F}.

\begin{figure}
\begin{center}
\includegraphics[width=8cm]{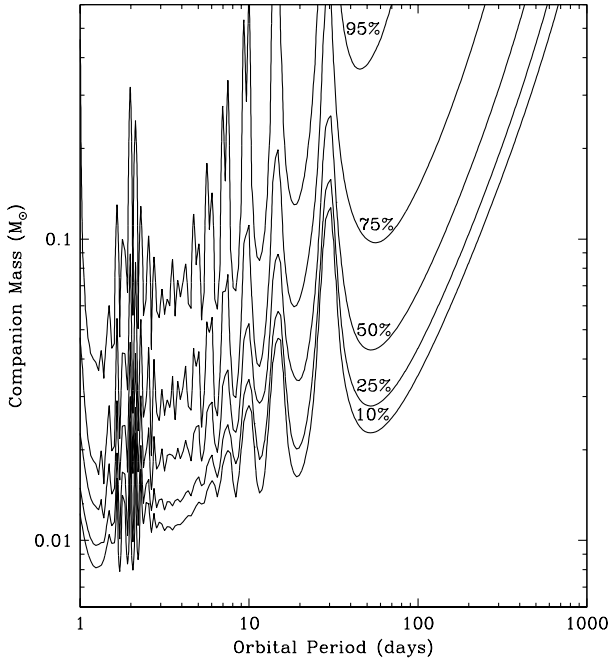}
\caption[Detection probabilities in ($M_2$, $P$) space]{Detection probabilities for a given companion mass and orbital period for typical T~Tauri stars in our sample ($0.6\,M_\sun$; $v\,\sin\,i = 15$\,km\,s$^{-1}$) observed four times at epochs $t = 0$, $0.2$, $2$ and $30$\,days and assuming a measurement uncertainty of $0.1$\,km\,s$^{-1}$. At each point in the ($M_2$, $P$) space, the inclination angle $i$ and the orbital phase $\phi$ were incremented over their full range. For each combination of orbital elements, $1000$ radial velocity curves were generated, and Gaussian noise with an rms value of $0.34$\,km\,s$^{-1}$ was added to each curve in Monte Carlo fashion to represent the typical intrinsic noise in our actual sample. The detection criteria used are the $\chi^2$ and minimum velocity scatter conditions described in \S\ref{sec:SB1}.}
\label{fig:MvsP}
\end{center}
\end{figure}

To derive the probability of detection for a given point in the $(M_2,\,P)$ parameter space, we simulate binaries in circular orbits with systematic combinations of $i$ selected incrementally from a $\cos i$ distribution, and $\phi$ selected incrementally from a uniform distribution spanning their full ranges. For each combination of orbital elements, we generate $1000$ sinusoidal radial velocity curves, and add random errors drawn from a normal distribution centered on zero and with a chosen rms value $\sigma_{\rm noise} = 0.34$\,km\,s$^{-1}$ which is equal to the typical intrinsic noise in our actual stars. Furthermore, like our typical observations of actual stars, we sample each simulated radial velocity curve four times (e.g., at epochs $t = 0, 0.2, 2$ and $30$\,days), and we adopt as measurement uncertainty the approximate median value of $0.10$\,km\,s$^{-1}$ from our actual data. The $\chi^2$ and minimum velocity scatter conditions described in \S\ref{sec:SB1} are used as detection criteria for the simulations. For a given $(M_2,\,P)$ pair, we define the detection probability as the fraction of successful detections over the number of trials for that point. The result of these simulations in the $(M_2,\,P)$ plane for a typical T~Tauri star in our sample ($0.6$\,$M_{\sun}$; $v\,\sin\,i = 15$\,km\,s$^{-1}$) is shown in Fig.~\ref{fig:MvsP} as iso-probability curves of detection. We see for example that we have a 75\% probability of detecting a companion of mass $0.1$\,$M_{\sun}$ with a period of $60$\,days if the typical $0.6$\,$M_{\sun}$ primary star has been observed four times with our $30$\,day baseline. From Fig.~\ref{fig:MvsP}, we also deduce that detection biases vary significantly for orbital periods $\lesssim\!50$\,days. This is due to aliasing from the sampling times.

The selection criteria for our study contain biases which affect a determination of the multiplicity rate. Specifically, a magnitude-limited sample favors the inclusion of SB2s \citep{1976ApJ...210..392B}. This is especially the case for young stars where the dependence of luminosity on mass is comparatively gradual, i.e., $L \propto M^{\alpha}$, with $\alpha\!\sim\!2$ for pre-main sequence stars rather than $\alpha\!\sim\!4$ for stars on the main sequence.

\section{Results and Discussion}
\label{sec:Results}

In our entire sample of $212$ members of Cha~I and Tau-Aur, we have identified a total of $12$ systems that show acceleration due to close companions (three SB1s, five SB2s and four SB3s) and $7$ systems for which we suspect this is the case. We will discuss these systems individually in \S\ref{sec:NotesIndividualSources}. To reduce sampling bias in our close companion statistics, we trim our sample to include only targets that have observations spanning multiple observing runs ($>\!25$\,days). This ensures that all targets included for statistics have observations with time baselines sufficiently long to capture the radial velocity scatter from most SB1s. With this condition, there are $11$ systems with close companions and $6$ systems with suspected close companions from a final statistical sample of $174$ systems. Of the $11$ systems with close companions, three are SB1s, five are SB2s, and three are SB3s.

Below, we will first look for differences between our two regions, with primary mass and accretion, and with stars in the field. Second, we will look at possible non-companion contributions to radial velocity variations. Third, we will discuss $13$ somewhat puzzling stars that have mean velocities that deviate from the velocity of the cluster, yet appear to be members from all other indicators.

\subsection{Comparison of Binary Populations}
\label{sec:BinaryPopulations}

We first compare the multiplicity fraction (MF) in Cha~I and in Tau-Aur with each other. The Chamaeleon~I and Taurus-Auriga star-forming regions are both T associations of similar age. Previous surveys using speckle and direct imaging show a similar multiplicity fraction for both regions \citep{2008ApJ...683..844L}. The multiplicity fractions ${\rm MF} = ({\rm SB1} + {\rm SB2} + \cdots)/({\rm Singles} + {\rm SB1} + {\rm SB2} + \cdots)$ we find, using high-resolution spectroscopy, in our sample are $0.07_{-0.03}^{+0.05}$ ($4/61$) for Cha~I, and $0.06_{-0.02}^{+0.03}$ $(7/113)$ for Tau-Aur.

Initially, to compare the multiplicity fractions, we used the binomial distribution comparison method described in \citet{2006ApJ...652.1572B}. However, that method could not reject the hypothesis that the close multiplicity probabilities of Cha I and Tau-Aur are equal, with any significance. Therefore, we instead compare the multiplicity fractions formally from first principles as follows. For random variables $Q$ and $R$, the likelihood $P$ that $Q$ is greater than $R$ is given by the formula
\\
\begin{equation}
\label{equ:ProbabilityQgtrR}
P[Q > R] = \int_{-\infty}^{+\infty} \left( f_Q(x) \int_{-\infty}^{x} f_R(y)\,dy \right) dx
\end{equation}
\\
where $f_Q$ and $f_R$ are the probability distribution functions of $Q$ and $R$, respectively. In our case, the random variables are the underlying multiplicity fractions of stellar populations. By sampling a population, we will find $k$ binaries out of $n$ targets with the outcome governed by the binomial distribution $B(n, p)$ where $p$ is the true multiplicity fraction of the population. Consequently, the probability distribution function of $p$ given $k$ observed binaries out of $n$ targets is
\\
\begin{equation}
\label{equ:BinomialDistributionPDF}
f(p | k, n) = (n + 1) \left(\begin{array}{c}
n\\
k\end{array} \right)
p^k (1 - p)^{n - k}
\end{equation}

Applying this function to Eq.~\ref{equ:ProbabilityQgtrR}, we get the likelihood that the multiplicity fraction of population $Q$ is greater than that of population $R$ is
\\
\begin{equation}
P[p_Q > p_R] = \int_{0}^{1} \left( f(p_Q | k_Q, n_Q) \int_{0}^{p_Q} f(p_R | k_R, n_R)\,dp_R \right) dp_Q
\end{equation}

By substituting the MF results from Cha~I and Tau-Aur into this formula, we find the likelihood that the MF of Cha~I is greater than that of Tau-Aur is only $58\%$. Therefore, the two regions have similar MF values. Subsequently, we will use the combined results when making comparisons with those of other populations, and between various sub-populations in our sample determined by physical characteristics.

To investigate the change in close multiplicity as a function of primary mass, we divide our sample into two bins of nearly equal number and calculate the MF in each bin. We find multiplicity fractions of $0.06_{-0.02}^{+0.03}$ $(6/98)$ for F--K spectral type targets, and $0.07_{-0.03}^{+0.04}$ $(5/76)$ for M spectral type targets. There is no clear dependence in the spectroscopic regime of multiplicity fraction on primary mass: the likelihood that the multiplicity fraction of the F--K type stars is greater than that of M-type stars is only $44\%$. This is in stark contrast to the results of imaging surveys \citep[e.g.,][]{2008ApJ...683..844L} which show, for Chamaeleon~I, Taurus, Upper~Scorpius~A and field stars, a marked increase in wide companions with increasing primary mass.

To address the question of whether close companions are responsible for the observed difference between classical and weak-lined T~Tauri stars from \S\ref{sec:Introduction}, we measure the full width of H$\alpha$ at $10\%$ of the peak (hereafter, H$\alpha$\,$10\%$\,width) of our targets to differentiate accretors and non-accretors. As suggested by previous spectroscopic studies \citep[e.g.,][]{2003ApJ...592..282J}, we classify accretors or CTTSs as stars with H$\alpha\,10\%$\,widths larger than $200$\,km\,s$^{-1}$ after subtracting rotational broadening. We measured H$\alpha\,10\%$\,widths previously for a subsample of our targets \citep{2009ApJ...695.1648N}, and apply the same technique here. Five targets in our statistical sample did not have reliable measurements for H$\alpha$\,$10\%$\,width. Of the remaining $169$ targets, we find a close binary fraction of $0.08_{-0.03}^{+0.04}$ $(8/104)$ and $0.05_{-0.02}^{+0.04}$ $(3/65)$ for non-accretors and accretors, respectively. The populations of WTTS and CTTS have statistically similar close binary fractions: the likelihood from our data that the true multiplicity fraction of WTTSs is greater than that of CTTSs is $75\%$ which is approximately $1\sigma$. Therefore, we cannot say confidently that close companions are the main source of the attenuated accretion in weak-lined T~Tauri stars.

Finally, we compare our multiplicity fractions with values for field stars from the G dwarf survey of \citet{1991AA...248..485D}, the M dwarf surveys of \citet{1992ApJ...396..178F}, and the nearby solar-type ($\sim$F6--K3) dwarf survey of \citet{2010ApJS..190....1R}. Comparison between our sample and those of the field population is not straightforward because the surveys have different time baselines and companion mass sensitivities. Specifically, we observed each of our stars typically at four epochs with baselines of up to several months whereas the aforementioned field dwarf surveys generally had more than a dozen observations per star spanning a four year interval. Furthermore, the field dwarf surveys are sensitive to companions having masses less than the primary mass, extending to the hydrogen burning limit, $0.08$\,$M_{\sun}$ whereas our survey should detect $\!\sim\!50\%$ of substellar companions with masses of $0.05$\,$M_{\sun}$\,for orbital periods $\!\lesssim\!100$\,days.

A meaningful comparison of multiplicity fractions requires that we correct for the differences in the detection completeness of the surveys. To accomplish this, we restrict our statistical comparison to common orbital periods and companion mass ranges. Since our stars are typically observed four times, to determine the likely range of companion masses and orbital periods for our survey we make use of probability theory.

Let ${\mathcal M}(m,p)$ be the a priori distribution of multiplicity rates for our star population in the companion mass and orbital period $(M_2,\,P)$ space. In the same space, let ${\mathcal D}(m,p)$ be the detection success rates of our observing scheme, i.e., the results of our simulations in \S\ref{sec:DetectionLimits}. Using Bayes' theorem \citep{1763RSTL..53..370B}, from the prior distribution ${\mathcal M}(m,p)$, and the likelihood function ${\mathcal D}(m,p)$, we calculate the probability ${\mathcal P}$ that a detected companion is in the given intervals $[M_{2,{\rm min}},\,M_{2,{\rm max}}]$ and $[P_{\rm min},\,P_{\rm max}]$,
\\
\begin{equation}
\label{equ:BayesTheorem}
{\mathcal P} = \frac{\sum_{m = M_{2,{\rm min}}}^{M_{2,{\rm max}}} \sum_{p = P_{\rm min}}^{P_{\rm max}} {\mathcal M}(m,p) {\mathcal D}(m,p)}{\sum_{m} \sum_{p} {\mathcal M}(m,p) {\mathcal D}(m,p)}
\end{equation}

\begin{figure}
\begin{center}
\includegraphics[width=8cm]{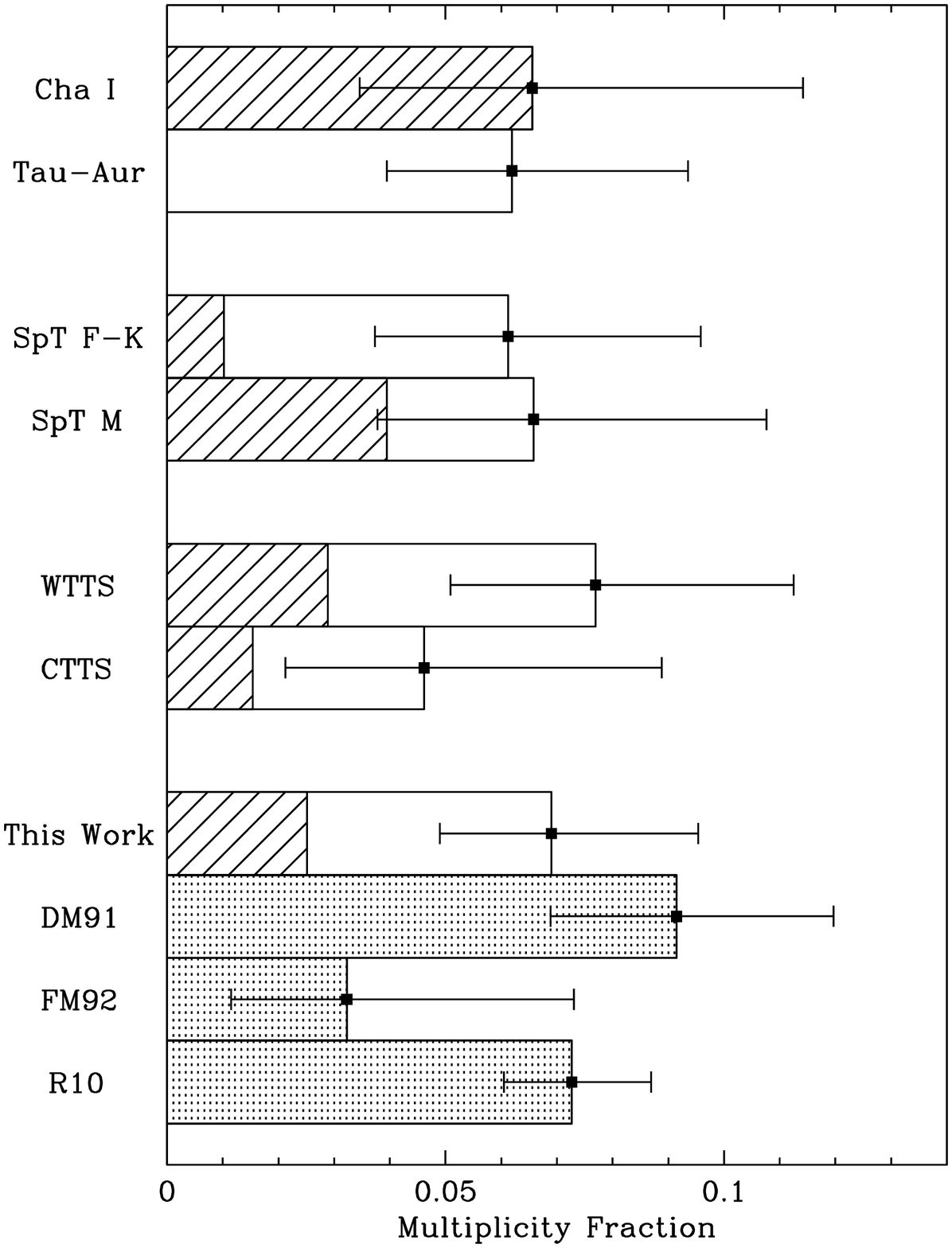}
\caption[Cha~I \& Tau-Aur: Comparisons of multiplicity fractions for various sub-populations and field dwarfs]{Comparisons of the multiplicity fractions of various sub-populations in Cha~I and Tau-Aur, and between this work and those of \citet{1991AA...248..485D} (DM91), \citet{1992ApJ...396..178F} (FM92), and \citet{2010ApJS..190....1R} (R10). The contribution from the Cha~I sample is denoted by striped areas. The unshaded areas represent the contribution from the Tau-Aur sample. The results from the field dwarf surveys of DM91, FM92 and R10 are stippled, and count only companions with stellar masses and orbital periods between $1$ and $200$\,days which is comparable to the detection coverage of this work. For the comparison with the field dwarf surveys, the multiplicity fraction for this work has been corrected for detection sensitivity.}
\label{fig:HistogramMF}
\end{center}
\end{figure}

Since we have no prior knowledge about the multiplicity rates, we use the most ignorant prior distribution ${\mathcal M}(m,p) = 1$, and assume companions can have masses up to that of the primary star. Furthermore, we do not expect orbital periods of our binary systems to be much less than a day (see Appendix \ref{sec:ShortestPeriods}). From our simulation results, we estimate that $95\%$ of our detected companions have stellar masses with orbital periods up to $\sim\!200$\,days. In this range, the mean detection probability is $0.87$. Therefore, we estimate a corrected MF of $(11/174) \times 0.95 \div 0.87 = 0.07_{-0.02}^{+0.03}$ in the companion mass and orbital period intervals $[0.08\,M_{\sun},\,0.6\,M_{\sun}]$ and $[1{\rm\,d},\,200{\rm\,d}]$, respectively. In the same range, counting the companion detections of \citet{1991AA...248..485D} and \citet{1992ApJ...396..178F}, we find an MF of $0.09_{-0.02}^{+0.03}$ and $0.03_{-0.02}^{+0.04}$ for field G and M dwarfs, respectively. Furthermore, for the nearby solar-type dwarfs of \citet{2010ApJS..190....1R}, we find an MF of $0.073_{-0.012}^{+0.014}$. Thus, we do not find a significant difference with the field. The results are illustrated graphically in Fig.~\ref{fig:HistogramMF}.

\subsection{Mass Dependence}
\label{sec:MassDependence}

The observed fraction of T~Tauri stars with spectroscopic companions in our sample does not depend on stellar mass. This result contrasts those of imaging multiplicity surveys \citep[e.g.,][]{2007ApJ...662..413K,2008ApJ...683..844L} where there is an observed trend of increasing wide binary fraction with increasing primary mass. One possibility is that primary mass is important for binaries with wide separations because of dynamical interactions within the cluster. Dynamical interactions can only eject companions if there is sufficient kinetic energy. Moreover, the kinetic energy of a passing star must be greater than the gravitational binding energy of the binary for break up to occur. Binding energy is proportional to primary mass, so higher mass stars are more likely than their lower mass counterparts to hold onto their companions. However, companions in tight orbits are so strongly bound to their primary star that disruption from dynamical interaction is unlikely regardless of the primary mass.

Primary mass becomes important at binary separations where the total energy of the binary system is comparable to the kinetic energy of a passing star. By equating the energies, we get the following relation for this transitional binary separation distance $r_\ast$
\\
\begin{equation}
\label{equ:BinarySeparation}
r_\ast = \frac{M_1}{M_3} \frac{G M_2}{v^2}
\end{equation}
\\
where $M_1$, $M_2$ and $M_3$ are the masses of the primary, secondary and passing star, and $v$ is the relative velocity of the encounter. We can estimate the typical encounter velocity from the velocity dispersion of the clusters. From our single stars, we estimate that the radial velocity dispersions $\sigma_{\rm RV}$ of Chamaeleon~I and Taurus-Auriga are $1.7{\rm\,km\,s}^{-1}$ and $3.6{\rm\,km\,s}^{-1}$, respectively. If we assume three-dimensional symmetry, then the average velocity between stars is $(4\, / \sqrt{\pi})\,\sigma_{\rm RV}$ which amounts to $\sim\!3.8{\rm\,km\,s}^{-1}$ and $\sim\!8.1{\rm\,km\,s}^{-1}$ for Cha~I and Tau-Aur, respectively. Applying these values to Eq.~\ref{equ:BinarySeparation} and assuming $M_1 / M_3\!\sim\!1$, we find primary mass becomes important to ejection at binary separations of $\sim\!63\,(M_2 / M_{\sun})$\,AU in Chamaeleon~I, and $\sim\!14\,(M_2 / M_{\sun})$\,AU in Taurus-Auriga. These separations are in the regime of imaging surveys.

Although scattering events are possible for binaries with separations of tens of AU in Cha~I and Tau-Aur, the likelihood of an encounter depends on the stellar densities of the star-forming regions. Moreover, the interaction timescale $\tau$ is roughly equal to $(\sigma n v)^{-1}$ where $\sigma$ is the encounter cross-sectional area $(\pi r_\ast^2)$, $n$ is the stellar density, and $v$ is the average relative velocity between stars. For a typical wide binary in Cha~I and Tau-Aur, at current densities of $\lesssim\!100{\rm\,pc}^{-3}$ \citep{1999ApJ...526..336O,2006AA...448..655J}, the interaction timescale is of order a few Gyr or longer. The interaction timescale only becomes comparable to the current cluster age for binaries with separations of $\gtrsim\!10^4$\,AU. Visual binaries with separations much less than this show a mass dependence. Therefore, if dynamical interaction is a major source of the mass dependence of wide binaries, then those stars must have emerged from subgroups with stellar densities several orders of magnitude higher than their current sparse environment.

There is observational evidence that stellar density may not solely influence the frequency of wide binaries. The wide binary frequency in the periphery of the ONC is similar to that in the core, where stellar densities are two orders of magnitude higher \citep[]{2006AA...458..461K}. While this trend is not observed by \citet{2007AJ....134.2272R}, both wide binary frequencies in the ONC are observed to be several times lower than those of Chamaeleon~I and Taurus-Auriga. If the initial binary fractions in these regions were similar, we would expect a much higher number of binaries in the ONC periphery than observed. Aside from the dense subgroup scenario mentioned above, one possible explanation is that the initial binary frequency differs significantly between star-forming regions. This suggests that the binary formation rate is influenced by environmental conditions.

\subsection{Radial Velocity Scatter}
\label{sec:RadialVelocityScatter}

One obvious source of stellar radial velocity fluctuations over time is a close companion. However, close companions are not the only plausible explanation for oscillatory radial velocity behaviour. Periodic variations in radial velocity, especially in young stars using measurements at optical wavelengths, can be caused by star spots \citep[e.g., see][]{2008ApJ...678..472H, 2008AA...489L...9H}. To gauge the maximum amplitude of this effect, let us consider a star spot on the equator of a star observed edge-on. For simplicity, we will model the star as a rigid unit sphere, and the star spot as a black spherical cap. A spherical cap is the region of a sphere which is cut off by a plane. We will ignore limb darkening, and assume the face of the star is a uniformly illuminated stellar disk. From this model, with the stellar rotation about the y-axis, and the line of sight to the observer along the z-axis, the observed net radial velocity shift caused by the star spot is
\\
\begin{equation}
\label{equ:StarSpotRV}
\Delta v = \frac{ \int \!\!\! \int_T v_0 x \,dy\,dx } { \int \!\!\! \int_S \,dy\,dx - \int \!\!\! \int_T \,dy\,dx}
\end{equation}
\\
where $S$ is the region of the stellar disk (a unit circle), $T$ is the region of the star spot projected onto the stellar disk, and $v_0$ is the linear rotational velocity of the star at the equator. Here, the numerator represents the radial velocity contribution integrated over the area of the star spot, and the denominator accounts for the observed surface area of the star, and effectively normalizes the radial velocity result.

Evaluating Eq.~\ref{equ:StarSpotRV} over all rotational phases, we find that a star spot covering $10\%$ of the stellar surface can produce a theoretical maximum RV oscillation semi-amplitude of $\sim\!0.20\,v_0$. This amplitude is similar to the RV scatter upper bound of $15\%~v\,\sin\,i$ observed in our sample. However, real star spots do not affect all spectral features equally nor do they have identical impact at all optical depths. A technique such as line bisector analysis, which probes radial velocity bias at different depths, is required to properly assess the contribution of star spots on variations in radial velocity.

\begin{figure}
\begin{center}
\includegraphics[width=8cm]{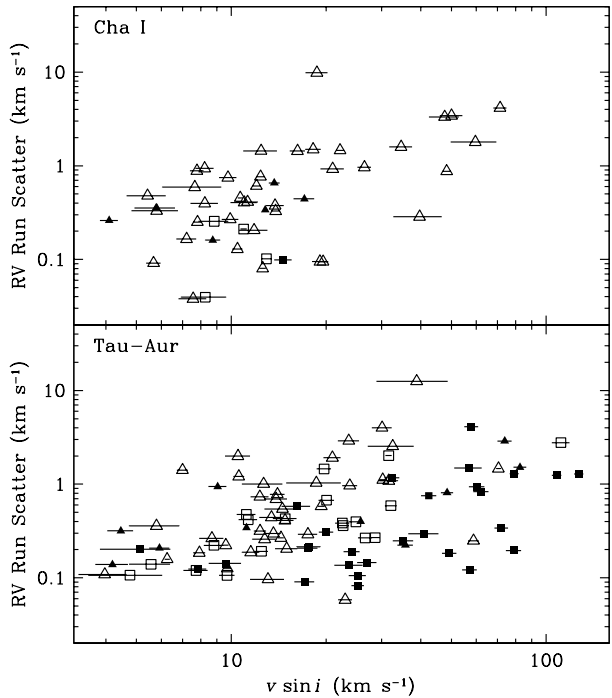}
\caption[Cha~I \& Tau-Aur: Radial velocity scatter within observing runs of single non-accretors as a function of $v\,\sin\,i$]{Radial velocity scatter within the observing runs for single non-accretors in Cha~I and Tau-Aur as a function of projected rotational velocity $v\,\sin\,i$. The stars have been grouped by H$\alpha$ equivalent width. Hollow triangles represent stars with H$\alpha$ EW $<\!-3$. Solid triangles indicate stars with H$\alpha$ EW in the range of $[-3, -1.5)$. Hollow squares represent stars with H$\alpha$ EW in the range of $[-1.5, 0)$. Solid squares indicate stars with H$\alpha$ EW $\geq 0$. There is an increase in radial velocity scatter with $v\,\sin\,i$. However, the groups by H$\alpha$ EW are not well separated from each other.}
\label{fig:RVRunWSDShortvsVsiniAndHaEW}
\end{center}
\end{figure}

The net effect of star spots on radial velocity measurements relies on the asymmetry that the spots produce on the stellar surface. In terms of contribution to radial velocity bias, multiple star spots can partially negate one another at different rotational phases. Therefore, since the total number of star spots, not their effective asymmetry, is connected to stellar activity, we do not necessarily expect a correlation between activity and radial velocity scatter. In Fig.~\ref{fig:RVRunWSDShortvsVsiniAndHaEW}, we show the radial velocity scatter over time for single non-accretors grouped by H$\alpha$ equivalent width, an activity indicator, in our sample and as a function of $v\,\sin\,i$. As expected, the upper bound of RV scatter increases with $v\,\sin\,i$. More importantly, there is no clear separation of stars by H$\alpha$ equivalent width. Thus, while spot-induced RV variability obviously requires some activity, it is perhaps not surprising that the level of activity does not correlate well with the amount of variability.

\subsection{Radial Velocity Outliers}
\label{sec:RVOutliers}

The targets in our original sample were identified as either members of Cha~I by \citet{2004ApJ...602..816L}, or members of Tau-Aur by \citet{1993AA...278..129L}, \citet{1993AJ....106.2005G}, \citet{1995ApJ...443..625S}, \citet{1998AA...331..977K}, \citet{2002ApJ...580..317B} and \citet{2004ApJ...617.1216L} based on a combination of one or more indicators such as Li-$\lambda$6708 absorption, reddening, emission lines, and IR excess emission. However, we find among the $151$ single stars in our statistical sample that five stars in Cha~I, and eight stars in Tau-Aur have overall radial velocities that deviate substantially from the velocities of their associated clusters, i.e., the weighted mean velocities of these targets are Tukey outliers with respect to the rest of the sample. Moreover, for identifying these outliers, we adopted the median cluster radial velocity of $15.3$\,km\,s$^{-1}$, and limits of $12.9$ and $17.2$\,km\,s$^{-1}$ for Cha~I. For Tau-Aur, we used a median radial velocity of $16.2$\,km\,s$^{-1}$, and limits of $10.3$ and $22.3$\,km\,s$^{-1}$. Stars with radial velocities that exceed the outlined ranges are indicated in Table~\ref{tbl:SingleStars}. We examine several possible sources for these abnormal mean radial velocities.

\begin{figure}
\begin{center}
\includegraphics[width=8cm]{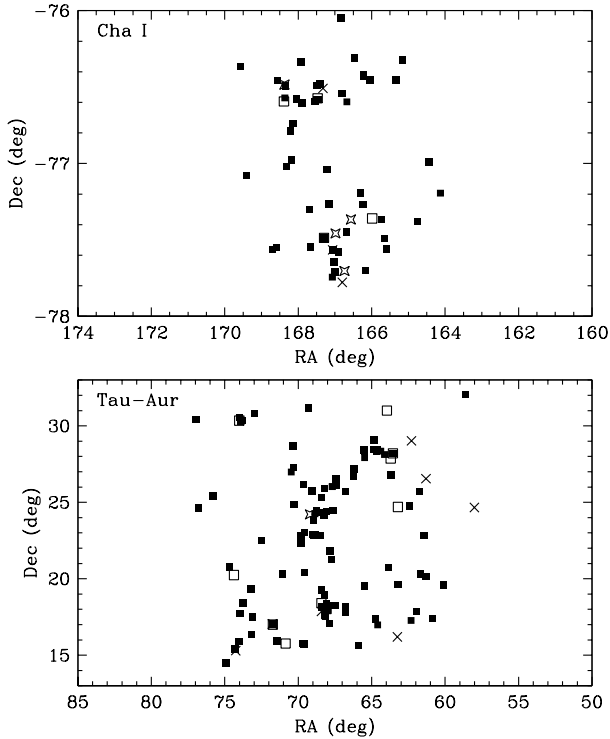}
\caption[Cha~I \& Tau-Aur: Spatial distribution of the sample]{The spatial distribution of targets in our sample of Cha~I and Tau-Aur. The solid symbols represent single stars that have mean radial velocities consistent with their respective clusters. Objects with mean radial velocities that are outliers with respect to their clusters are denoted by cross symbols for those moving faster away from us, and starred symbols for those moving more toward us. Spectroscopic binaries are plotted as hollow symbols. The distribution of radial velocity outliers shows no particular pattern.}
\label{fig:RADecRV}
\end{center}
\end{figure}

First, to see if localized phenomena in the star-forming regions are responsible for the radial velocity anomalies, we examine the positions of the radial velocity outliers on the sky. The projected spatial distributions of Cha~I and Tau-Aur are shown in Fig.~\ref{fig:RADecRV}. From the plots, we see no pattern in the spatial distribution of the radial velocity outliers, i.e., the radial velocity deviations are not confined to a particular area within the star-forming regions. Hence, it is unlikely that the radial velocity outliers in Cha~I and Tau-Aur are the result of peculiarities in specific parts of the clusters.

\begin{figure}
\begin{center}
\includegraphics[width=8cm]{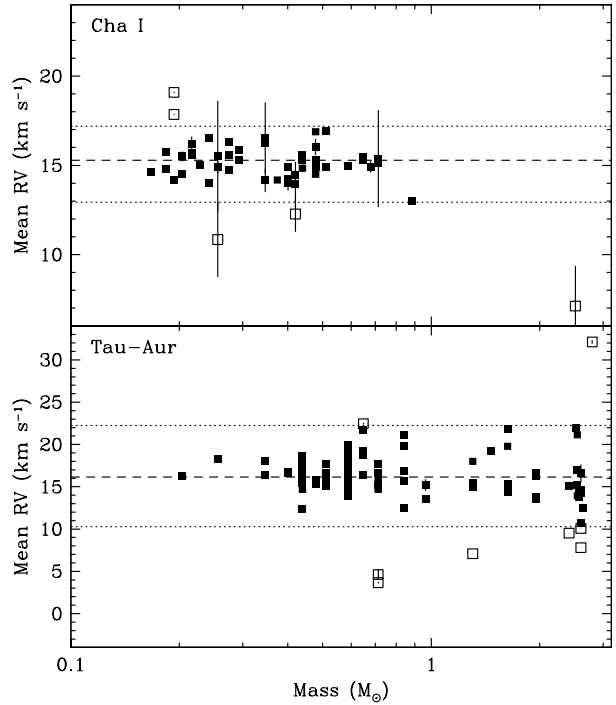}
\caption[Cha~I \& Tau-Aur: Mean radial velocity as a function of stellar mass]{Mean radial velocity as a function of stellar mass derived from the models of \citet{1997MmSAI..68..807D} for our sample from Cha~I and Tau-Aur $(\sim\!2$\,Myr old). The solid symbols represent single stars that have mean radial velocities consistent with their respective clusters. Objects with mean radial velocities that are outliers with respect to their clusters are denoted by hollow symbols. The dashed-lines show the median radial velocity of the single stars. The Tukey outlier threshold is delineated by the dotted-lines. There appears to be no mass preference for radial velocity outliers.}
\label{fig:RVvsM}
\end{center}
\end{figure}

Next, we examine the possibility that the targets with unusual mean radial velocity are the result of ejection from past multiple systems. If this is the case, we would expect the radial velocity outliers to have masses generally lower than the rest of the sample. The weighted mean radial velocity as a function of stellar mass derived from the models of \citet{1997MmSAI..68..807D} for our $\sim\!2$\,Myr old sample is shown in Fig.~\ref{fig:RVvsM}. We see from the plots that the targets with outlying radial velocities have masses spanning the full range of the sample. In addition, we expect that ejected stars from previous multiple systems should show little or no accretion because we do not expect that ejected stars would carry away with them a substantial portion of the material surrounding the original system. However, we find that $3$ out of the $13$ radial velocity outliers are CTTSs. Therefore, from the accretion and mass distribution findings, we conclude that ejection is not likely the main source of radial velocity outliers in our sample.

Finally, we check if the anomalous radial velocities could be caused by unseen companions. The lower bound of flux ratio for our identified SB2 candidates was $\sim\!0.1$. Using the models of \citet{1997MmSAI..68..807D} and assuming a stellar age of $2$\,Myr, if our lowest mass RV outlier $(0.19\,M_{\sun})$ had a companion of mass $\lesssim\!0.08\,M_{\sun}$, we would not detect it as an SB2. This is also the case if we use instead the models of \citet{1998AA...337..403B}. Furthermore, our SB1 detection success rate for such a binary system from Monte Carlo simulations assuming a circular orbital period $>\!1$\,yr is less than $2\%$. For a one-year period, the primary star would have an expected semi-amplitude of $\sim\!5$\,km\,s$^{-1}$, which is consistent with the radial velocity differences we observe of the low mass outliers in our sample. Similarly, if our highest mass RV outlier $(2.8\,M_{\sun})$ had a $2\,M_{\sun}$ companion in a circular orbit with a period of $>\!1000$\,days, we would not likely identify it as a binary. For a $1000$-day period, the primary star would have a semi-amplitude of up to $\sim\!15$\,km\,s$^{-1}$ which is consistent with the radial velocity deviations we observe in the high mass outliers. The results for the rest of the radial velocity outliers are similar. Thus, some of the single stars with an overall radial velocity inconsistent with their cluster velocity may be long period SB1s, i.e., they would not affect our comparisons with, e.g., the field population.

\section{Notes on individual sources}
\label{sec:NotesIndividualSources}

For each object in our survey, we provide brief comments and plots in Fig.~\ref{fig:Cha__T55} through \ref{fig:Tau__Hubble4}. Each figure consists of four main sections as follows:

\begin{enumerate}
\item In the upper left, basic data and measurements are listed: these are the star-forming region, object name, coordinates, spectral type, estimated S/N at H$\alpha$, H$\alpha$ 10\% width, $v\,\sin\,i$, \ion{Ca}{2}-$\lambda$8662 equivalent width, and {\it Spitzer Space Telescope} [3.6] - [8.0] IRAC color magnitude  \citep{Fazio:2004p4374}.

\item In the upper right, radial velocity curves are plotted. The error bars are the weighted standard deviation of the radial velocity estimates over all echelle orders for a given epoch. The dotted-line is the weighted mean of the radial velocities. For SB2s and SB3s, the radial velocities of the secondary and tertiary objects are plotted as hollow squares and diagonal crosses, respectively.

\item In the lower panels, there are five columns of plots with each row representing a different observed epoch: from left to right, these are the broadening function (see \S\ref{sec:BroadeningFunctions}), H$\alpha$, Li-$\lambda$6708, \ion{Ca}{2}-$\lambda$8662, and \ion{Mg}{1}-$\lambda$8807. The dotted-line in the broadening function shows the observer's rest frame velocity, and is used primarily to check for moonlight and twilight sky contamination. The H$\alpha$ and \ion{Ca}{2}-$\lambda$8662 plots are provided to gauge activity. Stellar youth is indicated by Li-$\lambda$6708 absorption. The \ion{Mg}{1}-$\lambda$8807 line is shown because it should be present in stars spanning the spectral type range of our sample, and shows graphically the overall quality of the data for that epoch.

\item In the caption, we outline our reasons for classifying the object as with or without spectroscopic companions, and provide a short discussion on the object. This discussion makes note of, but is not limited to, possible contaminations from moonlight, twilight sky, or nearby visual companions, and support for cluster membership for radial velocity outliers and objects without clear Li-$\lambda$6708 absorption.

\end{enumerate}

Among our complete sample, we found three SB1s, five (short-period) SB2s, and four SB3s. We will discuss individual sources grouped in reverse order below.

\subsection{SB3: Triple-lined Spectroscopic Binaries}
\label{sec:NotesSB3}

\paragraph{T55 (Ass\,Cha\,T\,2-55, CHXR 61)} (Fig.~\ref{fig:Cha__T55}) This is a previously unknown SB3 consisting of two rapidly accelerating sources and a stable brighter component. This system had not previously been studied spectroscopically, and no resolved companions were found by \citet{2008ApJ...683..844L}.

For the close orbit, from the four reliable sets of velocity measurements and assuming a circular orbit, we infer a mass ratio of $q\!\sim\!0.8$ and a center-of-mass velocity of $\gamma\!\sim\!12{\rm\,km\,s^{-1}}$. The maximum observed velocity separation is $\Delta v\!\simeq\!150{\rm\,km\,s^{-1}}$, and the rapid change between the third and fourth epoch suggests a short period, of at most a few days. We do not have sufficient measurements to determine the period uniquely, but we can derive some constraints. Assuming a total radial velocity amplitude $v_{\rm tot}\!\simeq\!\Delta v$, one can estimate the total mass as $M\sin^3 i\!\simeq\!v_{\rm tot}^3P/2\pi G\!\simeq\!0.3(P/1{\rm\,d})\,M_\odot$ and the projected semi-major axis as $a\sin i=v_{\rm tot}P/2\pi=3(P/1{\rm\,d})\,R_\odot$.  We also have the following constraints. First, given the M4.5 spectral type, the expected mass of the brighter, stationary component, is $M_A\!\sim\!0.2\,M_\odot$. The other stars should be less massive, so the period cannot be much longer than one day. Second, the radius of the brighter star should be around $1\,R_\odot$ and, given the flux ratios of $0.4$ and $0.2$, components B and C should have radii around $0.6$ and $0.5\,R_\odot$, respectively. For these components to fit in the orbit, the period can also not be much shorter than $1$\,day. Independent of the precise period, the orbit is sufficiently close-in such that it should be circularized, and the rotation of the stars synchronized. From the observed rotational broadenings of $34$ and $31{\rm\,km\,s^{-1}}$, one infers rotation periods of $P\sin i=2\pi R\sin i/v\sin i\!\simeq\!0.8\,$d. From all evidence, we thus conclude that the orbital period is $P\!\sim\!0.8\,$d. Given the mass ratio of $q\!\sim\!0.8$, the individual masses are $M_{B,C}\sin^3i\!\simeq\!(0.14,0.12)\,M_\odot$.

For the outer orbit, from the fact that we do not see variations larger than a few ${\rm\,km\,s^{-1}}$ on our $\sim\!300\,$day baseline, we can only limit the period to be longer than $\gtrsim\!2\,$yr. A long period is also suggested by the velocity separation of $\lesssim\!5{\rm\,km\,s^{-1}}$ between the close binary center-of-mass and component~A.

\paragraph{RX\,J0412.8+2442 (V1198\,Tau)} (Fig.~\ref{fig:Tau__RXJ0412.8+2442}) This is a previously unknown SB3 consisting of two rapidly accelerating sources (with a maximum velocity difference of at least 115\,km\,s$^{-1}$), and a stable fainter component, which is likely a less massive star in a wide orbit. The B and C components have a flux ratio of $0.88\pm0.12$ and $0.36\pm0.04$ to the A component, respectively. Based on the spectral type of G9 for the primary and the flux ratios, we find that the secondary is likely of G9/K0 type, and the tertiary a K1. Given the high theoretical mass of the inner binary ($2.23\,M_{\odot} + 2.16\,M_{\odot}$), its maximum separation is $\mathrm{max}(a_{\mathrm{tot}}) = 0.3$\,AU. On the other hand, the tertiary shows a velocity difference to the A--B center-of-mass that is smaller than 3\,km\,s$^{-1}$, implying a maximum possible separation of 600 AU between the C component and the A--B mass center. \citet{1998AA...331..977K} observed this system with high spatial resolution using speckle interferometry, but found no spatially resolved companions with a limiting contrast ratio of 0.13 at 0\farcs13 (corresponding to the projected distance of 18\,AU at 140\,pc). The flux ratio between A+B and C is estimated to be 0.19 in $R$ and 0.23 in $K$; thus, it should be detected easily outside 0\farcs13 by \citet{1998AA...331..977K}. The possibilities remain that we are seeing the tertiary in a strongly inclined orbit, in a special position in its orbit, or both. Note, the systemic radial velocity of this SB3 is $\sim\!30$\,km\,s$^{-1}$, which deviates from the cluster radial velocity of $\sim\!16$\,km\,s$^{-1}$, and makes this an RV outlier.

\paragraph{V773\,Tau (HD\,283447, HIP\,19762, HBC\,367)} (Fig.~\ref{fig:Tau__V773TauA+B}) This is an SB3 and previously known quadruple system \citep{2003ApJ...592..288D,2003AA...406..685W}. The fourth component is an infrared companion, which explains its absence from our spectra. Of the other three, two are in a tight 51-day (2.8\,mas, 0.38\,AU) orbit \citep{2007ApJ...670.1214B}, and the third in a 46\,yr orbit around A--B \citep{2003ApJ...592..288D}. The radial velocities of the A and B components computed by \citep{2007ApJ...670.1214B} do not match well with the radial velocities we estimate from decomposing the derived broadening functions of the system. This is because the broadening functions have broad and blended peaks, which are difficult to decompose reliably to estimate radial velocities.

\paragraph{LkCa\,3\,A (HBC\,368, V1098\,Tau)} (Fig.~\ref{fig:Tau__LkCa3A+Ba}) This is an SB3 consisting of three sources, all of which show acceleration over our one month baseline. The source is listed as an SB1 in the review of \citet{1994ARAA..32..465M}, with period $P=12.941\,$d, eccentricity $e=0.2$, projected semi-major axis of relative orbit $a\sin i=0.032\,$AU, and center-of-mass velocity $\gamma = 14.9{\rm\,km\,s^{-1}}$. However, it is not clear how reliable the numbers are given the additional components we see. It was found to have a resolved companion with a separation of $0\farcs\!47$--$0\farcs\!491$, Lk\,Ca\,3\,B, by \citet{1993AA...278..129L} and \citet{1993AJ....106.2005G}. We see no direct evidence for this component in our broadening function, even though it should contribute given the flux ratio, and in epoch 2 none of the other components overlap with its expected velocity (roughly the cluster velocity). Perhaps, it is fainter in $R$ than estimated, it rotates relatively rapidly, or the resolved companion could itself be a close spectroscopic binary which would not necessarily match the system velocity. Nevertheless, we should caution that especially the velocity of the slowest and brightest component, Lk\,Ca\,3\,Aa, may be somewhat biased.

\subsection{SB2: Double-lined Spectroscopic Binaries}
\label{sec:NotesSB2}

\paragraph{CHXR\,12} (Fig.~\ref{fig:Cha__CHXR-12}) This is a previously unknown SB2 with no spatially detected companions \citep{2008ApJ...683..844L}. The measured spectral type is M3.5 for the primary, and the flux ratio $0.2\pm0.08$ implies M6 for the secondary, assuming an age of 2\,Myr. This corresponds to a system mass of $0.2\,M_{\odot} + 0.14\,M_{\odot} = 0.34\,M_{\odot}$. This total mass is close to that calculated using the estimated mass ratio of $\sim\!0.43$ derived from the method outlined in \citet{1941ApJ....93...29W}. Given the maximum observed velocity separation is $\Delta v\!\sim\!33{\rm\,km\,s}^{-1}$, we find the widest possible spatial separation to be 0.3\,AU, well within the 13\,AU limit reported by \citet{2008ApJ...683..844L}.

\paragraph{T42 (Ass\,Cha\,T\,2-42, FM\,Cha, HBC\,579)} (Fig.~\ref{fig:Cha__T42}) This is a previously unknown SB2 with no spatially detected companions \citep{2008ApJ...683..844L}. The measured spectral type is K5 for the primary, and the flux ratio to the secondary is $0.8\pm0.3$, corresponding to a system mass of $0.70\,M_{\odot} + 0.65\,M_{\odot} = 1.35\,M_{\odot}$. This total mass is similar to the value calculated from the mass ratio of $\sim\!0.81$ estimated using the method of \citet{1941ApJ....93...29W}. Given the maximum observed velocity separation is $\Delta v\!\sim\!40{\rm\,km\,s}^{-1}$, we find the widest possible spatial separation to be 0.8\,AU, undetectable by \citet{2008ApJ...683..844L}.

\paragraph{V826\,Tau (HBC\,400)} (Fig.~\ref{fig:Tau__V826TauA+B}) This is an SB2, and the first spectroscopic binary pre-main sequence (PMS) star to be confirmed \citep{1983ApJ...269..229M}. \citet{1990AA...235..197R} refined the orbital elements and found a period of 3.9 days. Furthermore, the extrapolated radial velocities from the orbital elements are consistent to within a few km~s$^{-1}$ of our radial velocities estimates, which is well-matched given very short-period and the radial velocity semi-amplitudes of $\sim\!18$\,km\,s$^{-1}$. By comparing $M\sin^3 i$ dynamical masses to masses from theoretical evolutionary tracks, they infer the inclination $i = 13\arcdeg\pm1\arcdeg$. Because of the cubic dependence on inclination, this estimate is not sensitive to the determined mass, and using more modern models by \citet{1997MmSAI..68..807D} only changes the best fit inclination to 14\fdg9. Assuming that the rotations of the stars are tidally locked together with their measured $v\,\sin\,i = 4.2{\rm\,km\,s}^{-1}$, they derive stellar radii of 1.44\,R$_{\odot}$. This is consistent with the 1.4\,R$_{\odot}$ expected from a 2\,Myr old K7 star \citep{1997MmSAI..68..807D}, but inconsistent with our measurement of $v\,\sin\,i\!\sim\!8.9{\rm\,km\,s}^{-1}$, which would double the radii. However, the broadening function technique we used to measure $v\,\sin\,i$ of SB2 components has a lower limit of $\sim\!9{\rm\,km\,s}^{-1}$ for our spectra (see \S\ref{sec:RadialVelocities}). Therefore, we cannot refute the previously published rotational velocities for this binary, and the stars could very well be in synchronous rotation with their orbit. Unfortunately, the close to pole-on inclination of the system makes it potentially difficult to measure periods from photometric modulation by star spots. We measure a flux ratio of $0.88\,\pm\,0.06$ which is in fair agreement with the mass ratio of $1.0$ estimated using the method of \citet{1941ApJ....93...29W}. Using high-resolution speckle interferometry, \citet{1993AA...278..129L} did not find any additional companions.

\paragraph{DQ\,Tau (HBC\,72)} (Fig.~\ref{fig:Tau__DQTau}) This is a known SB2, first reported by \citet{1997AJ....113.1841M} to have a period of 15.8\,days, and a circumbinary disk. The radial velocities extrapolated from the orbital elements given by \citet{1997AJ....113.1841M} are consistent to within a few km~s$^{-1}$ of our measured radial velocities, which is not a bad match given the estimated radial velocity semi-amplitudes of $\sim\!20$\,km\,s$^{-1}$. No spatially resolved companions were found in a high-resolution search using speckle interferometry by \citet{1993AA...278..129L}, but \citet{2009ApJ...696L.111B} resolved the spectroscopic binary to a 0.96 mas semi-major axis orbit, corresponding to 0.13\,AU at the 140\,pc distance to the system.

\paragraph{HBC\,427 (V397\,Aur)} (Fig.~\ref{fig:Tau__HBC427}) This is an SB2 in our data with a flux ratio between the components of $0.157\pm0.014$. The system was first noted as an SB1 by \citet{1988AJ.....96..297W}, and subsequently monitored to find an orbital solution. An astrometric-spectroscopic orbital solution was derived by combining 58 RV measurements distributed over 14\,yr (using the \textit{Center for Astrophysics Digital Speedometers}) with 14 spatially resolved astrometric measurements spanning 3.3\,yr (using the \textit{Fine Guidance Sensor} on the \textit{Hubble Space Telescope}) by \citet{2001AJ....122..997S}. They determined for the binary system an orbital period of $6.913\pm0.033$\,yr, and dynamical masses of $1.45\pm0.19$\,M$_{\odot}$ and $0.81\pm0.09$\,M$_{\odot}$. Furthermore, they find a $V$-band flux ratio of 0.11, which translates to an $R$-band flux ratio of 0.17 assuming colors of an K5 and M2 atmosphere for the primary and secondary, consistent with our finding of $0.157\,\pm\,0.014$. Given their orbital elements, our observing epochs range between orbital phase 0.951 and 0.963. This is close to the maximum RV separation of the components at the heliocentric RV $4{\rm\,km\,s}^{-1}$ and $35{\rm\,km\,s}^{-1}$, respectively, which are close to our derived velocities of $4$--$6{\rm\,km\,s}^{-1}$ and $32$--$35{\rm\,km\,s}^{-1}$.

\paragraph{T31 A (VW\,Cha)} (Fig.~\ref{fig:Cha__T31n}) T31\,A is a suspected SB2. The entire system T31 is a known hierarchical quadruple system with a wide companion (C) at 16\farcs8 \citep{2006AA...459..909C} and a tight 0\farcs1 binary (Ba,Bb) located 0\farcs7 from the primary A \citep{2001ApJ...561L.199B}. A previously reported companion (D) at 2\farcs7 \citep{1997ApJ...481..378G} has not been confirmed \citep{2001ApJ...561L.199B,2006AA...459..909C}. Our spectra of the primary show evidence for two components (Aa,Ab) separated by 20\,km\,s$^{-1}$, with a tentative flux ratio of $0.7\pm0.3$. The near-equal brightness of the components and good seeing conditions at the time of observations precludes the observed spectral components to be due to contamination by the known spatially resolved companions. \citet{2003AA...410..269M} reports the presence of three components in the cross-correlation function, one which they attribute to contamination by component B. \citet{2007AA...467.1147G} speculate that their observed erratic RV variability, and the asymmetric shape of the cross-correlation function are due to B contributing a varying amount of light, depending on the precise placement of the 2\farcs0 wide optical fiber. They do not rule out T31\,A being SB2, however, and we think this a more likely explanation given the small flux contribution by B, which has a A--B $\Delta K\!\sim\!1.74$ or an R-band flux ratio of $\sim\!0.03$.

For the primary A, we derive a spectral type of K3, which is slightly earlier than the K5/K7 found by \citet{1997AA...321..220B}, but later than K2 found by \citet{1989AARv...1..291A}. For Ba+Bb, we find a spectral type of K7, same as \citet{1997AA...321..220B}, while we did not observe component D. With a radial velocity difference of 24\,km\,s$^{-1}$, the widest possible separation would be 3\,AU, which corresponds to observing the 2\,M$_{\odot}$ system edge-on. AO observations rule out any projected separation greater than 75\,mas, which corresponds to 12\,AU at an assumed distance of 160 pc \citep{2008ApJ...683..844L}.

\subsection{SB1: Single-lined Spectroscopic Binaries}
\label{sec:NotesSB1}

\paragraph{T39 B (Ass\,Cha\,T\,2-39, CHXR\,36)} (Fig.~\ref{fig:Cha__T39Be}) This is an SB1 that is part of a known visual triple system, T39, consisting of an inner 1\farcs2 binary (Aa,Ab) of spectral types K7 and M1.5, and an outer 4\farcs5 component (B) of spectral type M1.5 \citep{1993AA...278...81R}. We obtained separate spectra of all three components (for T39\,Aa \& Ab). Components Aa and Ab show no evidence of additional companions. Therefore, since component B is an SB1, T39 is a quadruple system. The observed change in radial velocity of T39\,B is 2.6\,km\,s$^{-1}$ over 0.8\,yr.

\paragraph{RX\,J0415.8+3100 (V952\,Per)} (Fig.~\ref{fig:Tau__RXJ0415.8+3100}) This is an SB1 with a substantial RV variability amplitude of at least 70\,km\,s$^{-1}$ during a period on the order of two days. Recall, the mass function equation $(m_2 \sin i)^3 / (m_1 + m_2)^2 = K_1^3 P (1 - e^2)^{3/2} / 2 \pi G$ where $m_1$ and $m_2$ are the primary and secondary masses, $i$ is the inclination angle, $K_1$ is the radial velocity semi-amplitude of the primary mass, $P$ is the orbital period, $e$ is the orbital eccentricity, and $G$ is the gravitational constant. If we assume a primary star mass of 2.5\,M$_{\odot}$ (G6 type and 2\,Myr), an RV semi-amplitude of $35$\,km\,s$^{-1}$, an orbital period of 2\,d, and a circular orbit, then we estimate a companion mass of at least 0.42\,M$_{\odot}$ (M1 type). Given these mass estimates, the $R$-band flux ratio between the primary and an M1 companion would only be $\sim\!0.01$, which explains why the companion is not visible in our optical spectra. \citet{1998AA...331..977K} report on an additional companion at 0\farcs94, making this system triple.

\paragraph{RX\,J0457.5+2014 (V1354\,Tau)} (Fig.~\ref{fig:Tau__RXJ0457.5+2014}) This is an SB1 that shows a modest but significant 6\,km\,s$^{-1}$ increase in RV over 31 days. \citet{1998AA...331..977K} report a companion B at 6\farcs86, which is too distant to be responsible for the radial velocity change of component A. Therefore, this system is triple.

\section{Summary and Concluding Remarks}
\label{sec:SummaryAndConcludingRemarks}

Binary and higher-order multiple systems are prevalent in young clusters, and studying the statistical properties of these systems offers indirect constraints on star formation. Here, we present a spectroscopic study of mulitiplicity for young stars in the Chamaeleon~I and Taurus-Auriga star-forming regions. Our main results are as follows.

\begin{enumerate}

\item The spectroscopic multiplicity fractions for Cha~I and Tau-Aur are similar to each other, and to those of field stars in the same mass and period regime (for details, see \S\ref{sec:BinaryPopulations}). This finding implies that the overall fraction of short period stellar companions could stablize after initial formation.

\item Close multiplicity is not seen to depend on primary mass. The mass of the host star can become important in the separation regime of imaging surveys. For discussion, see \S\ref{sec:MassDependence}.

\item There is no strong correlation between accretion and the frequency of systems with close companions, and thus, close stellar companions are unlikely the principal source of the accretion cutoff observed in WTTS (see \S\ref{sec:BinaryPopulations}). Moreover, if close companions are responsible solely for the accretion difference between CTTSs and WTTSs then there should be a disparity in the proportion of spectroscopic binaries between these two populations, but this is not seen in our survey of Cha~I and Tau-Aur.

\end{enumerate}

By undertaking this extensive spectroscopic survey of T~Tauri stars, we have gained some insight that could be beneficial to future efforts. First and foremost, we now know the main limiting factor for a radial velocity study of young stars is the strong intrinsic noise present in some objects. This noise would need to be reduced in some way in order both to detect lower mass companions more effectively, and to extend the measurement to longer orbital periods. Our radial velocity precision is sufficient to detect, at the time of this writing, a few dozen of the known hot Jupiter planets, which have radial velocity semi-amplitudes of a few $100{\rm\,m\,s}^{-1}$. If such planets were to be detected around pre-main sequence stars, it would provide direct constraints on their formation and migration timescales. For young stars, one way to reduce the intrinsic noise is to observe at infrared wavelengths \citep{2006ApJ...644L..75M,2008AA...489L...9H}. Another option is to average out the noise by observing over its characteristic period, i.e., the rotation period of the noisy star. To some extent, we have already done this by taking multiple spectra of the same targets during an observing run. Ideally, one would multiplex observations, e.g., by using multiple fibers as in VLT/FLAMES. This solution may not be suitable for star-forming regions like Taurus-Auriga, which span a large area of the sky, but it would probably be quite useful for compact regions like Chamaeleon.

We have also learned that our high resolution and S/N were extremely beneficial. It was probably a major reason we could find so many SB2 candidates; being able to resolve the stellar rotation helped us to determine the stability of the broadening functions. Less clear is whether one would really need the large spectral range, or whether a few well-chosen echelle orders would suffice. This consideration is relevant since it would determine whether or not multiplexing is feasible.

\begin{acknowledgements}
This paper was significantly improved by the many detailed suggestions from an anonymous referee. We would like to thank David Lafreni\`{e}re, Robert Mathieu, Brian Lee, Russel White, and Eric Mamajek for enlightening discussions relating to the work presented in this paper. This paper includes data gathered with the 6.5 meter Magellan Telescopes located at Las Campanas Observatory, Chile. We would like to thank the Magellan staff for their tenacious effort and dedication in accommodating our aggressive observing program. This work was supported in part by Stockholm University grant dnr 301-3014-08 to DCN, Swedish National Space Board (contract 84/08:1) to AB, and NSERC grants to R.J. and M.H.vK., and by a fellowship to R.J. from the Radcliffe Institute for Advanced Study at Harvard University.
\end{acknowledgements}

\begin{appendices}
\section{Estimating Shortest Possible Orbital Periods}
\label{sec:ShortestPeriods}

We define a close binary system as two bodies in orbit about a common center of mass where there is no consistent exchange of material between them, i.e., each object does not extend beyond its Roche lobe. The effective radius $r_L$ of a Roche lobe for an object with mass $M_1$ in a binary system from the approximation of \citet{1983ApJ...268..368E} is
\\
\begin{equation}
\label{equ:RocheLobeRadius}
r_L = \frac{0.49\,q^{2/3}}{0.6\,q^{2/3} + \ln \left(1 + q^{1/3} \right)}\,\,\,,\,\,\,0\!<\!q\!<\!\infty
\end{equation}
\\
where $q$ is the mass ratio $M_1/M_2$, and $r_L$ is in units of the orbital separation. Furthermore, the orbital period $P$ in days for an object that just fills its Roche lobe is
\\
\begin{equation}
\label{equ:RocheLobePeriod}
P(q,\rho) = 0.1375 \left( \frac{q}{1+q} \right)^{1/2} r_L^{-3/2} \rho^{-1/2}
\end{equation}
\\
where $\rho$ is the mean density of the object in cgs units. Therefore, the theoretical shortest orbital period $P_{\rm min}$ of a binary system is
\\
\begin{equation}
\label{equ:ShortestPeriod}
P_{\rm min} = \min \{ \max \left[ P(q,\rho_2), P(1/q,\rho_1) \right]\,{\rm for}\,\,0\!<\!q\!\le\!1 \}
\end{equation}
\\
where $\rho_1$ and $\rho_2$ are the mean densities of the primary and secondary objects, respectively. Using the densities derived from the models of \citet{1997MmSAI..68..807D}, we estimate the shortest possible orbital period for a typical $\sim\!2$\,Myr old T~Tauri star in our survey ($0.6\,M_{\sun};\,1.4\,R_{\sun}$) with a stellar mass companion is approximately $0.8$\,days.
\end{appendices}

\bibliographystyle{apj}

\clearpage \begin{figure}\includegraphics[width=\textwidth]{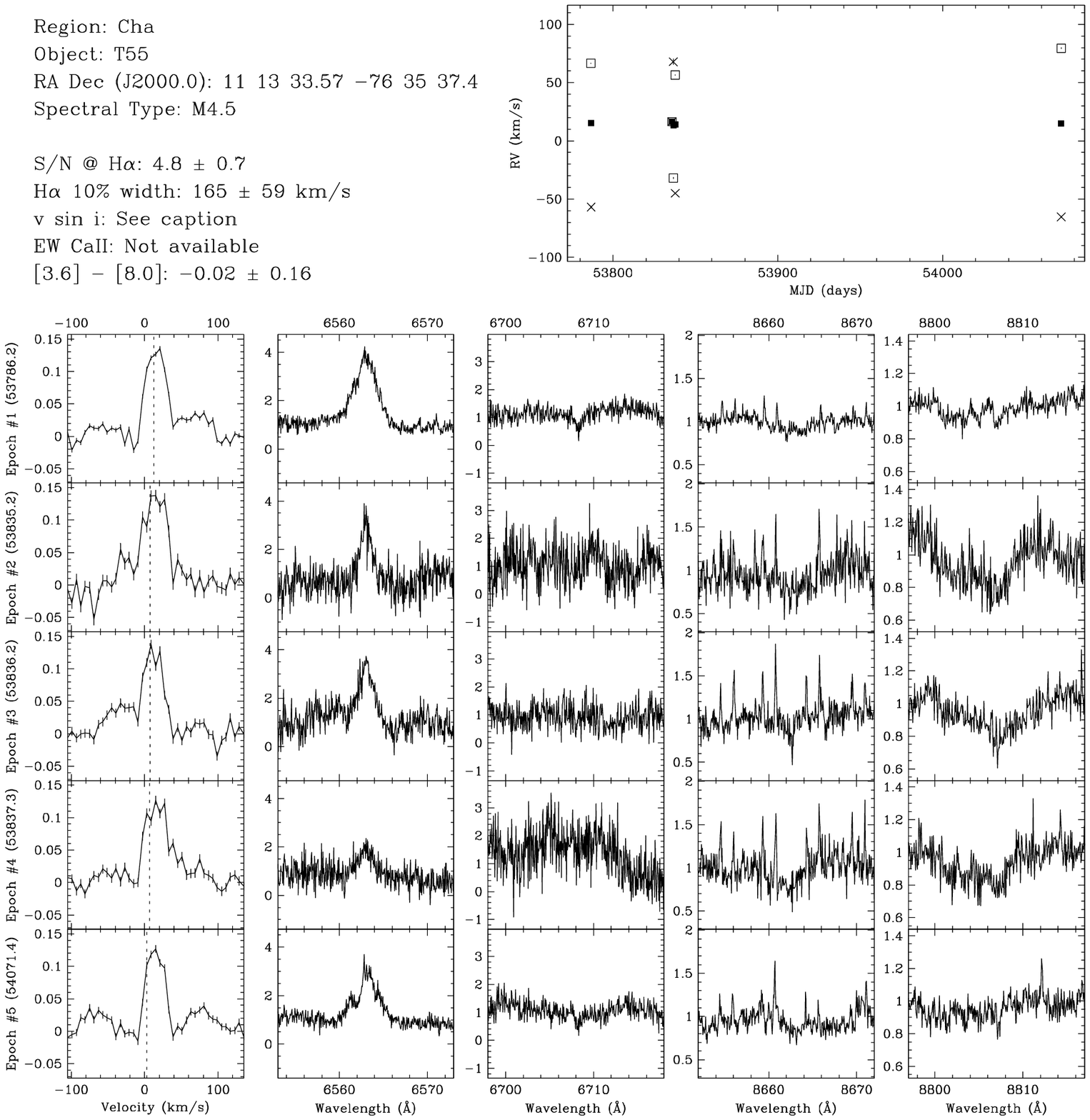}\caption[Plots and information on T55]{\footnotesize T55 is an SB3. The two fainter sources show large accelerations while the radial velocity of the brighter source is stable. This suggests a triple system comprised of a close binary in orbit with a farther away more massive star. By fitting the broadening function to three rotational broadening line profiles, we estimate the sources have an A--B flux ratio of $0.38\,\pm\,0.08$, an A--C flux ratio of $0.19\,\pm\,0.08$, and the A, B and C sources have $v\,\sin\,i$ of $23\,\pm\,2$\,km~s$^{-1}$, $34\,\pm\,4$\,km~s$^{-1}$ and $31.0\,\pm\,1.6$\,km~s$^{-1}$, respectively. The narrow emission lines in the spectra near 8662\,\AA\,are from the night sky. This target has been previously reported by \citet{2008ApJ...683..844L} to have no resolved companions. }\label{fig:Cha__T55}\end{figure}

\clearpage \begin{figure}\includegraphics[width=\textwidth]{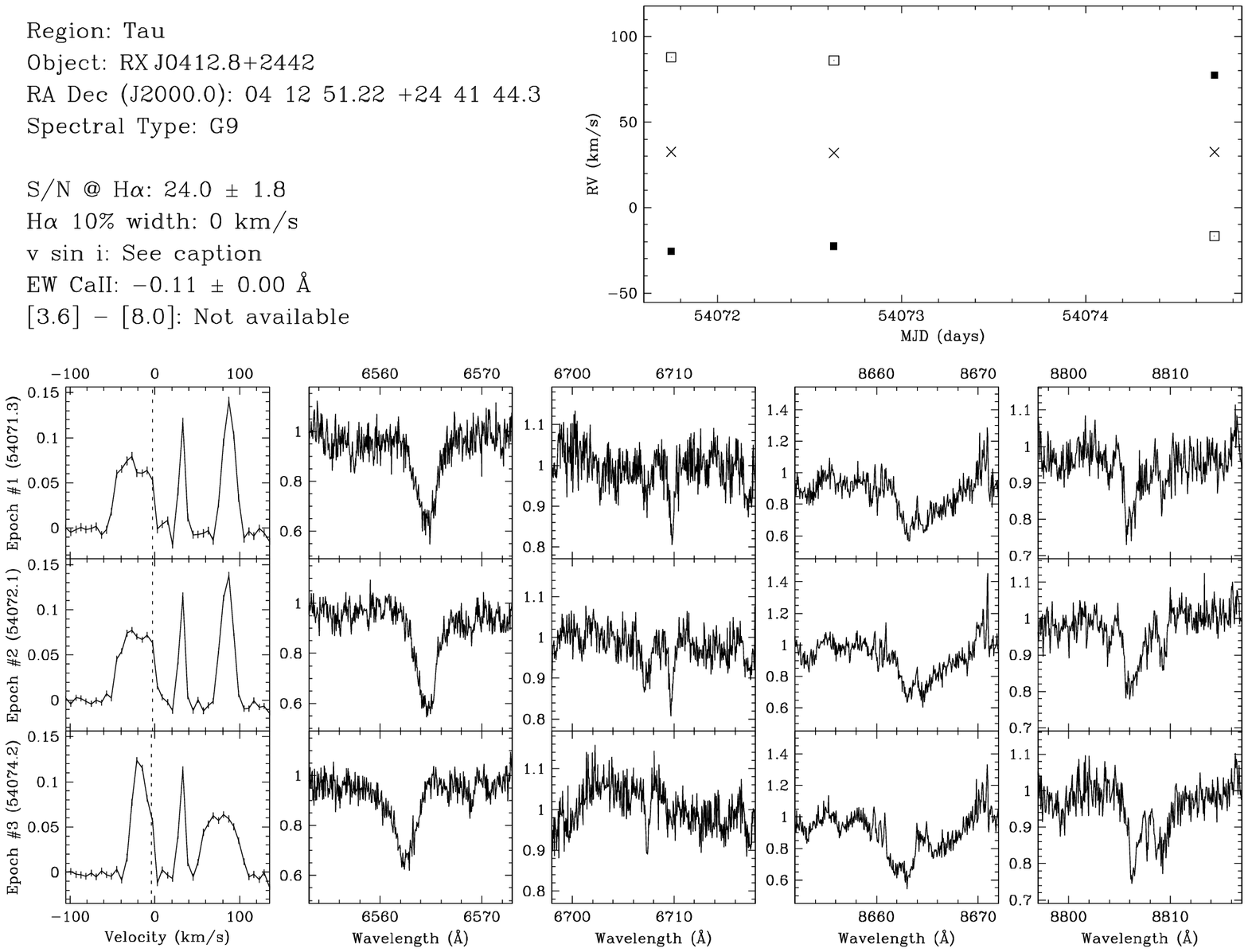}\caption[Plots and information on RX\,J0412.8+2442]{\footnotesize RX\,J0412.8+2442 is an SB3. The two of the sources show rapid radial acceleration. The radial velocity stable source has much less flux, and is likely a relatively farther away lower-mass companion. By fitting the broadening function to three rotational broadening line profiles, we estimate the sources have an A--B flux ratio of $0.88\,\pm\,0.12$, an A--C flux ratio of $0.36\,\pm\,0.04$, and the A, B and C sources have $v\,\sin\,i$ of $28.50\,\pm\,0.12$\,km~s$^{-1}$, $15.1\,\pm\,1.5$\,km~s$^{-1}$ and $7.4\,\pm\,0.3$\,km~s$^{-1}$, respectively. This target has been previously reported by \citet{1998AA...331..977K} to have no resolved companions. }\label{fig:Tau__RXJ0412.8+2442}\end{figure}

\clearpage \begin{figure}\includegraphics[width=\textwidth]{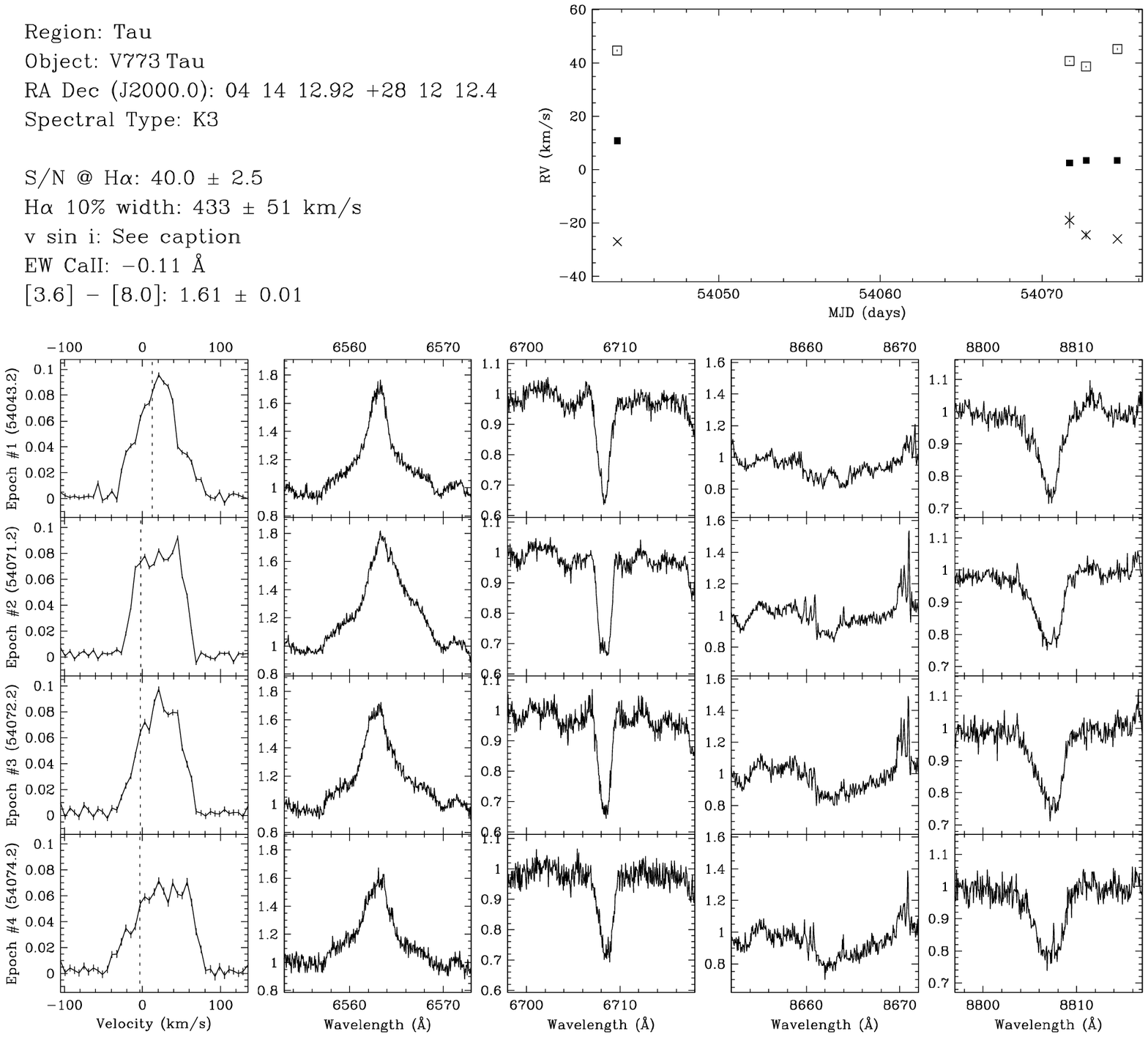}\caption[Plots and information on V773\,Tau]{\footnotesize V773\,Tau is an SB3. The broadening function appears to consist of three sources that fluctuate over time. By fitting the broadening function to three rotational broadening functions, we estimate the sources have an A--B flux ratio of $1.0\,\pm\,0.3$, an A--C flux ratio of $0.09\,\pm\,0.06$, and the A, B and C sources have $v\,\sin\,i$ of $27\,\pm\,4$\,km~s$^{-1}$, $28\,\pm\,4$\,km~s$^{-1}$ and $21\,\pm\,7$\,km~s$^{-1}$, respectively. However, these results and the derived radial velocities should be read with caution because of the broad and blended peaks in the broadening functions of this system which are difficult to decompose and fit reliably. This target has been previously reported by \citet{2003ApJ...592..288D}, and \citet{2003AA...406..685W} as a quadruple system consisting of a spectroscopic binary (V773~Tau~Aa,Ab) with a period of $\sim\!51$\,days and two wide companions (V773~Tau~B,C) with an A--B and A--C separations of $\sim\!0\farcs\!1$ ($\sim\!14$\,AU) and $\sim\!0\farcs\!2$ ($\sim\!28$\,AU) with an A--B $\Delta K$$\sim\!0.33--2.22$ and an A--C $\Delta K$$\sim\!1.84--2.85$. The radial velocities of the A and B components computed by \citep{2007ApJ...670.1214B} do not match well with the radial velocities we estimate because of our poor broadening function fits. }\label{fig:Tau__V773TauA+B}\end{figure}

\clearpage \begin{figure}\includegraphics[width=\textwidth]{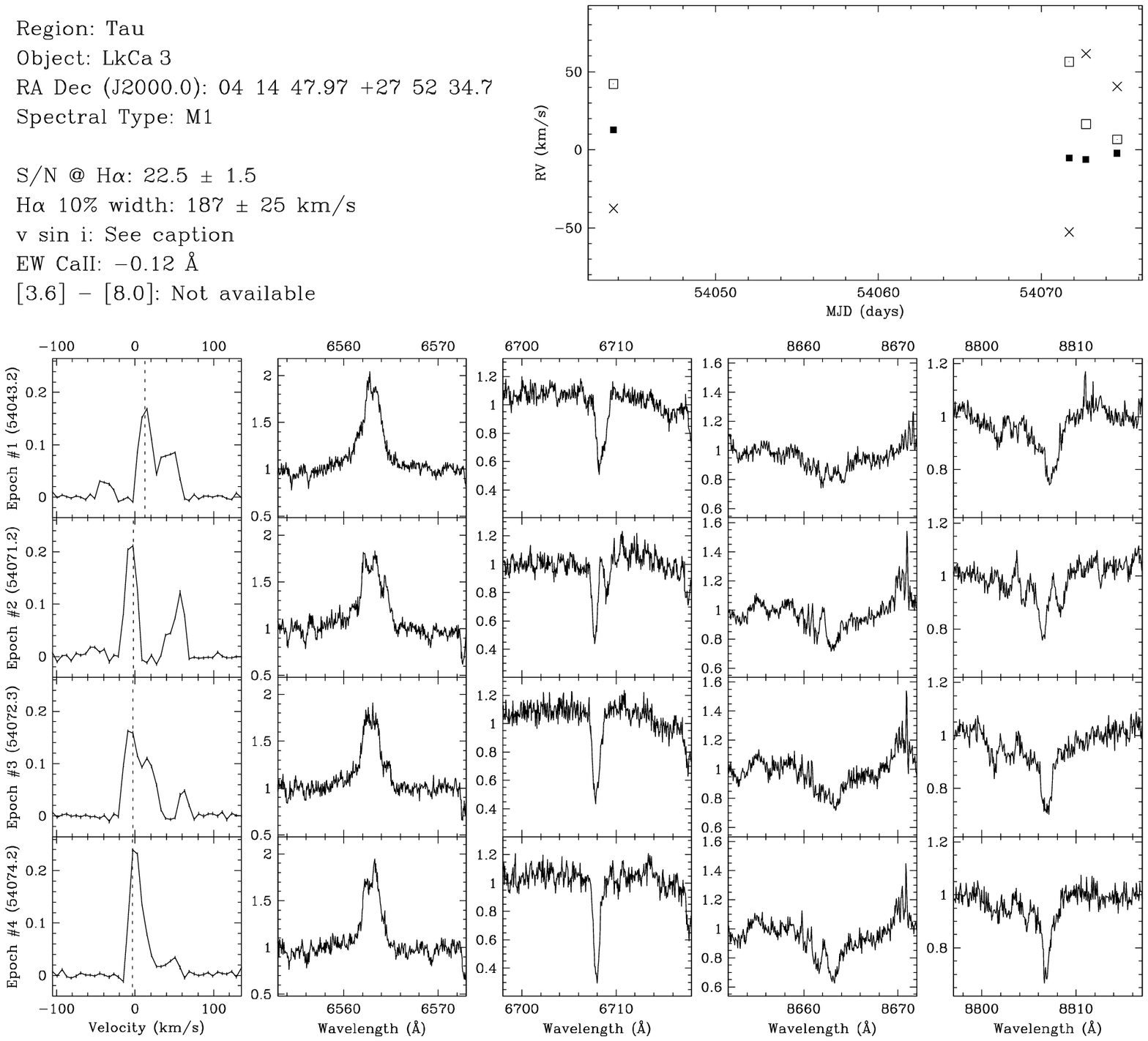}\caption[Plots and information on LkCa\,3]{\footnotesize LkCa\,3 is an SB3. The three sources also show radial acceleration and are very close companions. Therefore, the companions are not likely to be the previously resolved companion. By fitting the broadening function to three rotational broadening line profiles, we estimate the sources have an Aa--Ab1 flux ratio of $0.75\,\pm\,0.13$, an Aa--Ab2 flux ratio of $0.21\,\pm\,0.05$, and the Aa, Ab1 and Ab2 sources have $v\,\sin\,i$ of $12.0\,\pm\,1.0$\,km~s$^{-1}$, $16\,\pm\,2$\,km~s$^{-1}$ and $14\,\pm\,4$\,km~s$^{-1}$, respectively. This target has been previously reported by \citet{1993AA...278..129L}, and \citet{1993AJ....106.2005G} to have a resolved companion with a separation of $0\farcs\!47$--$0\farcs\!491$ ($66$--$69$\,AU) at a position angle of $77^\circ$--$78^\circ$, and an $R$-band flux ratio of $0.47$--$0.92$ (based on $\Delta K$$\sim\!0.05$--$0.42$). However, there is no clear evidence in the broadening function of the resolved companion. Since the resolved companion has an expected circular orbital speed of $\sim\!3$\,km\,s$^{-1}$, another profile could be obscured if the central star and the resolved companion have similar projected rotational velocities. Therefore, the system is possibly a quadruple system consisting of a spectroscopic binary in orbit with another spectroscopic companion and a resolved companion.}\label{fig:Tau__LkCa3A+Ba}\end{figure}

\clearpage \begin{figure}\includegraphics[width=\textwidth]{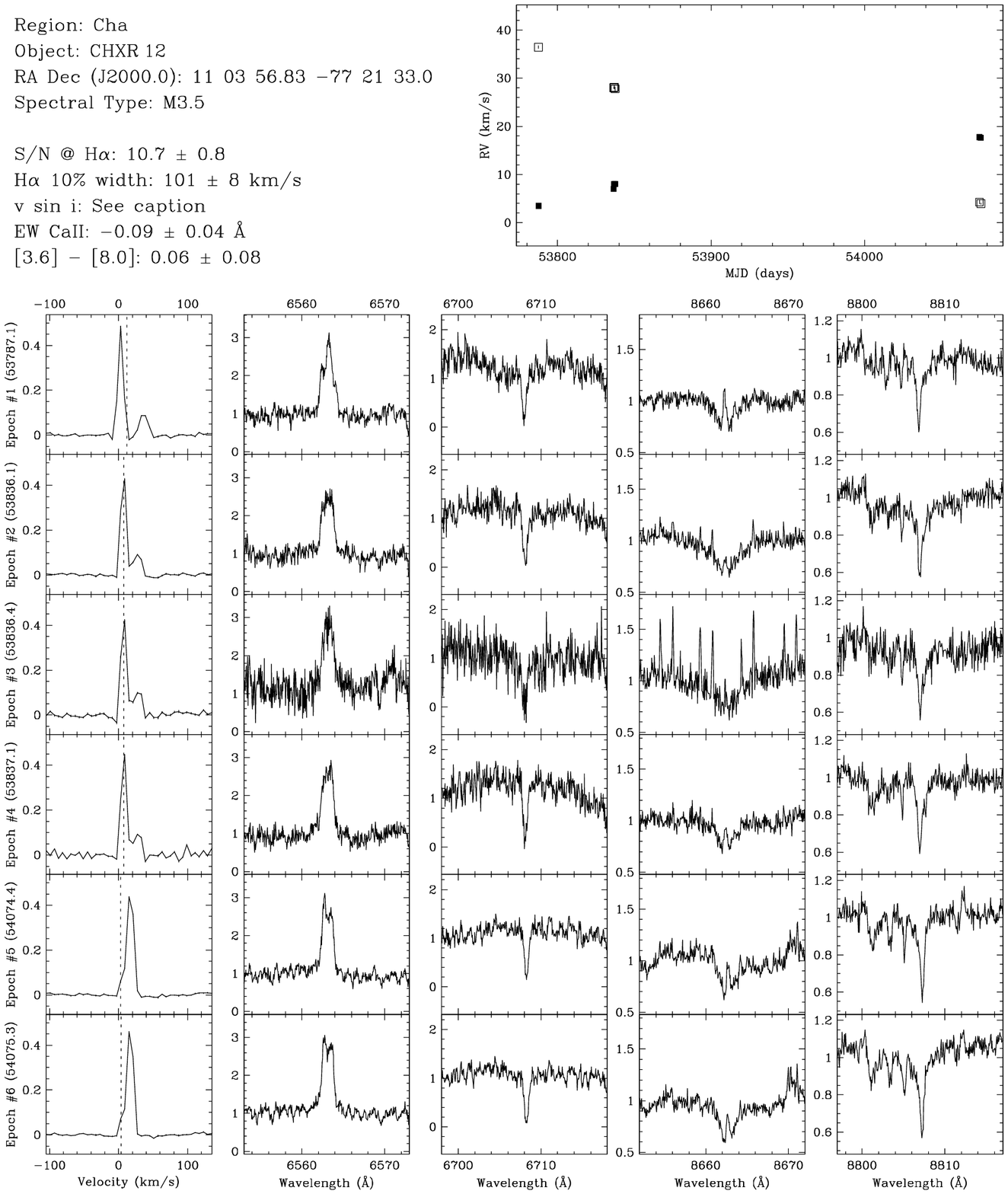}\caption[Plots and information on CHXR\,12]{\footnotesize CHXR\,12 is an SB2. Two sources can be seen in the broadening function, and both show acceleration. By fitting the broadening function to two rotational broadening line profiles, we estimate the two sources have a flux ratio of $0.20\,\pm\,0.08$, and the A and B sources have $v\,\sin\,i$ of $8.99\,\pm\,0.14$\,km~s$^{-1}$ and $8\,\pm\,2$\,km~s$^{-1}$, respectively. For epoch \#3, the narrow emission lines in the spectrum near 8662\,\AA\,are from the night sky. For epochs \#5 \& \#6, the radial velocities from fitting the blended broadening functions were fairly robust because of the strong S/N ratio and the low $v\,\sin\,i$ of the components. This target has been previously reported by \citet{2008ApJ...683..844L} to have no resolved companions. }\label{fig:Cha__CHXR-12}\end{figure}

\clearpage \begin{figure}\begin{center}\includegraphics[width=\textwidth]{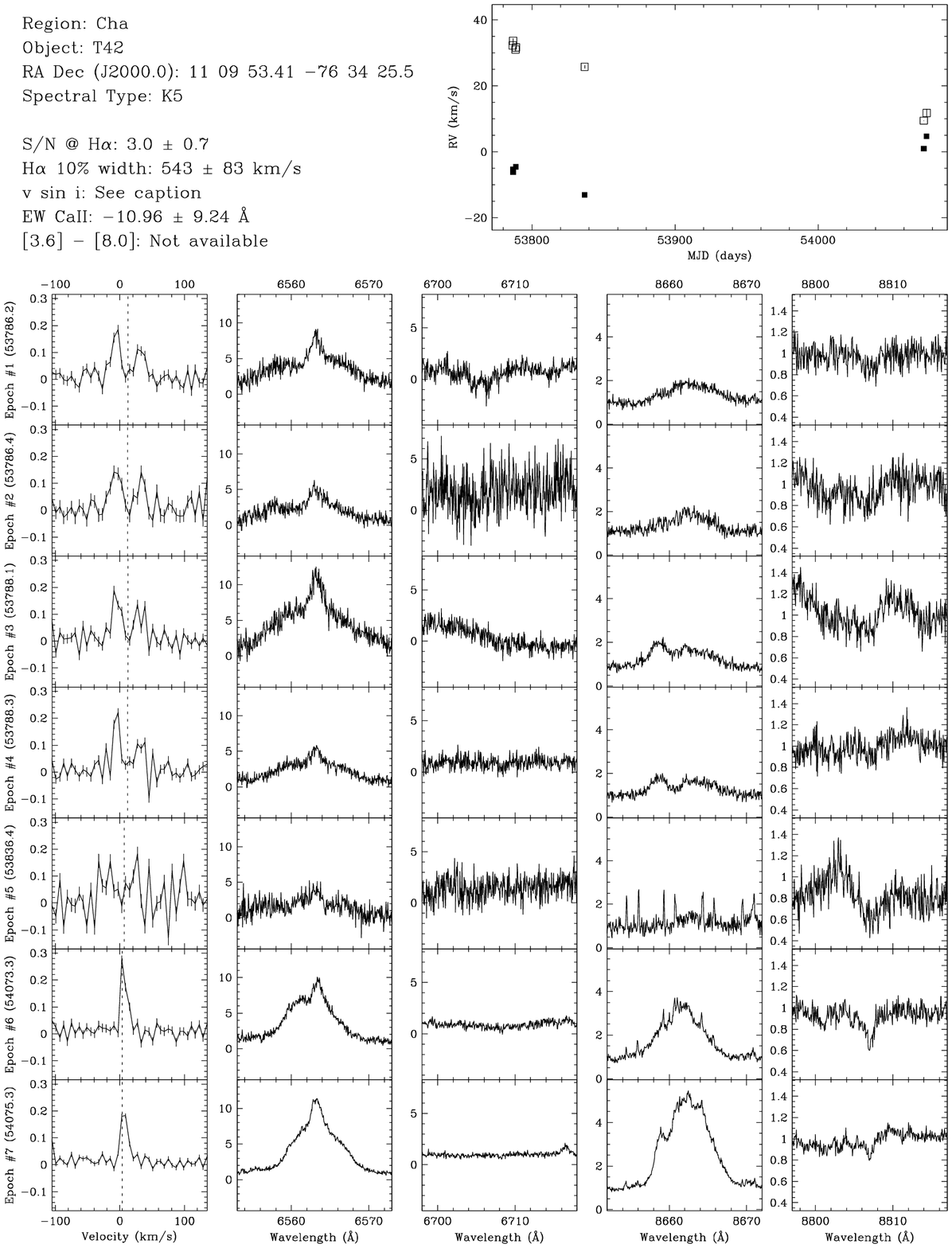}\caption[Plots and information on T42]{\footnotesize T42 is an SB2. The radial velocity estimate for epoch \#5 is unreliable due to poor S/N. By fitting the broadening function to two rotational broadening line profiles, we estimate the two sources have a flux ratio of $0.8\,\pm\,0.3$, and the A and B sources have $v\,\sin\,i$ of $11.4\,\pm\,1.1$\,km~s$^{-1}$ and $11.5\,\pm\,1.5$\,km~s$^{-1}$, respectively. For epoch \#5, the narrow emission lines in the spectrum near 8662\,\AA\,are from the night sky. For epochs \#6 \& \#7, the radial velocities from fitting the blended broadening functions were acceptable because of the decent S/N ratio. This target has been previously reported by \citet{2008ApJ...683..844L} to have no resolved companions. }\label{fig:Cha__T42}\end{center}\end{figure}

\clearpage \begin{figure}\includegraphics[width=\textwidth]{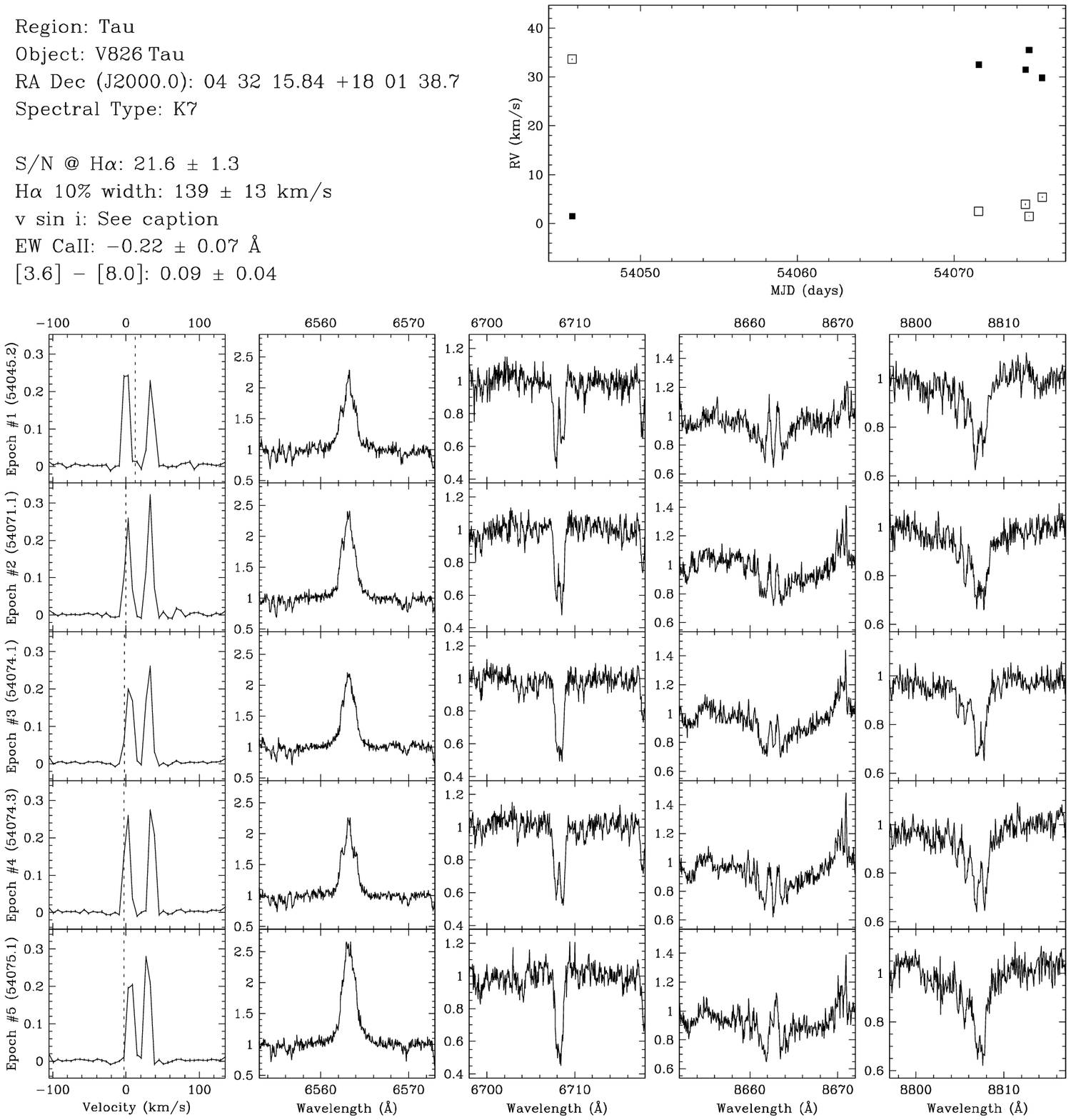}\caption[Plots and information on V826\,Tau]{\footnotesize V826\,Tau is an SB2. Two sources can be seen in the broadening function, and both show acceleration. By fitting the broadening function to two rotational broadening line profiles, we estimate the two sources have a flux ratio of $0.88\,\pm\,0.06$, and the A and B sources have $v\,\sin\,i$ of $8.5\,\pm\,0.5$\,km~s$^{-1}$ and $9.3\,\pm\,0.7$\,km~s$^{-1}$, respectively. This target has been previously reported by \citet{1990AA...235..197R} as a spectroscopic binary (V826~Tau~A+B) with a period of $\sim\!3.9$\,days. Furthermore, the extrapolated radial velocities from their orbital elements are consistent to within a few km~s$^{-1}$ of our radial velocities estimates, which is well-matched given very short-period and the radial velocity semi-amplitudes of $\sim\!18$\,km\,s$^{-1}$.}\label{fig:Tau__V826TauA+B}\end{figure}

\clearpage \begin{figure}\includegraphics[width=\textwidth]{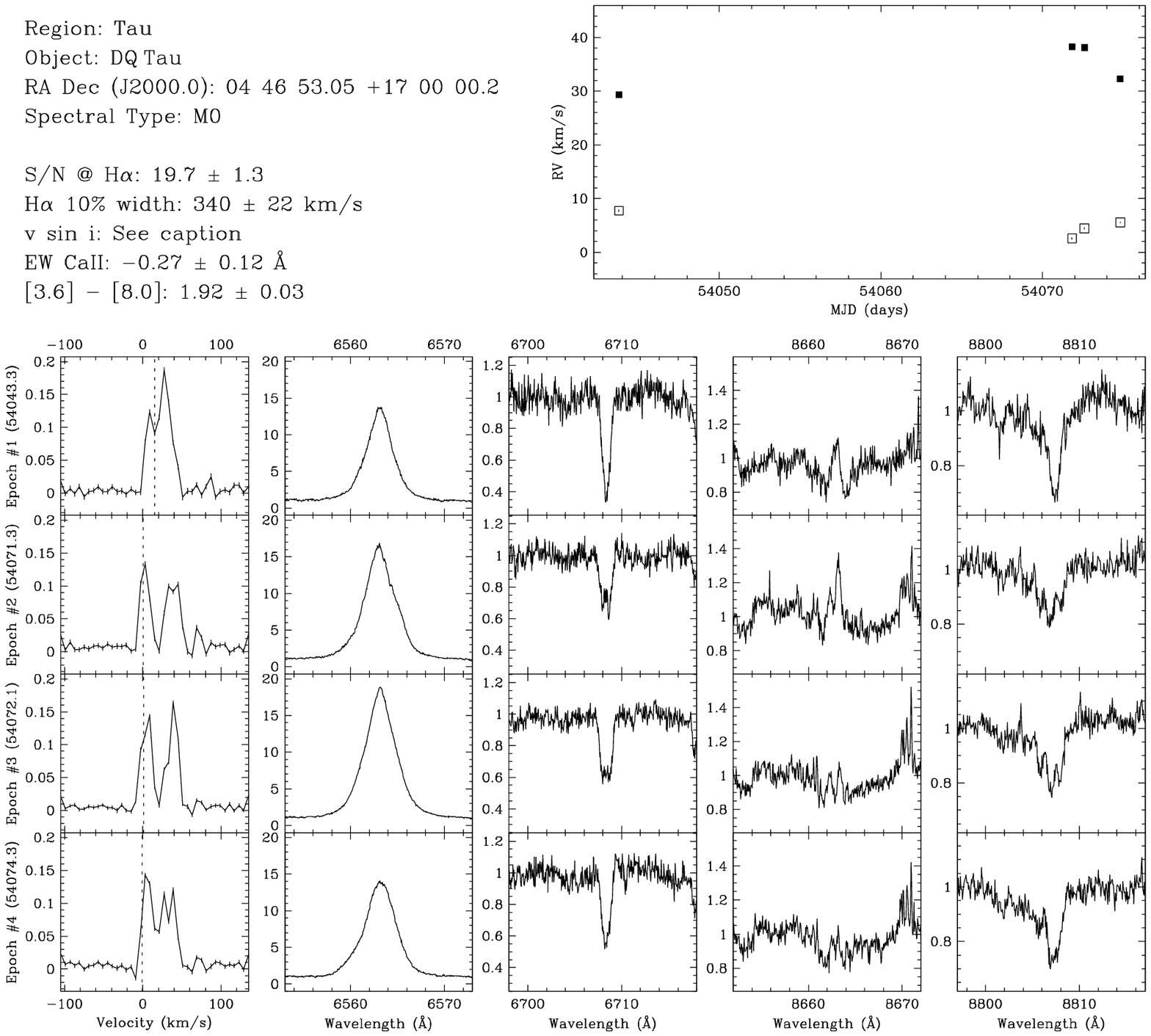}\caption[Plots and information on DQ\,Tau]{\footnotesize DQ\,Tau is an SB2. The two sources show radial acceleration. By fitting the broadening function to two rotational broadening line profiles, we estimate the two sources have a flux ratio of $0.78\,\pm\,0.18$, and the A and B sources have $v\,\sin\,i$ of $14.7\,\pm\,1.6$\,km~s$^{-1}$ and $11.3\,\pm\,0.7$\,km~s$^{-1}$, respectively. This target has been previously reported by \citet{1997AJ....113.1841M} as a spectroscopic binary with a period of $15.8$\,days. Furthermore, the radial velocities extrapolated from the orbital elements given by \citet{1997AJ....113.1841M} are consistent to within a few km~s$^{-1}$ of our measured radial velocities, which is not a bad match given the estimated radial velocity semi-amplitudes of $\sim\!20$\,km\,s$^{-1}$.}\label{fig:Tau__DQTau}\end{figure}

\clearpage \begin{figure}\includegraphics[width=\textwidth]{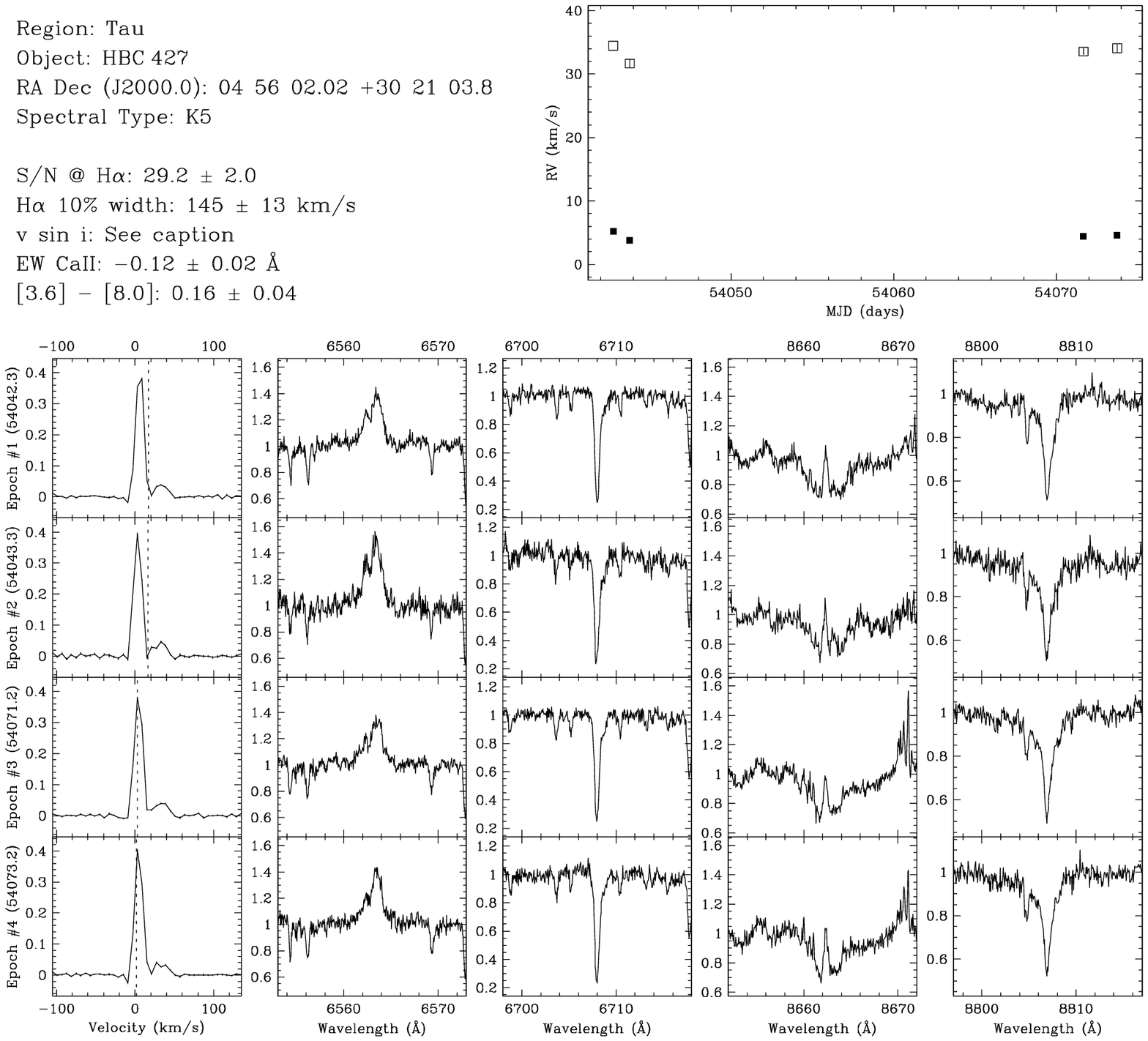}\caption[Plots and information on HBC\,427]{\footnotesize HBC\,427 is an SB2. Two sources can be seen in the profile: a peak at $\sim\!5$\,km\,s$^{-1}$ and another peak at $\sim\!30$\,km\,s$^{-1}$. By fitting the broadening function to two rotational broadening line profiles, we estimate the two sources have a flux ratio of $0.157\,\pm\,0.014$, and the A and B sources have $v\,\sin\,i$ of $9.9\,\pm\,0.3$\,km~s$^{-1}$ and $14.5\,\pm\,0.5$\,km~s$^{-1}$, respectively. This target has been previously reported by \citet{2001AJ....122..997S} as a spectroscopic binary with a period of $\sim\!2500$\,days.}\label{fig:Tau__HBC427}\end{figure}

\clearpage \begin{figure}\includegraphics[width=\textwidth]{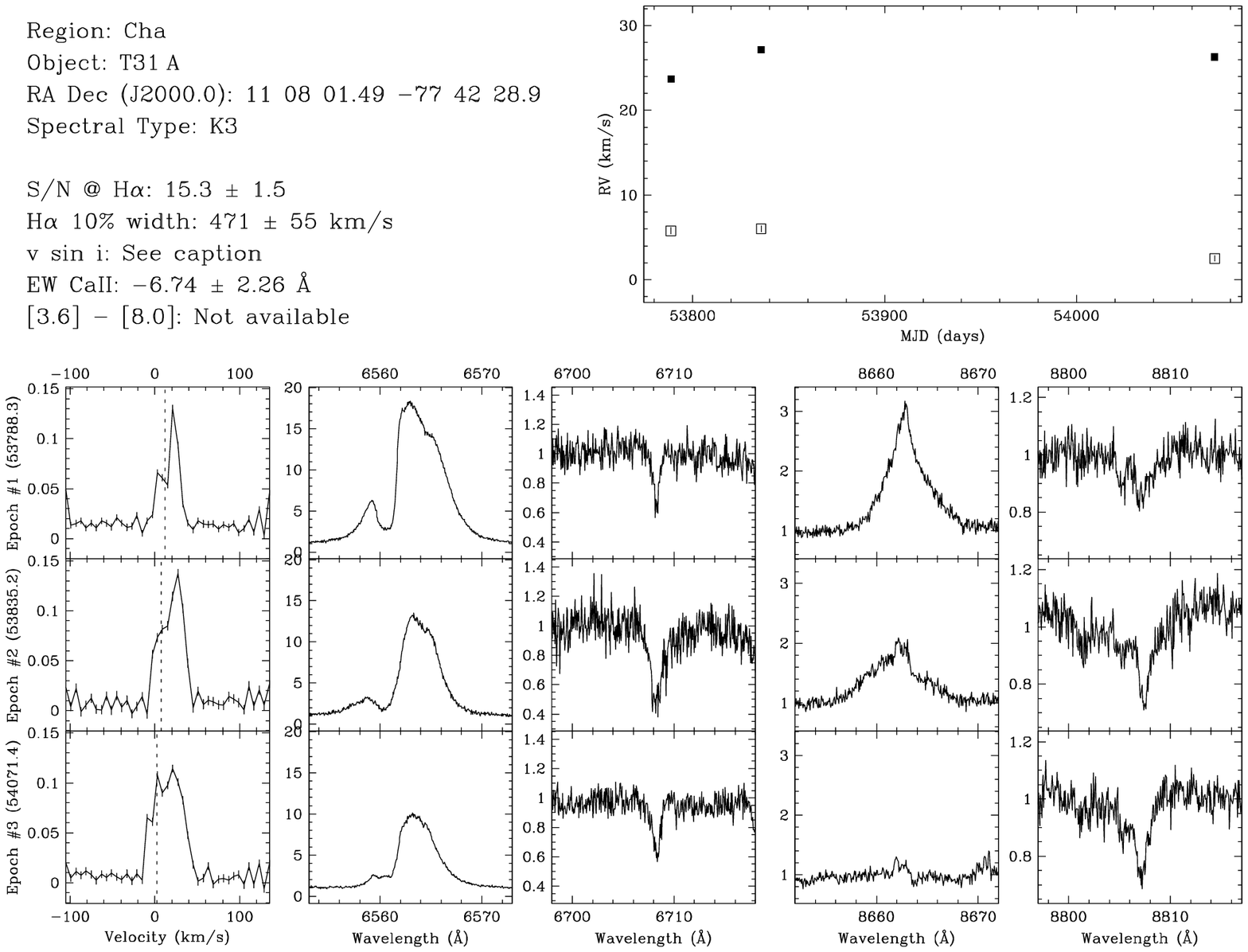}\caption[Plots and information on T31\,A]{\footnotesize T31\,A is a suspected SB2. Two sources can be seen in the profile: a peak at $\sim\!25$\,km\,s$^{-1}$ and another peak at $\sim\!5$\,km\,s$^{-1}$. By fitting the broadening function to two rotational broadening line profiles, we estimate the two sources have a flux ratio of $0.7\,\pm\,0.3$, and the a and b sources have $v\,\sin\,i$ of $13\,\pm\,2$\,km~s$^{-1}$ and $13\,\pm\,3$\,km~s$^{-1}$, respectively. This target has been previously reported by \citet{2008ApJ...683..844L} to have a resolved companion with a separation of $\sim\!0\farcs\!652$ ($\sim\!91$\,AU) at a position angle of $\sim\!178\fdg\!7$, and an $R$-band flux ratio of $\sim\!0.03$ (based on $\Delta K$$\sim\!1.74$). Given the flux ratio, the expected contribution to the broadening function from the resolved companion is negligible. The companion was resolved in the guide camera during good seeing conditions and was indeed faint, so it cannot have contributed to the profile.}\label{fig:Cha__T31n}\end{figure}

\clearpage \begin{figure}\includegraphics[width=\textwidth]{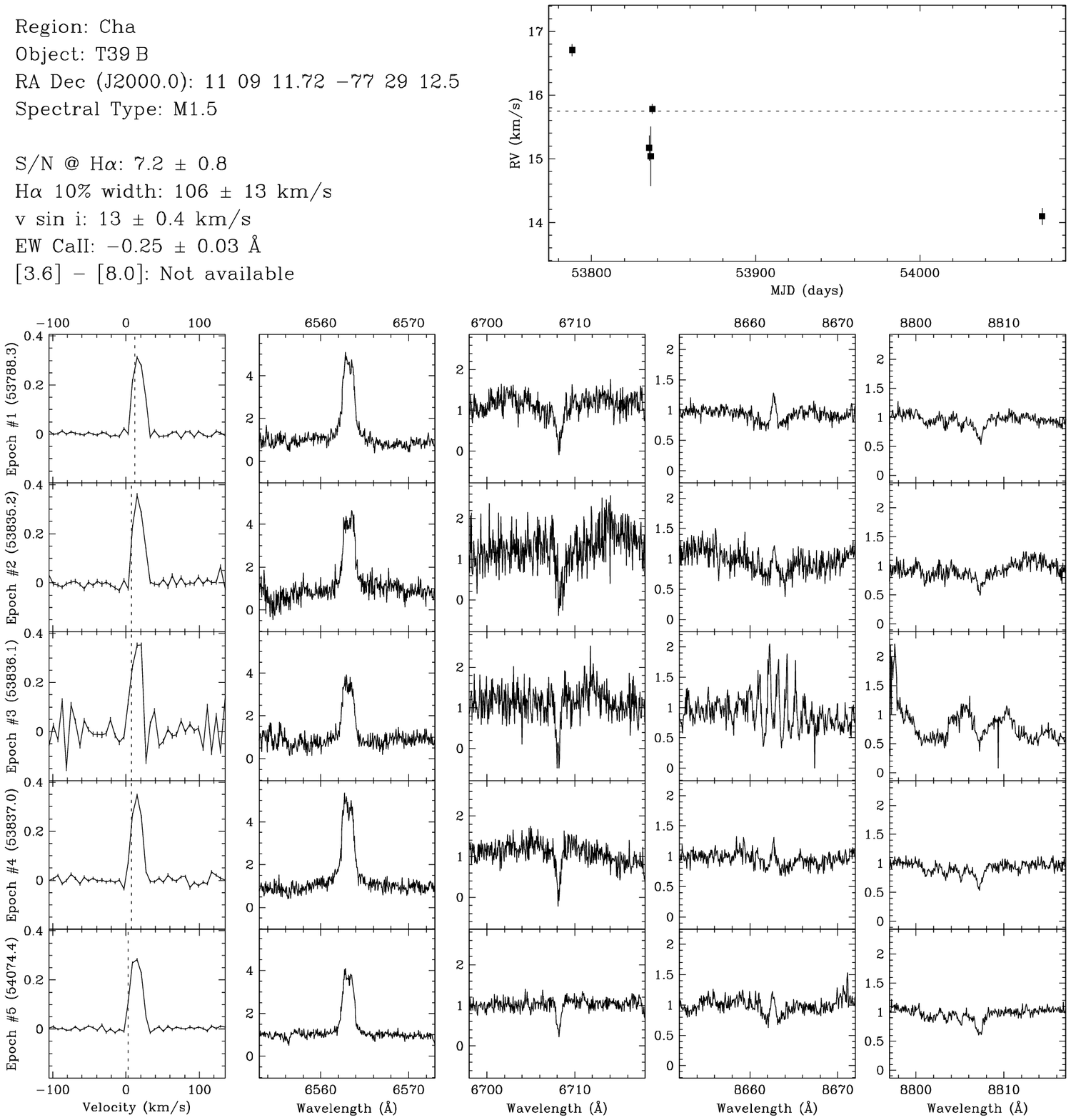}\caption[Plots and information on T39\,B]{\footnotesize T39\,B is an SB1. The overall radial velocity scatter is significant relative to the measurement uncertainties and to scatter observed within each observing run. The radial velocity scatter is larger than expected for a star with a $v\,\sin\,i$ of $13$\,km\,s$^{-1}$, and thus is unlikely to be due to star spots. The lack of radial velocity offset from the cluster velocity ($\sim\!16$\,km\,s$^{-1}$) implies a long orbital period and a small velocity amplitude. If star spots are responsible for the changes in radial velocity, based on observed $v~\sin~i$ ($\sim\!13$\,km\,s$^{-1}$) and model stellar radius ($\sim\!1.3\,R_{\odot}$), one would expect variations on a maximum timescale of $\sim\!5.1$\,days. From the stable results of the second observing run (epoch \#2, \#3 and \#4) which spans $2.8$\,days compared to the overall range of radial velocities, it is unlikely that the radial velocity trends are the result of star spots. This target has been previously reported by \citet{2008ApJ...683..844L} to be a component star of a resolved triple system (T39~Aa,Ab,B) with a separation of $\sim\!4\farcs\!497$ ($\sim\!630$\,AU) at a position angle of $\sim\!70\fdg\!8$, and a B--Aa $R$-band flux ratio of $\sim\!0.13$ (based on $\Delta K$$\sim\!0.77$). Given the separation, the expected contribution to the broadening function from the resolved companion is negligible.}\label{fig:Cha__T39Be}\end{figure}

\clearpage \begin{figure}\includegraphics[width=\textwidth]{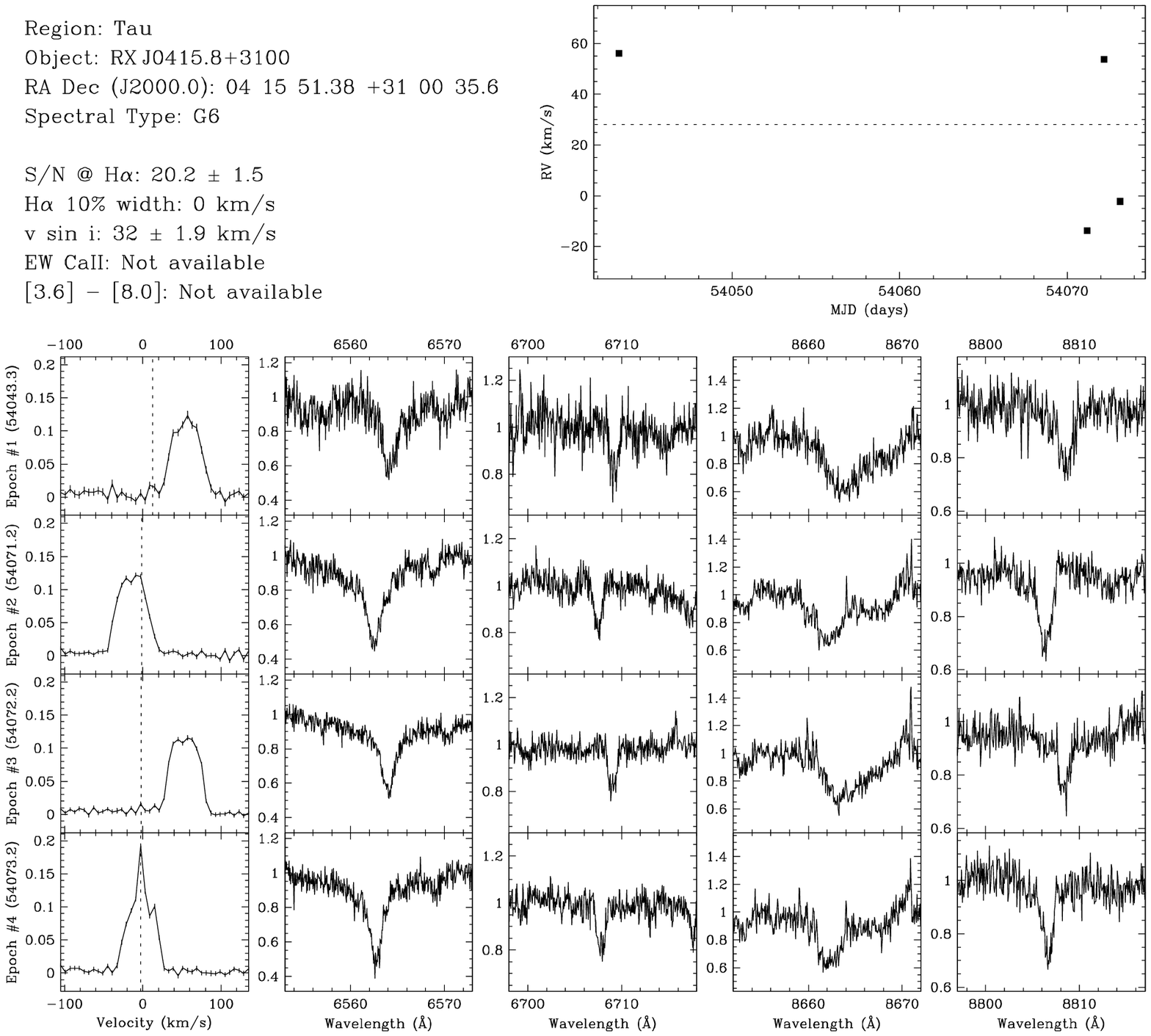}\caption[Plots and information on RX\,J0415.8+3100]{\footnotesize RX\,J0415.8+3100 is an SB1. The overall radial velocity scatter is significant relative to the measurement uncertainties. For epoch \#4, the strong sharp peak in the broadening function at the observer's rest frame is due to moonlight, and may have biased the radial velocity estimate toward the observer's rest frame. This target has been previously reported by \citet{1998AA...331..977K} to have a resolved companion with a separation of $\sim\!0\farcs\!94$ ($\sim\!130$\,AU) at a position angle of $\sim\!147\fdg\!2$, and an $R$-band flux ratio of $\sim\!0.19$ (based on $\Delta K$$\sim\!1.45$). Given the separation, the expected contribution to the broadening function from the resolved companion is small.}\label{fig:Tau__RXJ0415.8+3100}\end{figure}

\clearpage \begin{figure}\includegraphics[width=\textwidth]{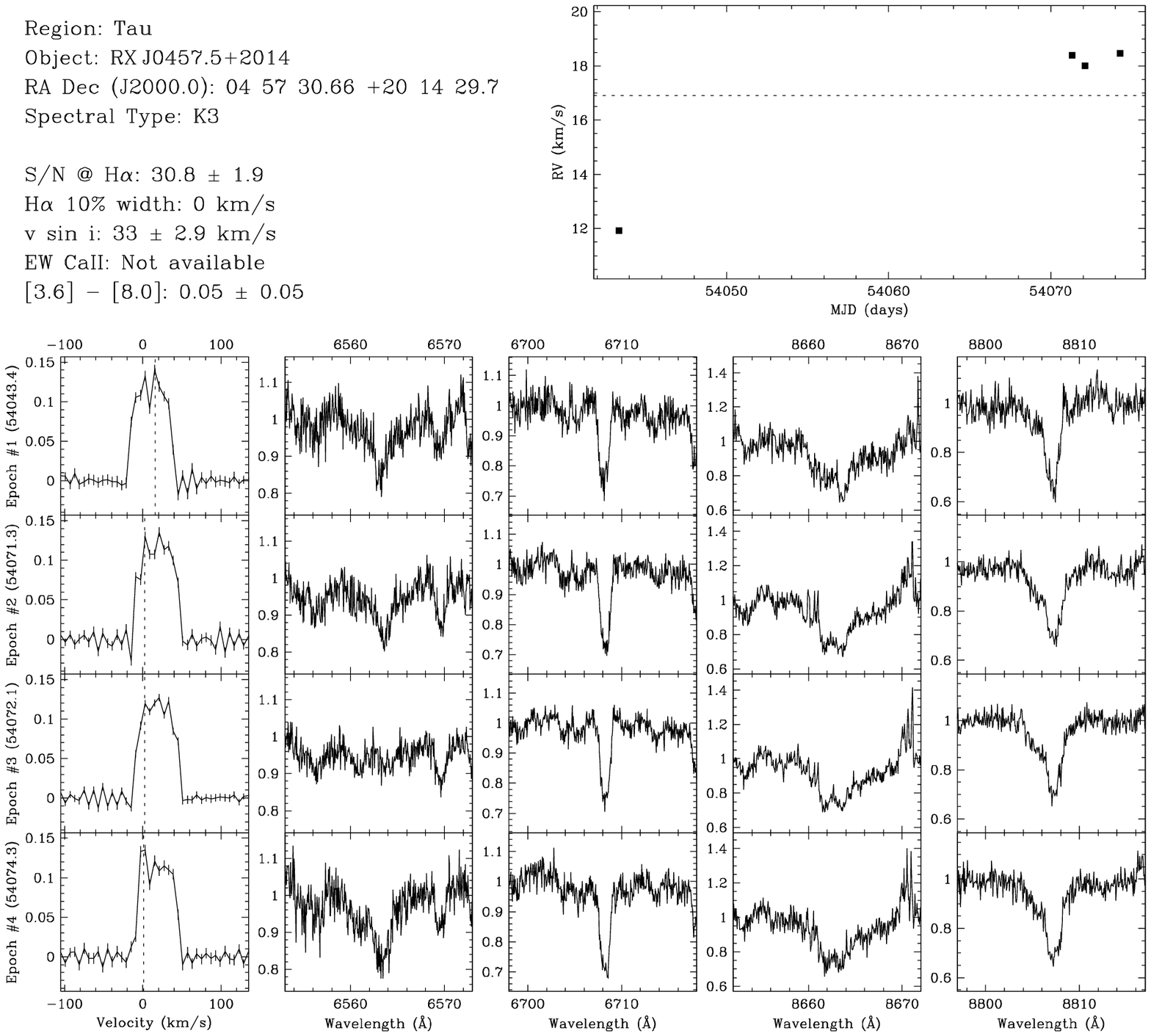}\caption[Plots and information on RX\,J0457.5+2014]{\footnotesize RX\,J0457.5+2014 is an SB1. The overall radial velocity scatter is significant relative to the measurement uncertainties and to scatter observed within each observing run. If star spots are responsible for the changes in radial velocity, based on observed $v\,\sin\,i$ ($\sim\!33$\,km\,s$^{-1}$) and model stellar radius ($\sim\!1.7\,R_{\odot}$), one would expect variations on a maximum timescale of $\sim\!2.6$\,days. From the stable results of the second observing run (epochs \#2, \#3 and \#4) which span $3.0$\,days, it is unlikely that the radial velocity trends are the result of star spots. For epoch \#4, the sharp peak in the broadening function at the observer's rest frame is due to moonlight, and may have biased the radial velocity estimate toward the observer's rest frame. This target has been previously reported by \citet{1998AA...331..977K} to have a resolved companion with a separation of $6\farcs\!865$--$6\farcs\!867$ ($\sim\!960$\,AU) at a position angle of $204\fdg\!8$--$205\fdg\!5$, and an $R$-band flux ratio of $\lesssim\!0.01$ (based on $\Delta K$$\sim\!2.20$--$2.42$). Given the flux ratio and separation, the expected contribution to the broadening function from the resolved companion is negligible.}\label{fig:Tau__RXJ0457.5+2014}\end{figure}

\clearpage \begin{figure}\includegraphics[width=\textwidth]{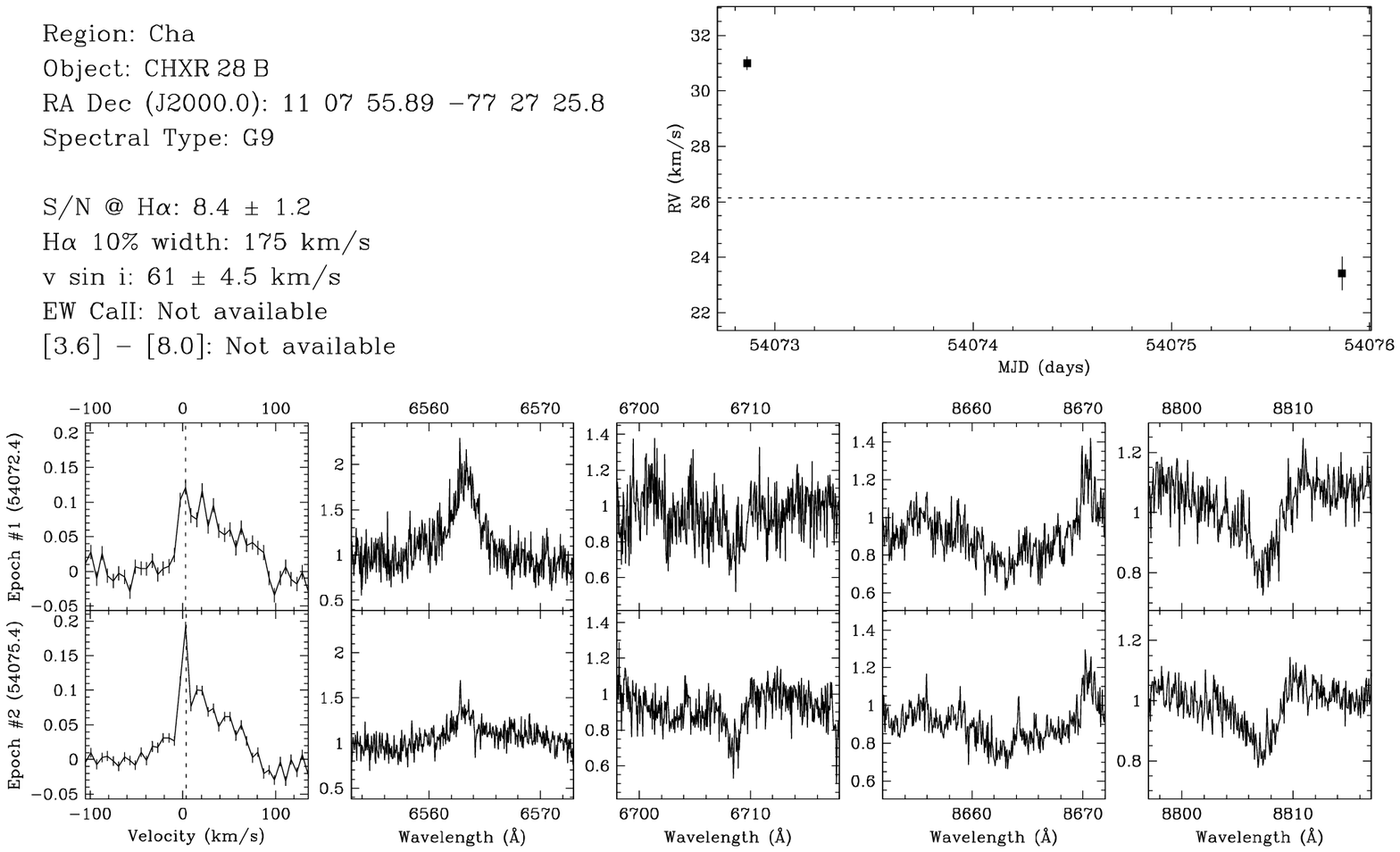}\caption[Plots and information on CHXR\,28\,B]{\footnotesize CHXR\,28\,B is a suspected SB1. The overall radial velocity scatter is larger than expected from non-companion influences, e.g., star spots. We derived the radial velocities from fits to the broadening function rather than from direct fits to the spectra (see \S\ref{sec:RadialVelocities}). The strong sharp peaks in the broadening functions at the observer's rest frame are due to dawn twilight, and was included in the broadening function fits. This target has been previously reported by \citet{2008ApJ...683..844L} to have a resolved companion with a separation of $\sim\!1\farcs\!818$ ($\sim\!250$\,AU) at a position angle of $\sim\!115\fdg\!9$, and an $R$-band flux ratio of $\sim\!0.68$ (based on $\Delta K$$\sim\!0.32$). Given the separation, the expected contribution to the broadening function from the resolved companion is negligible.}\label{fig:Cha__CHXR-28e}\end{figure}

\clearpage \begin{figure}\includegraphics[width=\textwidth]{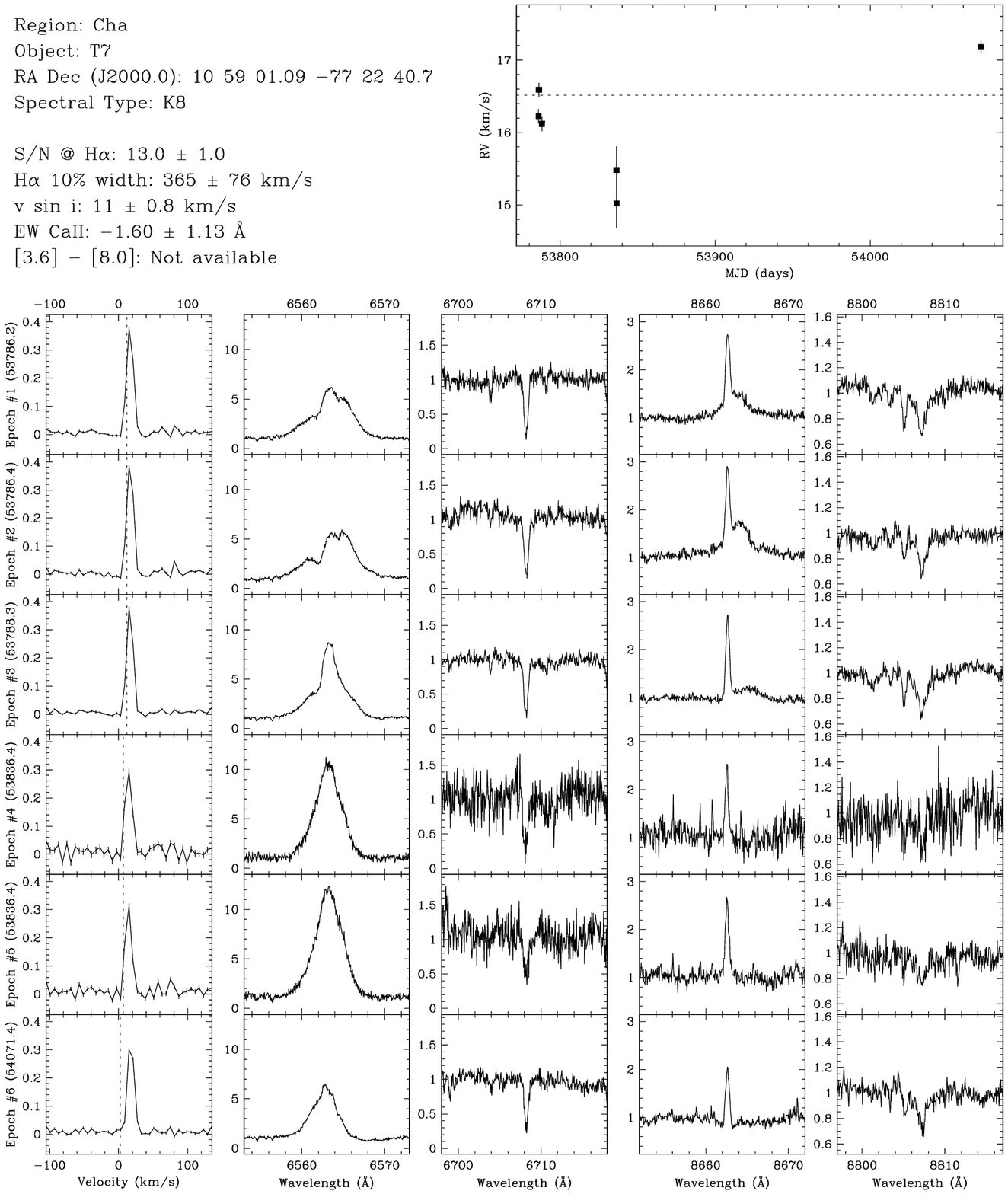}\caption[Plots and information on T7]{\footnotesize T7 is a suspected SB1. The overall radial velocity scatter is larger than expected from non-companion influences, e.g., star spots. If star spots are responsible for the changes in radial velocity, based on observed $v\,\sin\,i$ ($\sim\!11$\,km\,s$^{-1}$) and model stellar radius ($\sim\!1.4\,R_{\odot}$), one would expect variations on a maximum timescale of $\sim\!6.1$\,days. From the stable results of the first observing run (epoch \#1, \#2 and \#3) which spans $2.1$\,days, there is doubt that that the radial velocity trends are the result of star spots. This target has been previously reported by \citet{2008ApJ...683..844L} to have no resolved companions. }\label{fig:Cha__T7}\end{figure}

\clearpage \begin{figure}\includegraphics[width=\textwidth]{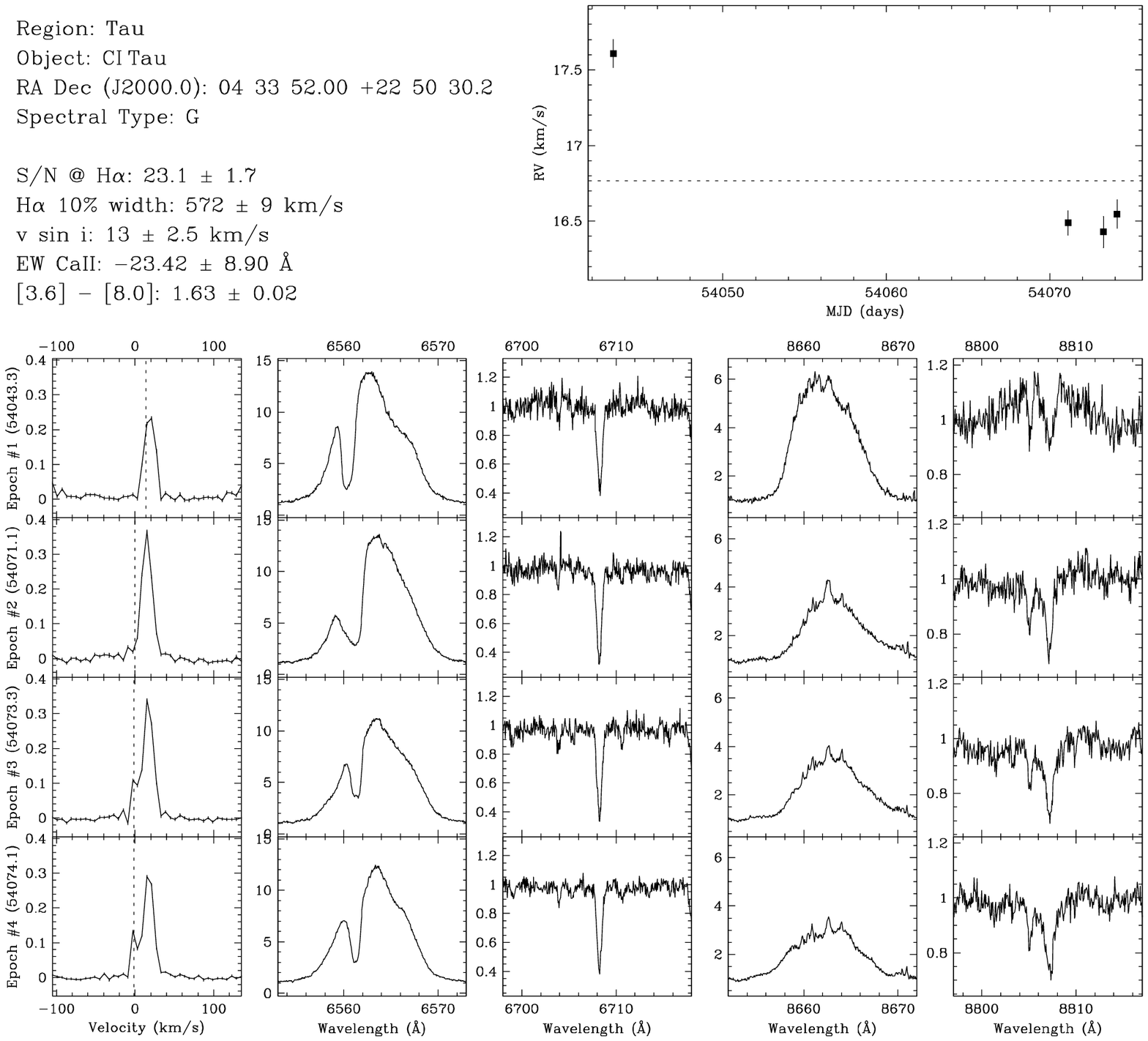}\caption[Plots and information on CI\,Tau]{\footnotesize CI\,Tau is a suspected SB1. The overall radial velocity scatter is larger than expected from non-companion influences, e.g., star spots. If star spots were responsible for the changes in radial velocity, based on observed $v~\sin~i$ ($\sim\!13$\,km\,s$^{-1}$) and model stellar radius ($\sim\!4.0\,R_{\odot}$), one would expect variations on a maximum timescale of $\sim\!15.5$\,days. From the stable results of the second observing run (epochs \#2, \#3 and \#4) which span $3.0$\,days, it is unlikely that the radial velocity trends are the result of star spots. For epoch \#3 and \#4, the peaks in the broadening function at the observer's rest frame are due to moonlight, and may have biased the radial velocity estimates toward the observer's rest frame. This target has been previously reported by \citet{1993AA...278..129L}, \citet{1993AJ....106.2005G}, and \citet{1995ApJ...443..625S} to have no resolved companions. }\label{fig:Tau__CITau}\end{figure}

\clearpage \begin{figure}\includegraphics[width=\textwidth]{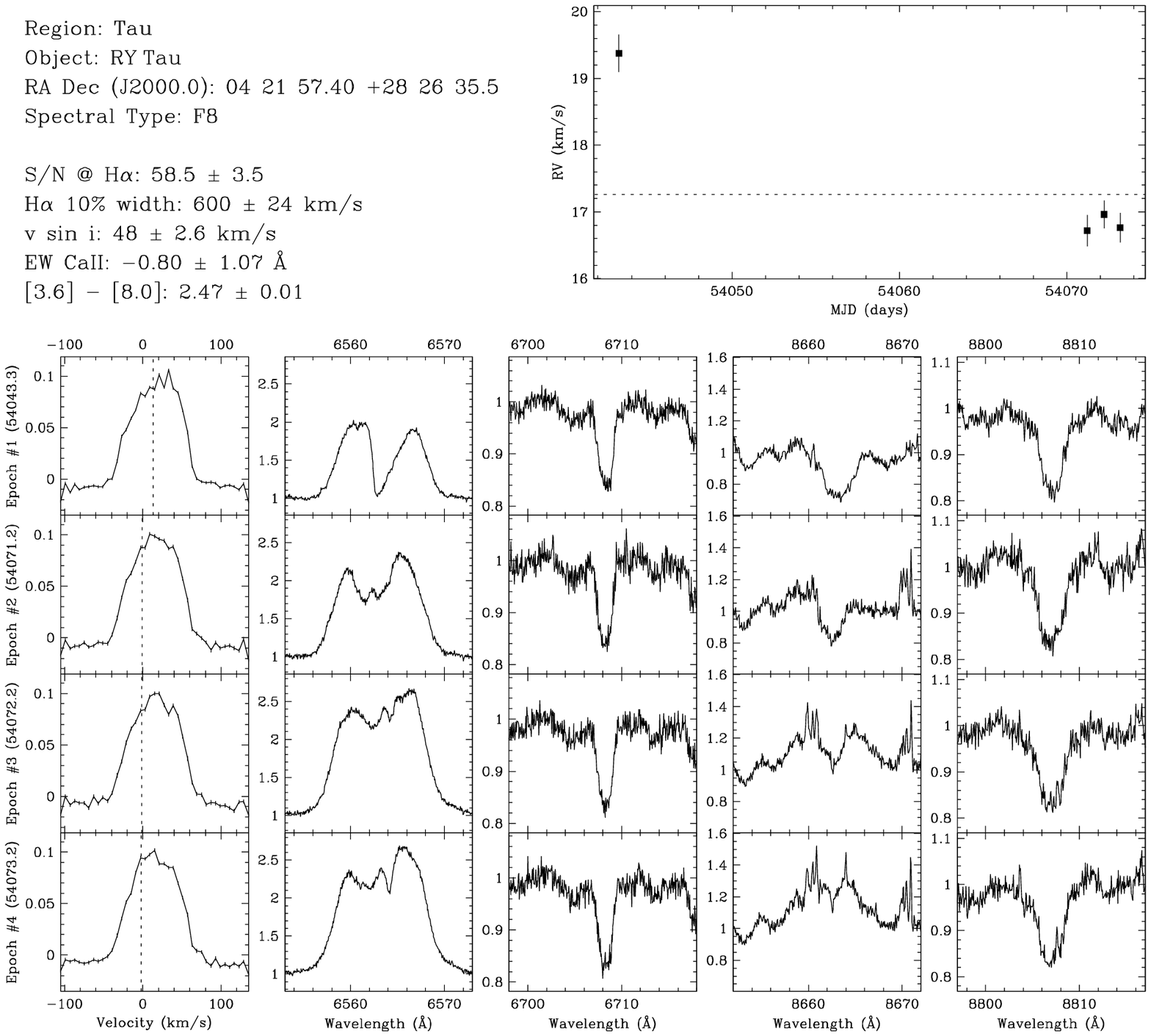}\caption[Plots and information on RY\,Tau]{\footnotesize RY\,Tau is a suspected SB1. The overall radial velocity scatter is larger than expected from non-companion influences, e.g., star spots. If star spots are responsible for the changes in radial velocity, based on observed $v\,\sin\,i$ ($\sim\!48$\,km\,s$^{-1}$) and model stellar radius ($\sim\!4.6\,R_{\odot}$), one would expect variations on a maximum timescale of $\sim\!4.9$\,days. From the stable results of the second observing run (epochs \#2, \#3 and \#4) which span $2.0$\,days, it is unlikely that the radial velocity trends are the result of star spots. This target has been previously reported by \citet{1993AA...278..129L}, and \citet{1993AJ....106.2005G} to have no resolved companions. }\label{fig:Tau__RYTau}\end{figure}

\clearpage \begin{figure}\includegraphics[width=\textwidth]{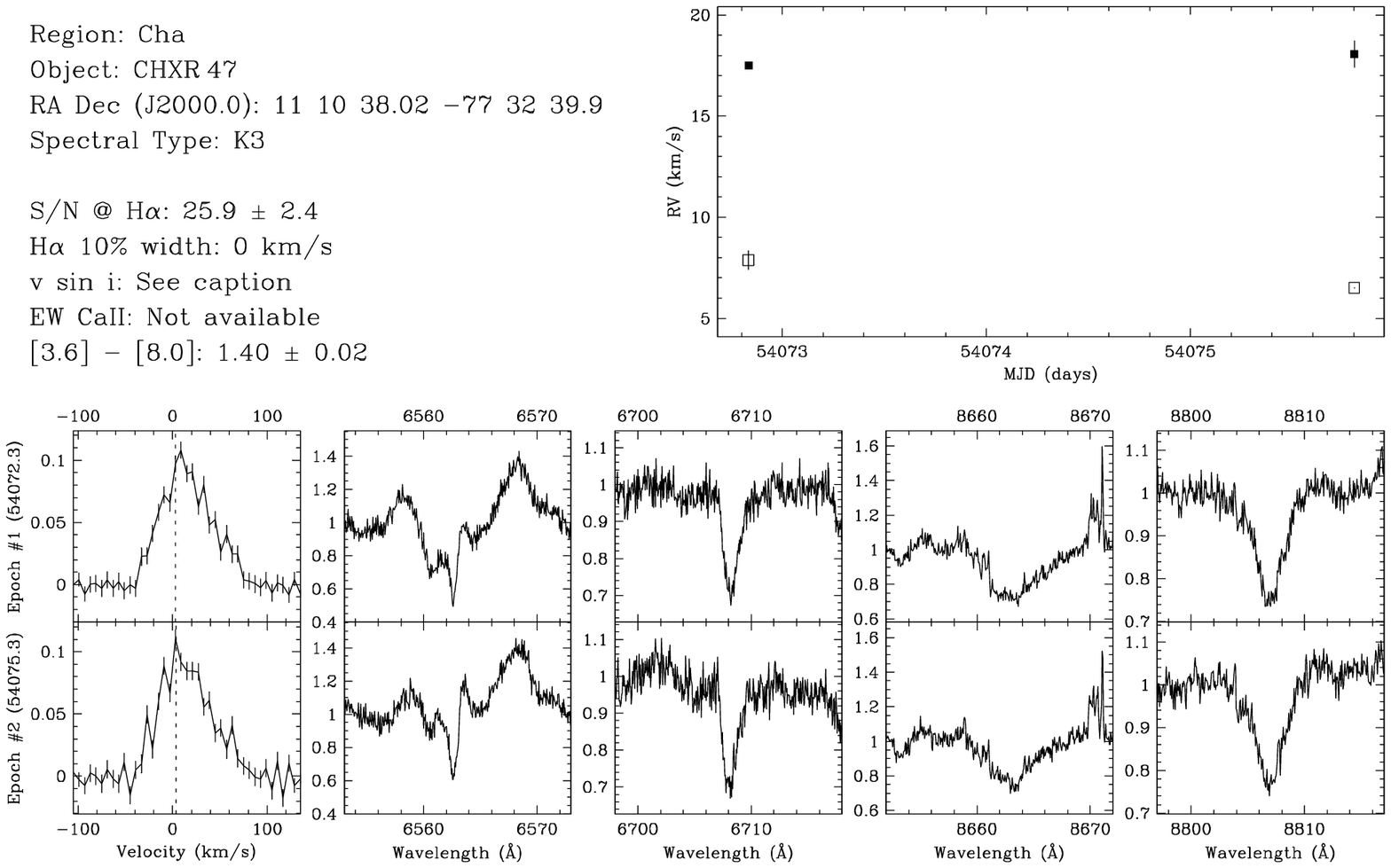}\caption[Plots and information on CHXR\,47]{\footnotesize CHXR\,47 is a long-period SB2. By fitting the broadening function to two rotational broadening line profiles, we estimate the two sources have a flux ratio of $0.24\,\pm\,0.06$, and the A and B sources have $v\,\sin\,i$ of $59.5\,\pm\,0.4$\,km~s$^{-1}$ and $24\,\pm\,3$\,km~s$^{-1}$, respectively. However, these results and the derived radial velocities should be read with caution because of the broad and blended peaks in the broadening functions of this system which are difficult to decompose and fit reliably. This target has been previously reported by \citet{2008ApJ...683..844L} to have a resolved companion with a separation of $\sim\!0\farcs\!175$ ($\sim\!25$\,AU) at a position angle of $\sim\!334\fdg\!8$, and an $R$-band flux ratio of $\sim\!0.49$ (based on $\Delta K$$\sim\!0.47$). This flux ratio is comparable to that estimated between the SB2 components. Furthermore, the resolved companion has an expected circular orbital speed of $\sim\!7$\,km\,s$^{-1}$ which is compatible with the estimated radial velocity separations of the SB2 component stars. Therefore, the resolved companion is likely the SB2 secondary star.}\label{fig:Cha__CHXR-47}\end{figure}

\clearpage \begin{figure}\includegraphics[width=\textwidth]{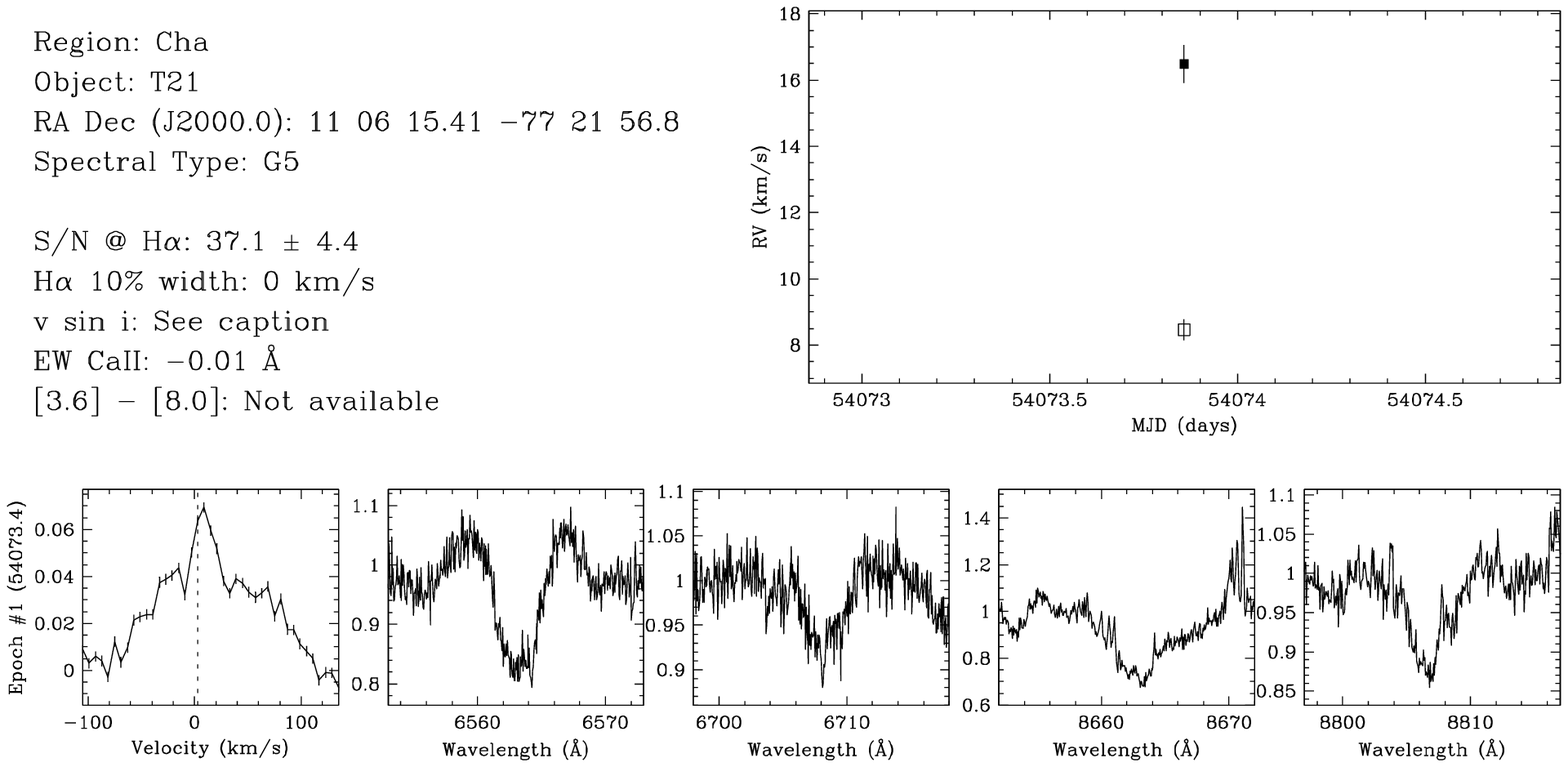}\caption[Plots and information on T21]{\footnotesize T21 is a long-period SB2. Two sources can be seen in the broadening function: a strong peak and a shallower broad feature. By fitting the broadening function to two rotational broadening line profiles, we estimate the two sources have a flux ratio of $0.101\,\pm\,0.005$, and the A and B sources have $v\,\sin\,i$ of $94.1\,\pm\,0.7$\,km~s$^{-1}$ and $14.5\,\pm\,0.3$\,km~s$^{-1}$, respectively. This target has been previously reported by \citet{2008ApJ...683..844L} to have a resolved companion with a separation of $\sim\!0\farcs\!14$ ($\sim\!20$\,AU) at a position angle of $\sim\!126\fdg\!1$, and an $R$-band flux ratio of $\sim\!0.08$ (based on $\Delta K$$\sim\!2.16$). This flux ratio is similar to that estimated between the SB2 components. Furthermore, the resolved companion has an expected circular orbital speed of $\sim\!13$\,km\,s$^{-1}$ which is compatible with the radial velocity separations of the SB2 components star if we consider the large uncertainties due to rapid rotation. Therefore, the resolved companion is likely the SB2 secondary star.}\label{fig:Cha__T21}\end{figure}

\clearpage \begin{figure}\includegraphics[width=\textwidth]{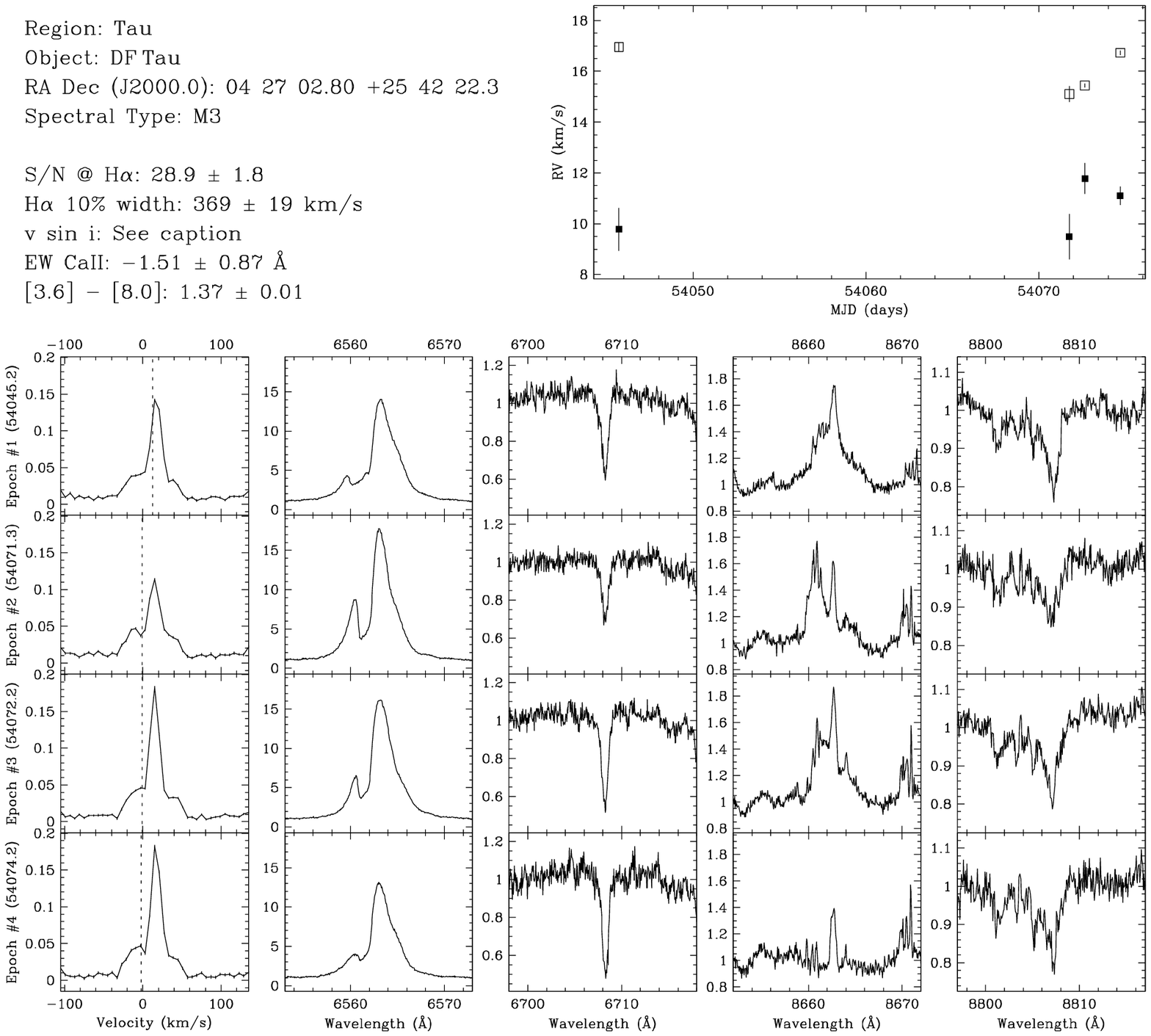}\caption[Plots and information on DF\,Tau]{\footnotesize DF\,Tau is a long-period SB2. Two sources can be seen in the broadening function: a strong peak and a shallower broad feature. By fitting the broadening function to two rotational broadening line profiles, we estimate the two sources have a flux ratio of $0.49\,\pm\,0.11$, and the A and B sources have $v\,\sin\,i$ of $46.6\,\pm\,1.8$\,km~s$^{-1}$ and $9.8\,\pm\,0.6$\,km~s$^{-1}$, respectively. This target has been previously reported by \citet{1993AJ....106.2005G}, \citet{1995ApJ...443..625S}, and \citet{1997ApJ...490..353G} as a resolved binary (DF~Tau~A+B) with a separation of $0\farcs\!0871$--$0\farcs\!088$ ($\sim\!12$\,AU) at a position angle of $301\fdg\!2$--$329^\circ$, and an $R$-band flux ratio of $0.05$--$0.24$ (based on $\Delta K$$\sim\!0.41$--$0.90$). This flux ratio is comparable to that estimated between the SB2 components with we consider the uncertainties in the models. Furthermore, the resolved secondary star has an expected circular orbital speed of $\sim\!5$\,km\,s$^{-1}$ which is consistent with the radial velocity separations of the SB2 component stars. Therefore, the resolved secondary star is likely the SB2 secondary star.}\label{fig:Tau__DFTauA+B}\end{figure}

\clearpage \begin{figure}\includegraphics[width=\textwidth]{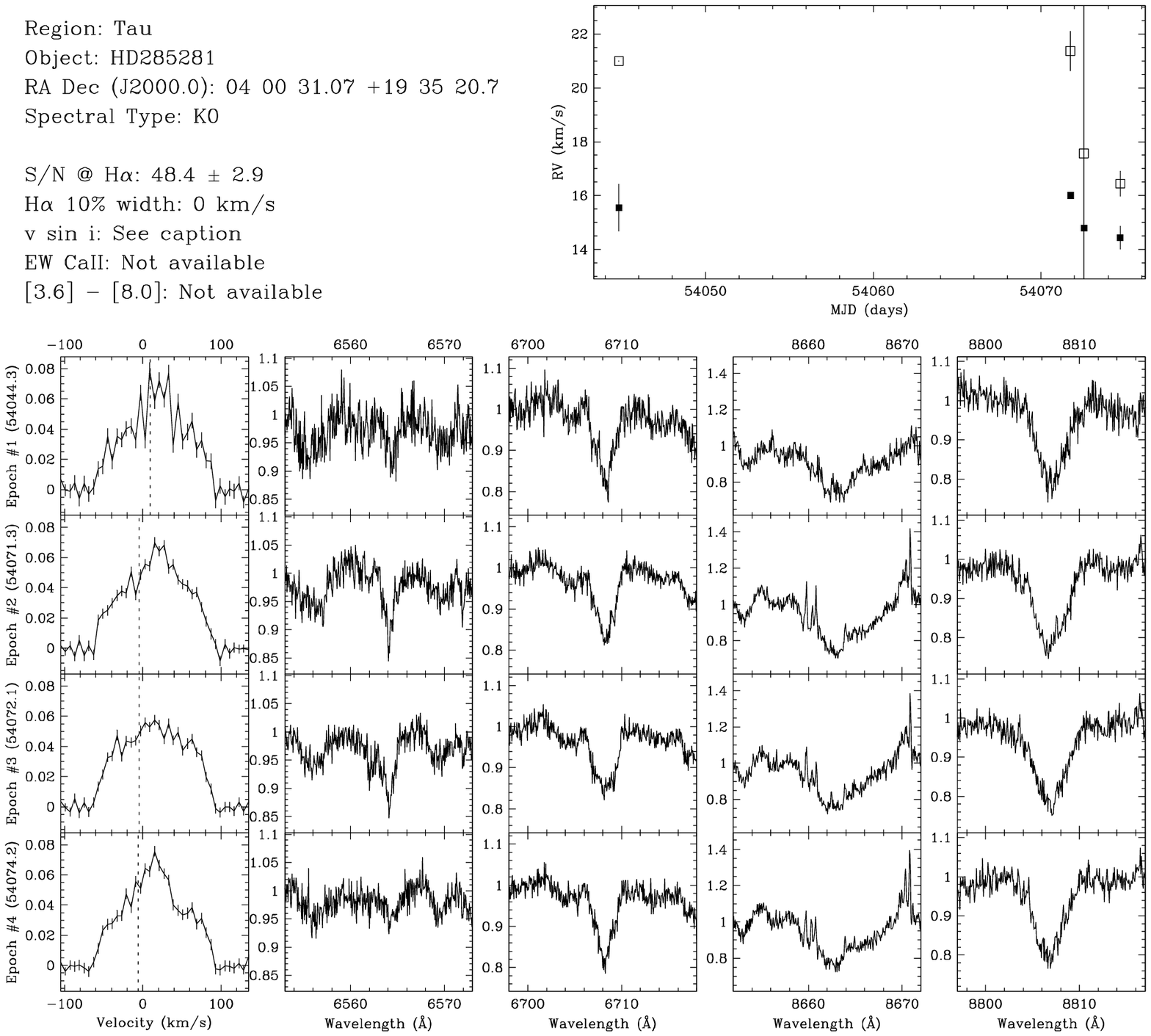}\caption[Plots and information on HD285281]{\footnotesize HD285281 is a long-period SB2. Two sources can be seen in the profile: a peak (at $\sim\!20$\,km\,s$^{-1}$) on top of a broad feature. By fitting the broadening function to two rotational broadening line profiles, we estimate the two sources have a flux ratio of $0.07\,\pm\,0.05$, and the A and B sources have $v\,\sin\,i$ of $78.0\,\pm\,0.3$\,km~s$^{-1}$ and $17.0\,\pm\,1.9$\,km~s$^{-1}$, respectively. The radial velocity estimates for epoch \#3 are inaccurate because the fitting routine could not delineate the two individual profiles. This target has been previously reported by \citet{1998AA...331..977K} to have a resolved companion with a separation of $\sim\!0\farcs\!773$ ($\sim\!110$\,AU) at a position angle of $\sim\!190\fdg\!7$, and an $R$-band flux ratio of $\sim\!0.21$ (based on $\Delta K$$\sim\!1.23$). This flux ratio is comparable to that estimated between the SB2 components. Furthermore, the resolved companion has an expected circular orbital speed of $\sim\!5$\,km\,s$^{-1}$ which is consistent with the radial velocity separations of the SB2 component stars. Therefore, the resolved companion is likely the SB2 secondary star.}\label{fig:Tau__HD285281}\end{figure}

\clearpage \begin{figure}\includegraphics[width=\textwidth]{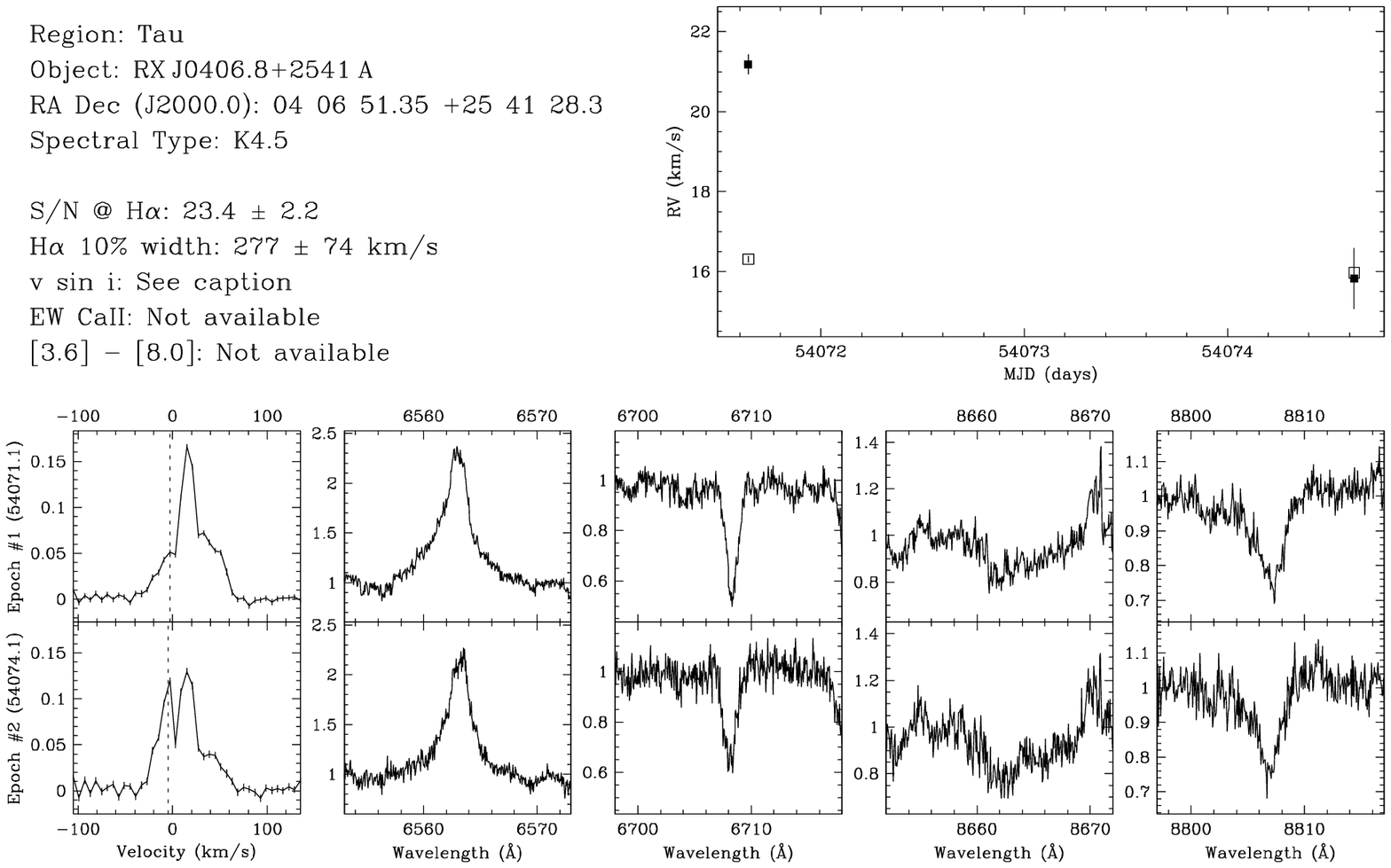}\caption[Plots and information on RX\,J0406.8+2541\,A]{\footnotesize RX\,J0406.8+2541\,A is a long-period SB2. Two sources can be seen in the broadening function: a strong peak and a shallower broad feature. By fitting the broadening function to two rotational broadening line profiles, we estimate the two sources have a flux ratio of $0.32\,\pm\,0.07$, and the a and b sources have $v\,\sin\,i$ of $47\,\pm\,4$\,km~s$^{-1}$ and $9.8\,\pm\,0.3$\,km~s$^{-1}$, respectively. For epoch \#2, the strong peak in the broadening function at the observer's rest frame is due to moonlight, and biased the radial velocity estimates in the fits. This target has been previously reported by \citet{1998AA...331..977K} to have a resolved companion (RX~J0406.8+2541~B) with a separation of $\sim\!0\farcs\!977$ ($\sim\!140$\,AU) at a position angle of $\sim\!12\fdg\!3$, and an $R$-band flux ratio of $\sim\!0.95$ (based on $\Delta K$$\sim\!0.04$). This flux ratio is comparable to that estimated between the SB2 components if one accounts for the diminished light from the companion at that distance from the slit. Furthermore, the resolved companion has an expected circular orbital speed of $\sim\!3$\,km\,s$^{-1}$ which is similar to the radial velocity separations of the SB2 component stars. Furthermore, the radial velocity estimates of the SB2 secondary star is consistent with that of RX~J0406.8+2541~B. Therefore, the resolved companion is likely the SB2 secondary star.}\label{fig:Tau__RXJ0406.8+2541s}\end{figure}

\clearpage \begin{figure}\includegraphics[width=\textwidth]{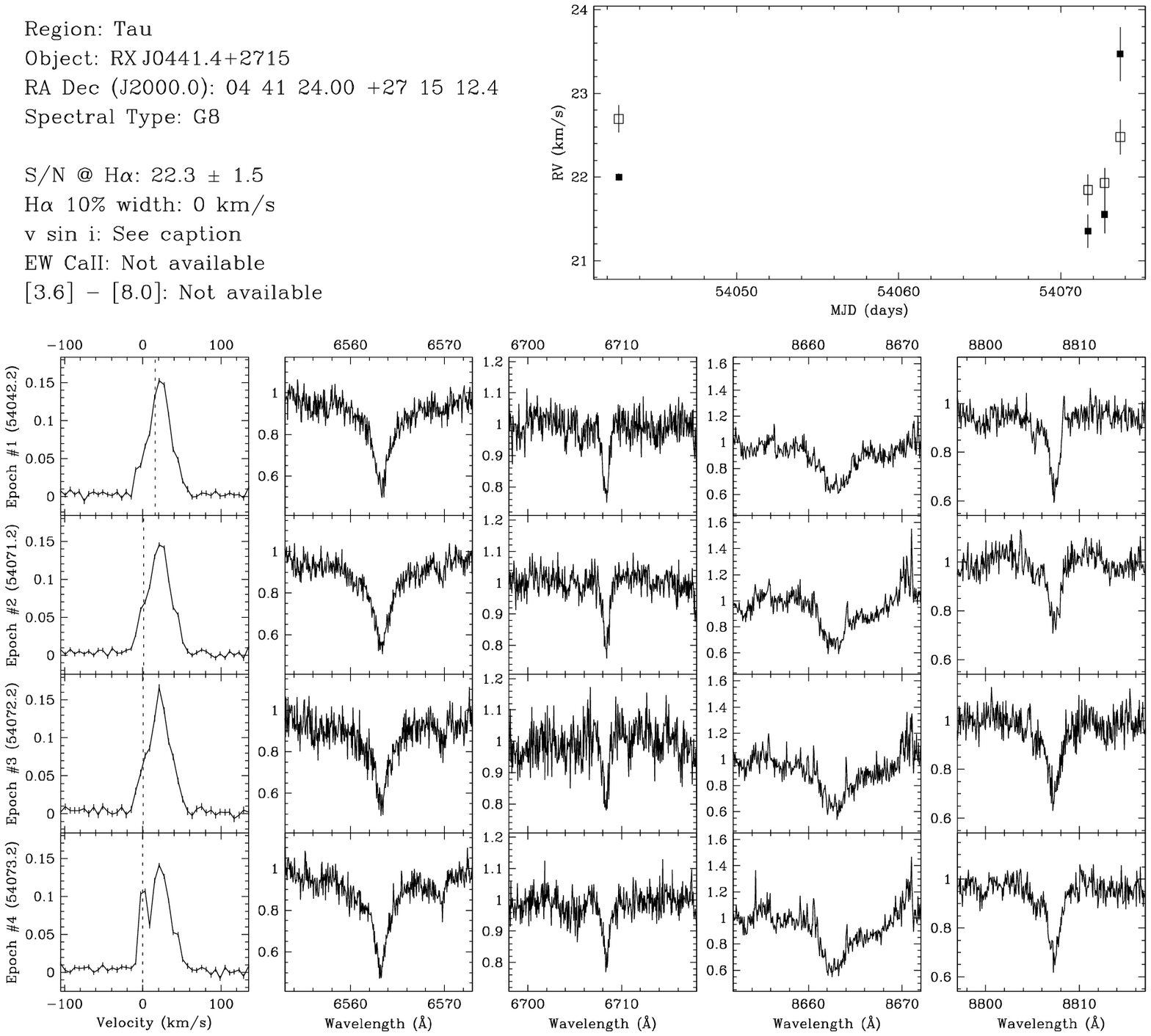}\caption[Plots and information on RX\,J0441.4+2715]{\footnotesize RX\,J0441.4+2715 is a long-period SB2. Two sources can be seen in the broadening function: a strong peak and a shallower broad feature. By fitting the broadening function to two rotational broadening line profiles, we estimate the two sources have a flux ratio of $0.32\,\pm\,0.07$, and the A and B sources have $v\,\sin\,i$ of $37.0\,\pm\,0.6$\,km~s$^{-1}$ and $12.6\,\pm\,1.5$\,km~s$^{-1}$, respectively. The overall radial velocity deviates from that of the star-forming region. However, cluster membership is supported by Li-$\lambda$6708 absorption. For epoch \#4, the sharp peak in the broadening function at the observer's rest frame is due to moonlight, and biased the radial velocity estimates in the fits. This target has been previously reported by \citet{1998AA...331..977K} to have a resolved companion with a separation of $\sim\!0\farcs\!065$ ($\sim\!9.1$\,AU) at a position angle of $\sim\!216^\circ$, and an $R$-band flux ratio of $\sim\!0.49$ (based on $\Delta K$$\sim\!0.63$). This flux ratio is comparable to that estimated between the SB2 components. Furthermore, the resolved companion has an expected circular orbital speed of $\sim\!21$\,km\,s$^{-1}$ which is consistent with the radial velocity separations of the SB2 component stars. Therefore, the resolved companion is likely the SB2 secondary star.}\label{fig:Tau__RXJ0441.4+2715}\end{figure}

\clearpage \begin{figure}\includegraphics[width=\textwidth]{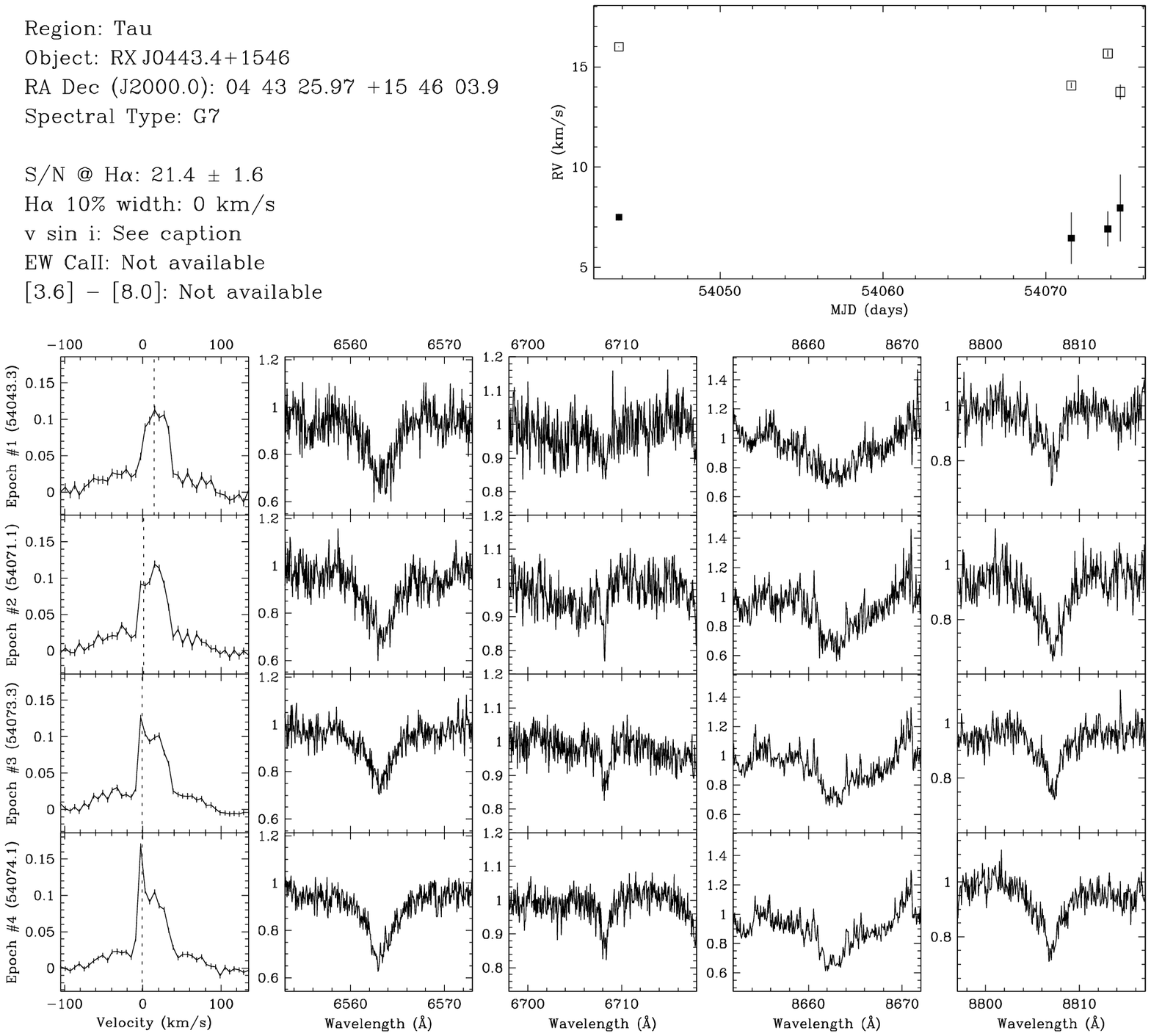}\caption[Plots and information on RX\,J0443.4+1546]{\footnotesize RX\,J0443.4+1546 is a long-period SB2. Two sources can be seen in the broadening function: a strong peak and a shallower broad feature. By fitting the broadening function to two rotational broadening line profiles, we estimate the two sources have a flux ratio of $0.83\,\pm\,0.10$, and the A and B sources have $v\,\sin\,i$ of $86.5\,\pm\,1.6$\,km~s$^{-1}$ and $24.0\,\pm\,0.8$\,km~s$^{-1}$, respectively. For epochs \#3 and \#4, the peaks in the broadening function at the observer's rest frame are due to moonlight, and have been accounted for in the radial velocity estimates. This target has been previously reported by \citet{1998AA...331..977K} to have no resolved companions. }\label{fig:Tau__RXJ0443.4+1546}\end{figure}

\clearpage \begin{figure}\includegraphics[width=\textwidth]{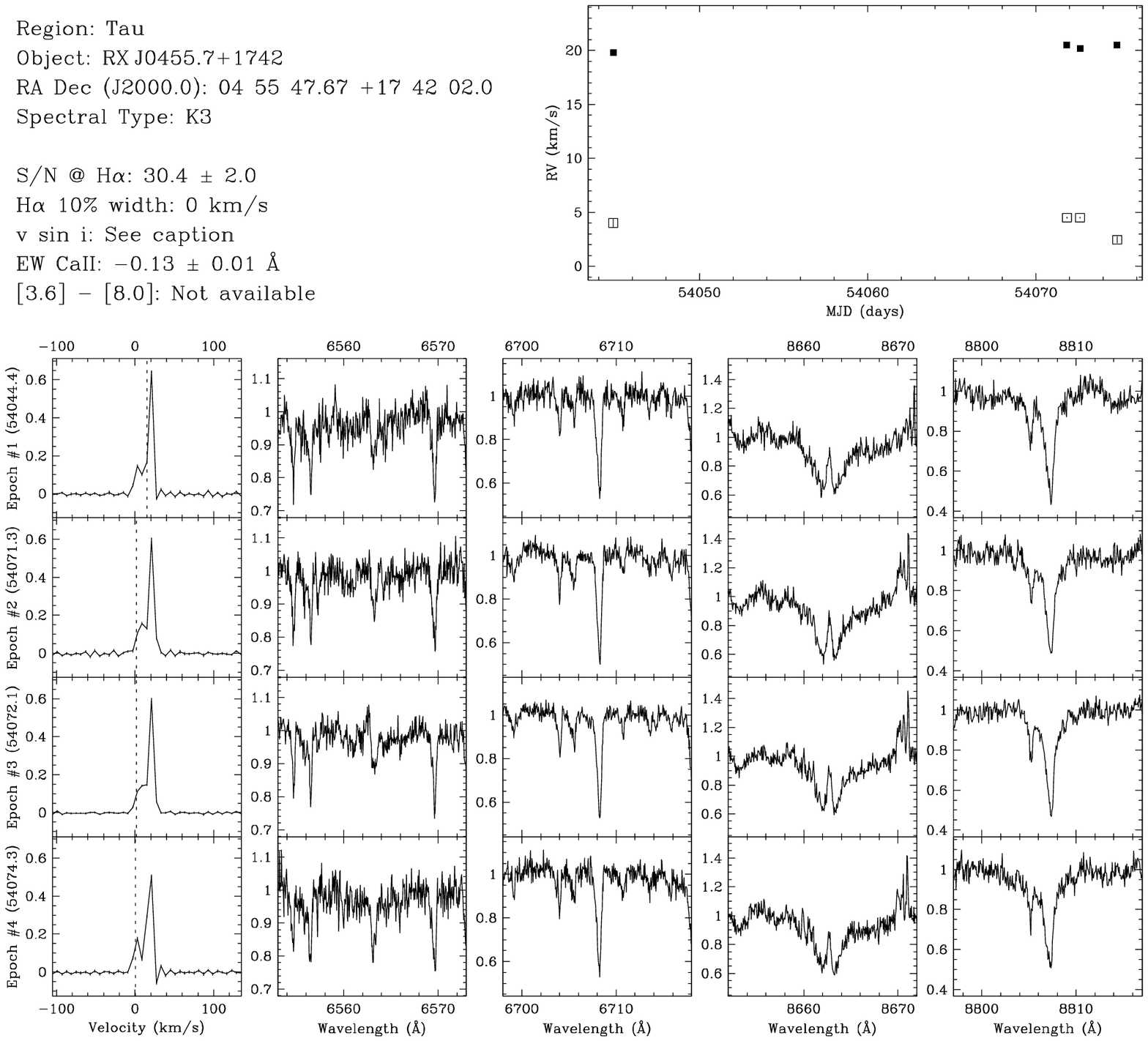}\caption[Plots and information on RX\,J0455.7+1742]{\footnotesize RX\,J0455.7+1742 is a long-period SB2. By fitting the broadening function to two rotational broadening line profiles, we estimate the two sources have a flux ratio of $0.337\,\pm\,0.014$, and the A and B sources have $v\,\sin\,i$ of $8\,\pm\,2$\,km~s$^{-1}$ and $9.5\,\pm\,0.4$\,km~s$^{-1}$, respectively. For epoch \#4, the small peak in the broadening function at the observer's rest frame has contribution from moonlight, and may have biased the radial velocity estimate toward the observer's rest frame. This target has been previously reported by \citet{1998AA...331..977K} to have a resolved companion with a separation of $\sim\!0\farcs\!093$ ($\sim\!13$\,AU) at a position angle of $\sim\!254\fdg\!6$, and an $R$-band flux ratio of $\sim\!0.53$ (based on $\Delta K$$\sim\!0.41$). This flux ratio is similar to that estimated between the SB2 components. Furthermore, the resolved companion has an expected circular orbital speed of $\sim\!11$\,km\,s$^{-1}$ which is comparable with the radial velocity separation of the SB2 component stars if we consider the large uncertainties due to rapid rotation. Therefore, the resolved companion is likely the SB2 secondary star.}\label{fig:Tau__RXJ0455.7+1742}\end{figure}

\clearpage \begin{figure}\includegraphics[width=\textwidth]{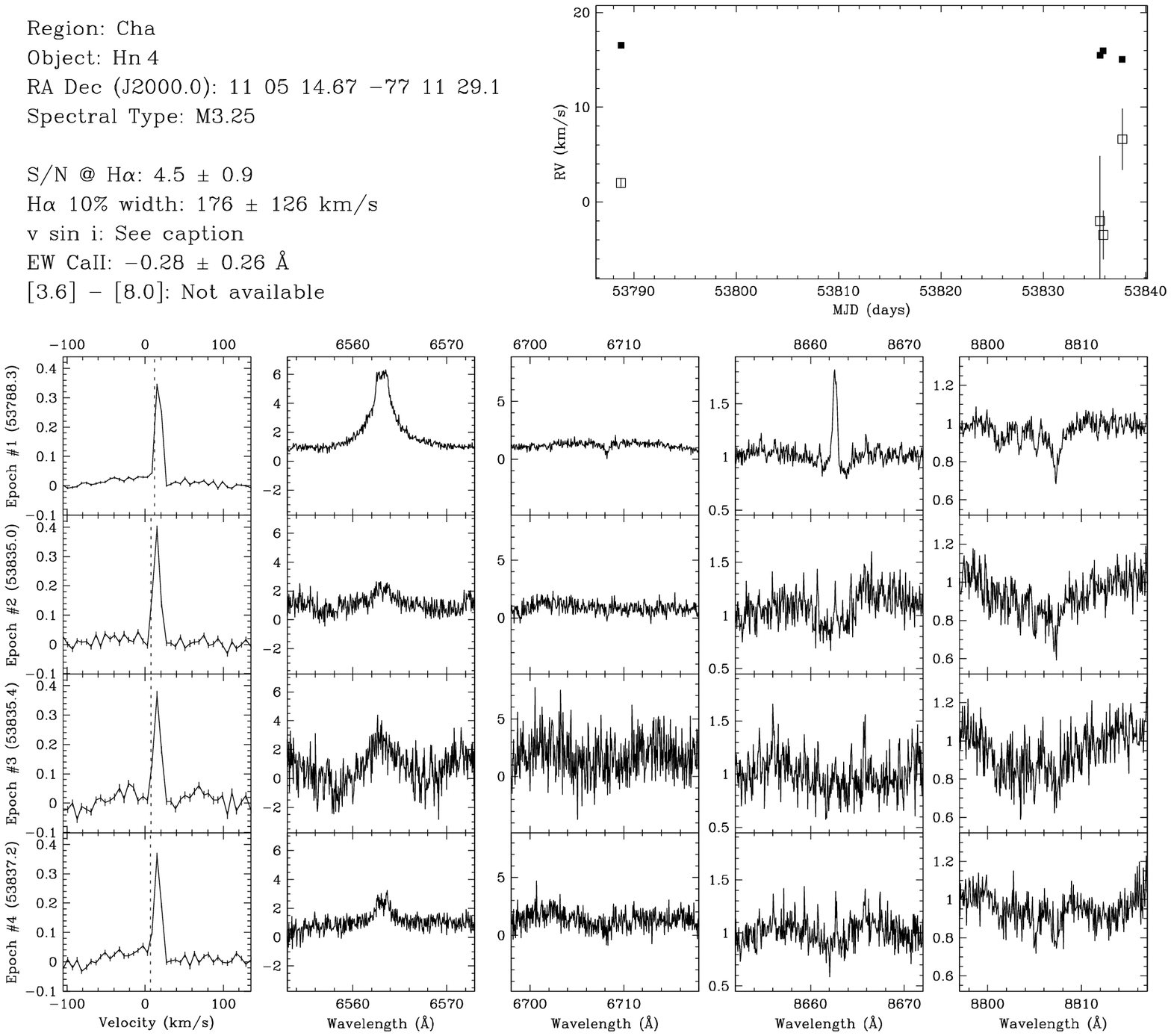}\caption[Plots and information on Hn\,4]{\footnotesize Hn\,4 is a suspected SB2. Two sources are hinted at in the broadening function: a strong peak and a very shallow broad feature. By fitting the broadening function to two rotational broadening line profiles, we estimate the two sources have a flux ratio of $0.62\,\pm\,0.09$, and the A and B sources have $v\,\sin\,i$ of $9.0\,\pm\,0.5$\,km~s$^{-1}$ and $90\,\pm\,15$\,km~s$^{-1}$, respectively. This target has been previously reported by \citet{2008ApJ...683..844L} to have a resolved companion with a separation of $\sim\!0\farcs\!211$ ($\sim\!30$\,AU) at a position angle of $\sim\!296\fdg\!1$, and an $R$-band flux ratio of $\sim\!0.89$ (based on $\Delta K$$\sim\!0.04$). This flux ratio is similar to that estimated between the SB2 components. Furthermore, the resolved companion has an expected circular orbital speed of $\sim\!4$\,km\,s$^{-1}$ which is compatible with the radial velocity separations of the SB2 components star if we consider the large uncertainties due to rapid rotation. Therefore, the resolved companion is likely the SB2 secondary star.}\label{fig:Cha__Hn-4}\end{figure}

\clearpage \begin{figure}\includegraphics[width=\textwidth]{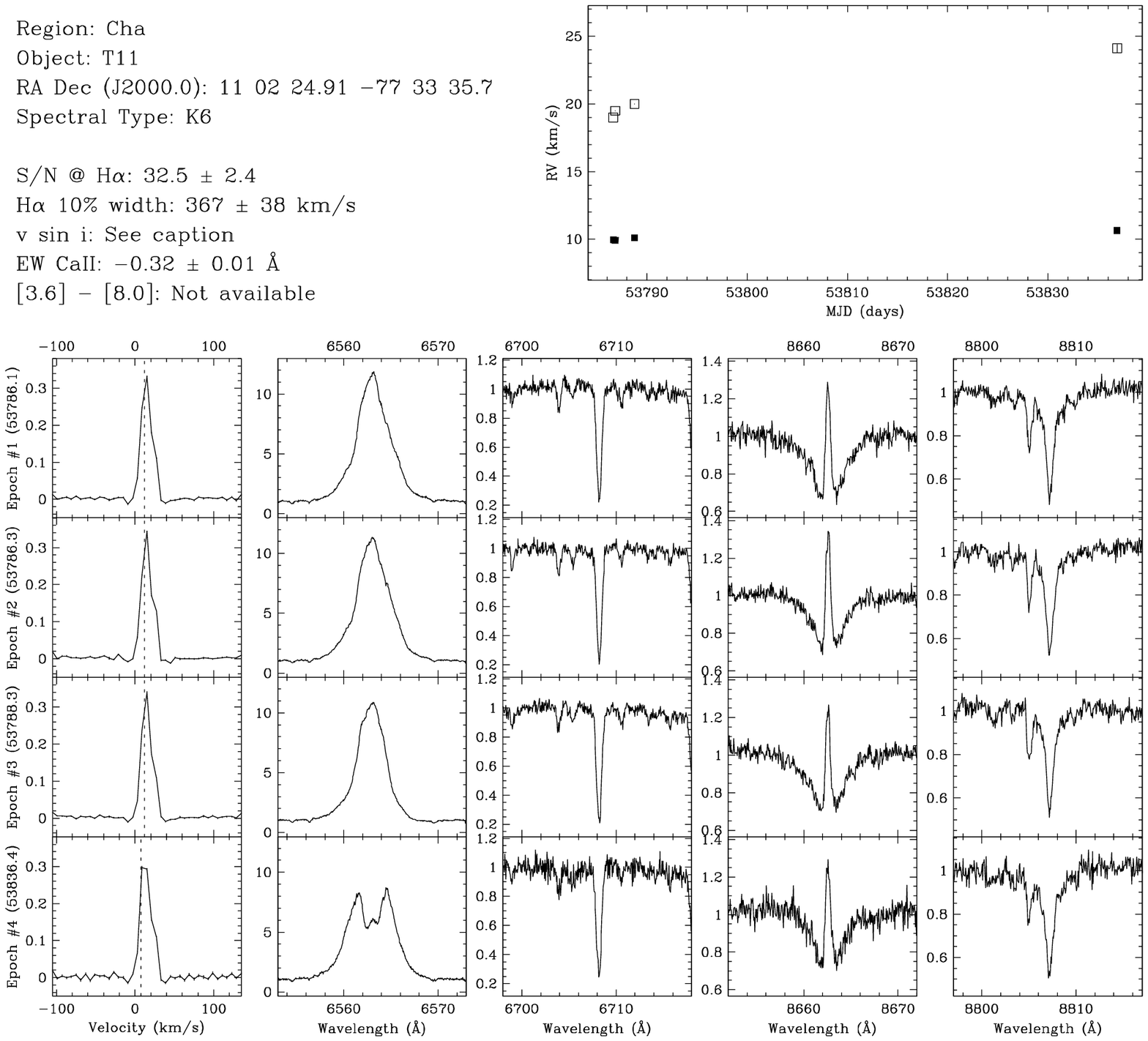}\caption[Plots and information on T11]{\footnotesize T11 is a suspected SB2. The broadening function is unusually asymmetrical, and has both a strong peak and a shallower slightly broader feature. If star spots are responsible for the asymmetries in the broadening function, based on observed $v\,\sin\,i$ ($\sim\!14$\,km\,s$^{-1}$) and model stellar radius ($\sim\!1.4\,R_{\odot}$), one would expect variations on a maximum timescale of $\sim\!5.1$\,days. From the constant broadening functions of the first observing run (epochs \#1, \#2 and \#3) which span 2.2\,days, it is unlikely that the broadening function irregularities are the result of star spots. By fitting the broadening function to two rotational broadening line profiles, we estimate the two sources have a flux ratio of $1.0\,\pm\,0.4$, and the A and B sources have $v\,\sin\,i$ of $11.2\,\pm\,0.3$\,km~s$^{-1}$ and $13.0\,\pm\,0.3$\,km~s$^{-1}$, respectively. This target has been previously reported by \citet{2008ApJ...683..844L} to have no resolved companions. }\label{fig:Cha__T11}\end{figure}

\clearpage \begin{figure}\includegraphics[width=\textwidth]{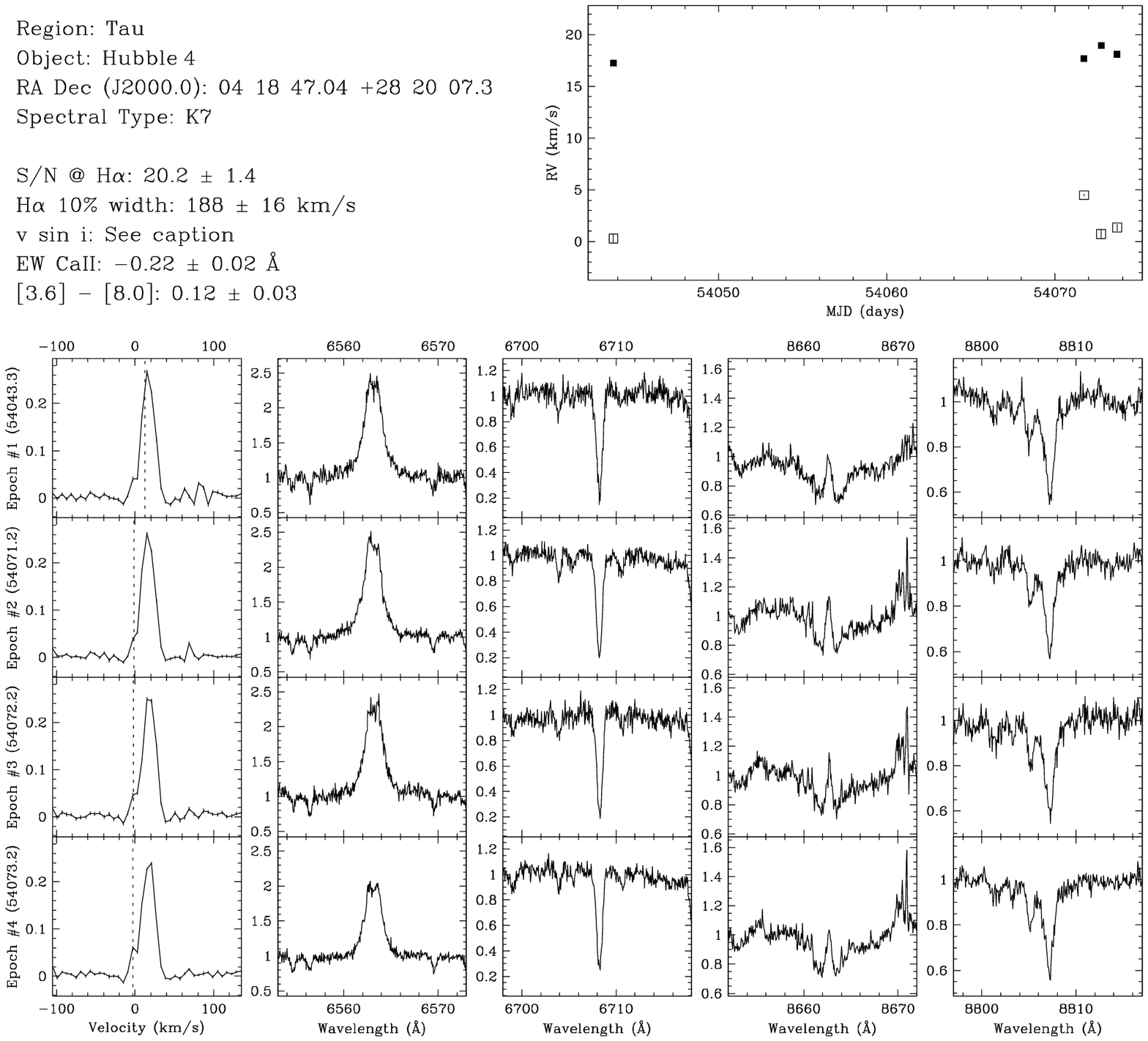}\caption[Plots and information on Hubble\,4]{\footnotesize Hubble\,4 is a suspected SB2. If star spots are responsible for the asymmetries in the broadening function, based on observed $v\,\sin\,i$ ($\sim\!16$\,km\,s$^{-1}$) and model stellar radius ($\sim\!1.4\,R_{\odot}$), one would expect variations on a maximum timescale of $\sim\!4.3$\,days. From the constant broadening functions of the second observing run (epochs \#2, \#3 and \#4) which span $2.0$\,days, it is unlikely that the broadening function irregularities are the result of star spots. By fitting the broadening function to two rotational broadening line profiles, we estimate the two sources have a flux ratio of $0.20\,\pm\,0.05$, and the A and B sources have $v\,\sin\,i$ of $12.1\,\pm\,0.3$\,km~s$^{-1}$ and $13.1\,\pm\,1.9$\,km~s$^{-1}$, respectively. This target has been previously reported by \citet{1993AA...278..129L}, and \citet{1993AJ....106.2005G} to have no resolved companions. }\label{fig:Tau__Hubble4}\end{figure}

\clearpage

\LongTables
\begin{landscape}
\begin{deluxetable}{lcccccccccc}
\tablecolumns{11}
\tabletypesize{\scriptsize}
\tablewidth{0pt}
\tablecaption{Measurements of Stars without Close Companions in Cha~I \& Tau-Aur\label{tbl:SingleStars}}
\tablehead{
\colhead{Object} & \colhead{R.A.} & \colhead{Dec.} & \colhead{SpT} & \colhead{TTS\tablenotemark{a}} & \colhead{$10\%$\,width\tablenotemark{bc}} & \colhead{$v~\!\sin~\!i$\tablenotemark{de}} & \colhead{$\overline{RV}\tablenotemark{f}$} & \colhead{$\sigma_{RV}\tablenotemark{g}$} & \colhead{$\sigma_{\rm N}\tablenotemark{h}$} & \colhead{\# of}\\
& \colhead{(J2000.0)} & \colhead{(J2000.0)} & & & \colhead{(km~s$^{-1}$)} & \colhead {(km~s$^{-1}$)} & \colhead{(km~s$^{-1}$)} & \colhead{(km~s$^{-1}$)} & \colhead{(km~s$^{-1}$)} & \colhead{obs.}
}
\startdata
\cutinhead{Cha~I}

T4\dotfill & $10~56~30.45$ & $-77~11~39.3$ & M0.5 & c & $341\,\pm\,28$ & $12.4\,\pm\,0.4$ & $15.30\,\pm\,0.03$ & $0.58$ & $0.77$ & $5$\\
T5\dotfill & $10~57~42.20$ & $-76~59~35.7$ & M3.25 & c & $324\,\pm\,30$ & $13.8\,\pm\,0.9$ & $16.55\,\pm\,0.08$ & $0.23$ & $0.38$ & $4$\\
T6\dotfill & $10~58~16.77$ & $-77~17~17.1$ & K0 & c & $284$ & $34\,\pm\,2$ & $14.99\,\pm\,0.08$ & \nodata & \nodata & $1$\\

T8\dotfill & $10~59~06.99$ & $-77~01~40.4$ & K2 & c & $347$ & $35\,\pm\,2$ & $14.99\,\pm\,0.06$ & \nodata & \nodata & $1$\\
T10\dotfill & $11~00~40.22$ & $-76~19~28.1$ & M3.75 & c & $252\,\pm\,24$ & $5.4\,\pm\,0.8$ & $15.69\,\pm\,0.03$ & $0.25$ & $0.48$ & $5$\\
CHXR\,9C\,A\dotfill & $11~01~18.75$ & $-76~27~02.5$ & M0.5 & w & $106\,\pm\,13$ & $12.8\,\pm\,0.2$ & $14.50\,\pm\,0.04$ & $0.32$ & $0.34$ & $5$\\
CHXR\,9C\,Ba+Bb\dotfill & $11~01~18.75$ & $-76~27~02.5$ & M1.5 & w & $132\,\pm\,12$ & $19.6\,\pm\,0.8$ & $14.89\,\pm\,0.09$ & $0.09$ & $0.09$ & $5$\\

CHXR\,71\dotfill & $11~02~32.65$ & $-77~29~13.0$ & M3 & w & $117\,\pm\,46$ & $17.1\,\pm\,1.3$ & $15.54\,\pm\,0.08$ & $0.40$ & $0.44$ & $4$\\
T12\dotfill & $11~02~55.05$ & $-77~21~50.8$ & M4.5 & c & $262\,\pm\,35$ & $10.7\,\pm\,0.2$ & $14.82\,\pm\,0.05$ & $0.53$ & $0.45$ & $4$\\

T14\dotfill & $11~04~09.09$ & $-76~27~19.4$ & K5 & c & $550\,\pm\,13$ & $7.8\,\pm\,0.2$ & $15.13\,\pm\,0.09$ & $0.25$ & $0.25$ & $4$\\
ISO\,52\dotfill & $11~04~42.58$ & $-77~41~57.1$ & M4 & w & $126\,\pm\,15$ & $9.9\,\pm\,0.6$ & $15.52\,\pm\,0.06$ & $0.18$ & $0.27$ & $4$\\
CHXR\,14N\dotfill & $11~04~51.00$ & $-76~25~24.1$ & K8 & w & $114\,\pm\,20$ & $13.7\,\pm\,0.6$ & $14.90\,\pm\,0.04$ & $0.58$ & $0.66$ & $5$\\
CHXR\,14S\dotfill & $11~04~52.85$ & $-76~25~51.5$ & M1.75 & w & $98\,\pm\,14$ & $5.7\,\pm\,0.3$ & $14.176\,\pm\,0.015$ & $0.089$ & $0.091$ & $4$\\
T16\dotfill & $11~04~57.01$ & $-77~15~56.9$ & M3 & w & $101$ & $11.3\,\pm\,0.9$ & $14.9\,\pm\,0.3$ & $0.4$ & $0.4$ & $4$\\

T20\dotfill & $11~05~52.61$ & $-76~18~25.6$ & M1.5 & w & $192\,\pm\,21$ & $48.3\,\pm\,1.4$ & $14.0\,\pm\,0.4$ & $0.9$ & $0.9$ & $4$\\

Hn\,5\dotfill & $11~06~41.81$ & $-76~35~49.0$ & M4.5 & c & $340\,\pm\,20$ & $7.8\,\pm\,0.4$ & $15.76\,\pm\,0.07$ & $0.51$ & $0.88$ & $4$\\
T22\dotfill & $11~06~43.47$ & $-77~26~34.4$ & M3 & w & $228$ & $60\,\pm\,10$ & $16\,\pm\,3$ & $1$ & $2$ & $4$\\
CHXR\,20\dotfill & $11~06~45.10$ & $-77~27~02.3$ & K6 & w & Absorp. & $14.6\,\pm\,0.9$ & $15.46\,\pm\,0.06$ & $1.18$ & $0.10$ & $3$\\
CHXR\,74\dotfill & $11~06~57.33$ & $-77~42~10.7$ & M4.25 & w & $97\,\pm\,7$ & $5.8\,\pm\,1.0$ & $17.85\,\pm\,0.03\ddagger$ & $0.25$ & $0.33$ & $4$\\
CHXR\,21\dotfill & $11~07~11.49$ & $-77~46~39.4$ & M3 & w & $135\,\pm\,61$ & $48\,\pm\,5$ & $11\,\pm\,2\ddagger$ & $3$ & $3$ & $4$\\
T24\dotfill & $11~07~12.07$ & $-76~32~23.2$ & M0.5 & c & $454\,\pm\,53$ & $10.5\,\pm\,0.4$ & $16.88\,\pm\,0.03$ & $0.09$ & $0.13$ & $4$\\
T25\dotfill & $11~07~19.15$ & $-76~03~04.8$ & M2.5 & c & $341\,\pm\,62$ & $12.6\,\pm\,0.3$ & $15.84\,\pm\,0.06$ & $0.11$ & $0.08$ & $4$\\
T26\dotfill & $11~07~20.74$ & $-77~38~07.3$ & G2 & c & $390$ & $32.7\,\pm\,1.6$ & $12.04\,\pm\,0.14\ddagger$ & \nodata & \nodata & $1$\\
CHXR\,76\dotfill & $11~07~35.19$ & $-77~34~49.3$ & M4.25 & w & $89\,\pm\,12$ & $9.8\,\pm\,0.6$ & $14.19\,\pm\,0.13$ & $0.64$ & $0.75$ & $4$\\
CHXR\,28\,Aa+Ab\dotfill & $11~07~55.89$ & $-77~27~25.8$ & K3.5 & ? & \nodata & $8.7\,\pm\,0.7$ & $16.76\,\pm\,0.05$ & $0.02$ & $0.02$ & $2$\\

ISO\,126\dotfill & $11~08~02.98$ & $-77~38~42.6$ & M1.25 & c & $381\,\pm\,19$ & $40\,\pm\,7$ & $14.0\,\pm\,1.3$ & $0.8$ & $0.3$ & $4$\\
T33\,A\dotfill & $11~08~15.10$ & $-77~33~53.2$ & K3.5 & w & $95\,\pm\,15$ & $12.9\,\pm\,0.5$ & $13.00\,\pm\,0.05$ & $0.23$ & $0.10$ & $4$\\
T33\,B\dotfill & $11~08~15.10$ & $-77~33~53.2$ & G7 & c & $318\,\pm\,21$ & $50\,\pm\,4$ & $7\,\pm\,2\ddagger$ & $3$ & $3$ & $4$\\
T34\dotfill & $11~08~16.49$ & $-77~44~37.2$ & M3.75 & w & $84\,\pm\,8$ & $5.8\,\pm\,0.8$ & $15.58\,\pm\,0.03$ & $0.21$ & $0.35$ & $5$\\
T35\dotfill & $11~08~39.05$ & $-77~16~04.2$ & K8 & c & $466\,\pm\,46$ & $21.0\,\pm\,1.8$ & $16.90\,\pm\,0.19$ & $1.07$ & $0.93$ & $6$\\
CHXR\,33\dotfill & $11~08~40.69$ & $-76~36~07.8$ & M0 & w & $153\,\pm\,3$ & $16.5\,\pm\,1.5$ & $18.71\,\pm\,0.08\ddagger$ & $0.30$ & $0.30$ & $2$\\
T38\dotfill & $11~08~54.64$ & $-77~02~13.0$ & M0.5 & c & $389\,\pm\,23$ & $18.7\,\pm\,1.5$ & $16.0\,\pm\,0.5$ & $4.2$ & $9.9$ & $4$\\
T39\,Aa\dotfill & $11~09~11.72$ & $-77~29~12.5$ & K7 & w & $131\,\pm\,27$ & $7.7\,\pm\,1.6$ & $14.97\,\pm\,0.04$ & $0.67$ & $0.59$ & $6$\\
T39\,Ab\dotfill & $11~09~11.72$ & $-77~29~12.5$ & M1.5 & w & $100\,\pm\,16$ & $4.1\,\pm\,0.3$ & $14.258\,\pm\,0.012$ & $0.191$ & $0.261$ & $5$\\

CHXR\,37\dotfill & $11~09~17.70$ & $-76~27~57.8$ & K7 & w & $178\,\pm\,11$ & $15.8\,\pm\,0.6$ & $14.82\,\pm\,0.06$ & $0.23$ & $0.23$ & $2$\\
CHXR\,79\dotfill & $11~09~18.13$ & $-76~30~29.2$ & M1.25 & w & $103\,\pm\,50$ & $12.4\,\pm\,1.5$ & $12.3\,\pm\,1.0\ddagger$ & $1.4$ & $1.4$ & $4$\\
T40\dotfill & $11~09~23.79$ & $-76~23~20.8$ & K6 & c & $498\,\pm\,78$ & $15.02\,\pm\,0.14$ & $11.0\,\pm\,1.0\ddagger$ & $2.7$ & $2.7$ & $2$\\
CHXR\,40\dotfill & $11~09~40.07$ & $-76~28~39.2$ & M1.25 & w & $104\,\pm\,18$ & $11.6\,\pm\,0.5$ & $14.46\,\pm\,0.02$ & $0.05$ & $0.01$ & $4$\\
Hn\,10E\dotfill & $11~09~46.21$ & $-76~34~46.4$ & M3.25 & c & $377\,\pm\,17$ & $8.2\,\pm\,0.5$ & $14.03\,\pm\,0.10$ & $0.68$ & $0.94$ & $4$\\

T43\dotfill & $11~09~54.08$ & $-76~29~25.3$ & M2 & c & $225\,\pm\,98$ & $18.2\,\pm\,1.0$ & $16.5\,\pm\,2.0$ & $1.4$ & $1.5$ & $4$\\
T45\dotfill & $11~09~58.74$ & $-77~37~08.9$ & K8 & c & $472\,\pm\,6$ & $6.6\,\pm\,0.3$ & $15.28\,\pm\,0.06$ & $0.97$ & $0.97$ & $2$\\
T44\dotfill & $11~10~00.11$ & $-76~34~57.9$ & K5 & c & $614\,\pm\,53$ & $71\,\pm\,3$ & $15\,\pm\,3$ & $6$ & $4$ & $4$\\

T45A\dotfill & $11~10~04.69$ & $-76~35~45.3$ & M0 & c & $340\,\pm\,77$ & $12.4\,\pm\,0.5$ & $16.94\,\pm\,0.06$ & $0.20$ & $0.20$ & $2$\\
T46\dotfill & $11~10~07.04$ & $-76~29~37.7$ & K8 & c & $437\,\pm\,27$ & $5.7\,\pm\,0.9$ & $14.79\,\pm\,0.04$ & $<\!0.01$ & $<\!0.01$ & $2$\\
ISO\,237\dotfill & $11~10~11.42$ & $-76~35~29.3$ & K5.5 & ? & \nodata & $19.8\,\pm\,1.3$ & $14.9\,\pm\,0.3$ & $0.8$ & $0.8$ & $4$\\

T47\dotfill & $11~10~49.60$ & $-77~17~51.8$ & M2 & c & $395\,\pm\,18$ & $16.2\,\pm\,0.9$ & $14.2\,\pm\,0.7$ & $0.7$ & $1.4$ & $4$\\
CHXR\,48\dotfill & $11~11~34.75$ & $-76~36~21.1$ & M2.5 & w & $110\,\pm\,12$ & $13.8\,\pm\,0.5$ & $15.29\,\pm\,0.05$ & $0.24$ & $0.33$ & $4$\\
T49\dotfill & $11~11~39.66$ & $-76~20~15.2$ & M2 & c & $280\,\pm\,25$ & $8.2\,\pm\,0.8$ & $16.26\,\pm\,0.05$ & $0.39$ & $0.40$ & $5$\\
CHX\,18N\dotfill & $11~11~46.32$ & $-76~20~09.2$ & K6 & w & $188\,\pm\,39$ & $26.5\,\pm\,1.3$ & $15.29\,\pm\,0.07$ & $0.91$ & $0.96$ & $5$\\
T50\dotfill & $11~12~09.85$ & $-76~34~36.6$ & M5 & c & $262\,\pm\,52$ & $12.0\,\pm\,0.4$ & $14.62\,\pm\,0.06$ & $0.74$ & $0.61$ & $4$\\
T51\dotfill & $11~12~24.41$ & $-76~37~06.4$ & K3.5 & c & $381\,\pm\,32$ & $32.7\,\pm\,1.1$ & $13.32\,\pm\,0.08$ & $0.81$ & $0.81$ & $2$\\
T52\dotfill & $11~12~27.72$ & $-76~44~22.3$ & G9 & c & $562$ & $28\,\pm\,3$ & $14.52\,\pm\,0.08$ & \nodata & \nodata & $1$\\
T53\dotfill & $11~12~30.93$ & $-76~44~24.1$ & M1 & c & $468\,\pm\,22$ & $22.2\,\pm\,0.6$ & $14.8\,\pm\,0.2$ & $1.5$ & $1.5$ & $5$\\
CHXR\,54\dotfill & $11~12~42.10$ & $-76~58~40.0$ & M1 & w & $120\,\pm\,22$ & $10.9\,\pm\,0.2$ & $15.58\,\pm\,0.03$ & $0.20$ & $0.21$ & $5$\\
T54\dotfill & $11~12~42.69$ & $-77~22~23.1$ & G8 & w & Absorp. & $11.3\,\pm\,2.0$ & $13.27\,\pm\,0.03$ & $0.74$ & $0.74$ & $2$\\
CHXR\,55\dotfill & $11~12~43.00$ & $-76~37~04.9$ & K4.5 & ? & \nodata & $14.8\,\pm\,0.8$ & $13.35\,\pm\,0.08$ & $0.91$ & $0.91$ & $2$\\
Hn\,17\dotfill & $11~12~48.61$ & $-76~47~06.7$ & M4 & w & $72\,\pm\,10$ & $8.7\,\pm\,0.5$ & $14.53\,\pm\,0.05$ & $0.14$ & $0.16$ & $4$\\
CHXR\,57\dotfill & $11~13~20.13$ & $-77~01~04.5$ & M2.75 & w & $100\,\pm\,15$ & $11.8\,\pm\,1.2$ & $16.30\,\pm\,0.03$ & $0.16$ & $0.20$ & $4$\\
Hn\,18\dotfill & $11~13~24.46$ & $-76~29~22.7$ & M3.5 & w & $121\,\pm\,24$ & $7.6\,\pm\,0.8$ & $15.01\,\pm\,0.04$ & $0.39$ & $0.04$ & $5$\\
CHXR\,59\dotfill & $11~13~27.37$ & $-76~34~16.6$ & M2.75 & w & $107\,\pm\,20$ & $11.0\,\pm\,1.1$ & $15.59\,\pm\,0.03$ & $0.41$ & $0.41$ & $4$\\
CHXR\,60\dotfill & $11~13~29.71$ & $-76~29~01.2$ & M4.25 & w & $95\,\pm\,12$ & $0.8\,\pm\,0.7$ & $19.09\,\pm\,0.04\ddagger$ & $0.12$ & $0.14$ & $4$\\

CHXR\,62\dotfill & $11~14~15.65$ & $-76~27~36.4$ & M3.75 & w & $158\,\pm\,33$ & $35\,\pm\,3$ & $16.2\,\pm\,0.4$ & $0.7$ & $1.6$ & $5$\\
Hn\,21W\dotfill & $11~14~24.54$ & $-77~33~06.2$ & M4 & c & $375\,\pm\,31$ & $19.1\,\pm\,1.1$ & $15.5\,\pm\,0.3$ & $0.4$ & $0.1$ & $4$\\
B53\dotfill & $11~14~50.32$ & $-77~33~39.0$ & M2.75 & w & $107\,\pm\,18$ & $8.3\,\pm\,1.3$ & $14.741\,\pm\,0.017$ & $0.070$ & $0.039$ & $4$\\
T56\dotfill & $11~17~37.01$ & $-77~04~38.1$ & M0.5 & c & $346\,\pm\,41$ & $7.2\,\pm\,0.5$ & $14.923\,\pm\,0.013$ & $0.189$ & $0.165$ & $6$\\
CHXR\,68\,Aa+Ab\dotfill & $11~18~20.24$ & $-76~21~57.6$ & K6 & w & $106\,\pm\,22$ & $8.8\,\pm\,1.0$ & $15.38\,\pm\,0.04$ & $0.23$ & $0.26$ & $4$\\
CHXR\,68\,B\dotfill & $11~18~20.24$ & $-76~21~57.6$ & M1 & w & $90\,\pm\,17$ & $8.3\,\pm\,0.3$ & $15.31\,\pm\,0.05$ & $0.11$ & $0.09$ & $4$\\

\cutinhead{Tau-Aur}

NTTS\,034903+2431\dotfill & $03~52~02.24$ & $+24~39~47.9$ & K5 & w & $229\,\pm\,31$ & $36\,\pm\,2$ & $3.66\,\pm\,0.10\ddagger$ & $0.23$ & $0.22$ & $4$\\
NTTS\,035120+3154SW\dotfill & $03~54~29.51$ & $+32~03~01.4$ & G0 & w & Absorp. & $62\,\pm\,3$ & $10.7\,\pm\,0.4$ & $1.0$ & $0.8$ & $4$\\

RX\,J0403.3+1725\dotfill & $04~03~24.95$ & $+17~24~26.2$ & K3 & ? & \nodata & $111\,\pm\,7$ & $15.2\,\pm\,0.7$ & $2.8$ & $2.8$ & $4$\\

RX\,J0405.1+2632\dotfill & $04~05~12.34$ & $+26~32~43.9$ & K2 & w & Absorp. & $17.5\,\pm\,1.3$ & $7.091\,\pm\,0.015\ddagger$ & $0.181$ & $0.207$ & $4$\\
RX\,J0405.3+2009\dotfill & $04~05~19.59$ & $+20~09~25.6$ & K1 & w & Absorp. & $24.1\,\pm\,1.4$ & $14.427\,\pm\,0.012$ & $0.187$ & $0.190$ & $4$\\
HD284135\dotfill & $04~05~40.58$ & $+22~48~12.2$ & G0 & w & Absorp. & $72\,\pm\,4$ & $14.23\,\pm\,0.10$ & $0.37$ & $0.34$ & $4$\\
HD284149\dotfill & $04~06~38.79$ & $+20~18~11.1$ & F8 & w & Absorp. & $27.0\,\pm\,1.9$ & $12.46\,\pm\,0.03$ & $0.30$ & $0.15$ & $5$\\

RX\,J0406.8+2541\,B\dotfill & $04~06~51.35$ & $+25~41~28.3$ & K6.5 & c & $295$ & $18\,\pm\,2$ & $16.01\,\pm\,0.20$ & \nodata & \nodata & $1$\\
RX\,J0407.8+1750\dotfill & $04~07~53.99$ & $+17~50~25.8$ & K4 & w & $115\,\pm\,17$ & $28.7\,\pm\,1.0$ & $12.47\,\pm\,0.05$ & $0.34$ & $0.27$ & $4$\\

RX\,J0409.1+2901\dotfill & $04~09~09.74$ & $+29~01~30.6$ & G8 & w & Absorp. & $24\,\pm\,2$ & $9.52\,\pm\,0.03\ddagger$ & $0.13$ & $0.14$ & $4$\\
RX\,J0409.2+1716\dotfill & $04~09~17.00$ & $+17~16~08.2$ & M1 & w & $223\,\pm\,23$ & $70.5\,\pm\,1.1$ & $16.1\,\pm\,0.5$ & $1.3$ & $1.5$ & $5$\\
RX\,J0409.8+2446\dotfill & $04~09~51.13$ & $+24~46~21.1$ & M1 & w & $83\,\pm\,8$ & $5.9\,\pm\,0.4$ & $12.330\,\pm\,0.012$ & $0.170$ & $0.208$ & $4$\\
RX\,J0412.8+1937\dotfill & $04~12~50.64$ & $+19~36~58.2$ & K6 & w & $96\,\pm\,16$ & $11.2\,\pm\,0.8$ & $16.37\,\pm\,0.02$ & $0.38$ & $0.47$ & $5$\\

HD285579\dotfill & $04~12~59.88$ & $+16~11~48.2$ & G0 & w & Absorp. & $9.6\,\pm\,1.1$ & $7.805\,\pm\,0.006\ddagger$ & $0.142$ & $0.142$ & $6$\\
LkCa\,1\dotfill & $04~13~14.14$ & $+28~19~10.8$ & M4 & w & $173$ & $30.9\,\pm\,1.1$ & $9.57\,\pm\,0.12\ddagger$ & \nodata & \nodata & $1$\\
RX\,J0413.4+3352\dotfill & $04~13~27.29$ & $+33~52~41.7$ & K0 & ? & \nodata & $16.0\,\pm\,1.7$ & $18.13\,\pm\,0.09$ & \nodata & \nodata & $1$\\

CW\,Tau\dotfill & $04~14~17.00$ & $+28~10~57.8$ & K3 & c & $647\,\pm\,7$ & $33\,\pm\,5$ & $13.60\,\pm\,0.10$ & $3.01$ & $2.55$ & $5$\\
FP\,Tau\dotfill & $04~14~47.31$ & $+26~46~26.4$ & M4 & c & $378\,\pm\,12$ & $32\,\pm\,2$ & $16.26\,\pm\,0.15$ & $0.97$ & $1.08$ & $4$\\
CX\,Tau\dotfill & $04~14~47.86$ & $+26~48~11.0$ & M2 & c & $319$ & $19.8\,\pm\,0.6$ & $16.63\,\pm\,0.12$ & \nodata & \nodata & $1$\\

RX\,J0415.3+2044\dotfill & $04~15~22.92$ & $+20~44~17.0$ & K0 & w & Absorp. & $35\,\pm\,3$ & $13.84\,\pm\,0.03$ & $0.37$ & $0.25$ & $5$\\

LkCa\,4\dotfill & $04~16~28.11$ & $+28~07~35.8$ & K7 & w & $198\,\pm\,30$ & $30\,\pm\,2$ & $15.77\,\pm\,0.11$ & $3.34$ & $4.00$ & $4$\\
CY\,Tau\dotfill & $04~17~33.73$ & $+28~20~46.9$ & M1.5 & c & $415\,\pm\,28$ & $10.6\,\pm\,0.4$ & $16.68\,\pm\,0.02$ & $1.06$ & $1.21$ & $4$\\
LkCa\,5\dotfill & $04~17~38.94$ & $+28~33~00.5$ & M2 & w & $163$ & $38.3\,\pm\,1.1$ & $15.83\,\pm\,0.16$ & \nodata & \nodata & $1$\\
NTTS\,041529+1652\dotfill & $04~18~21.47$ & $+16~58~47.0$ & K5 & w & Absorp. & $5.1\,\pm\,1.3$ & $15.818\,\pm\,0.020$ & $0.180$ & $0.202$ & $5$\\
V410\,Tau\dotfill & $04~18~31.10$ & $+28~27~16.2$ & K7 & c & $450\,\pm\,106$ & $83\,\pm\,4$ & $19.9\,\pm\,0.3$ & $1.3$ & $1.5$ & $4$\\

DD\,Tau\,A\dotfill & $04~18~31.13$ & $+28~16~29.0$ & M2 & c & $363$ & $11.5\,\pm\,1.4$ & $13.9\,\pm\,0.6$ & \nodata & \nodata & $1$\\

NTTS\,041559+1716\dotfill & $04~18~51.70$ & $+17~23~16.6$ & K7 & w & $210\,\pm\,29$ & $74\,\pm\,4$ & $15.5\,\pm\,0.3$ & $3.0$ & $2.9$ & $5$\\
BP\,Tau\dotfill & $04~19~15.84$ & $+29~06~26.9$ & K5 & c & $458\,\pm\,28$ & $13.1\,\pm\,1.6$ & $15.24\,\pm\,0.04$ & $0.09$ & $0.10$ & $4$\\
V819\,Tau\dotfill & $04~19~26.26$ & $+28~26~14.3$ & K7 & w & $166\,\pm\,41$ & $9.1\,\pm\,0.6$ & $16.64\,\pm\,0.02$ & $0.77$ & $0.94$ & $5$\\
LkCa\,7\dotfill & $04~19~41.27$ & $+27~49~48.5$ & K7 & w & $154\,\pm\,7$ & $14.7\,\pm\,1.2$ & $17.98\,\pm\,0.04$ & $0.21$ & $0.21$ & $2$\\
RX\,J0420.3+3123\dotfill & $04~20~24.12$ & $+31~23~23.7$ & K4 & ? & \nodata & $9.6\,\pm\,0.6$ & $13.99\,\pm\,0.03$ & \nodata & \nodata & $1$\\
DE\,Tau\dotfill & $04~21~55.64$ & $+27~55~06.1$ & M1 & c & $453\,\pm\,6$ & $9.7\,\pm\,0.3$ & $15.402\,\pm\,0.018$ & $0.105$ & $0.126$ & $4$\\

HD283572\dotfill & $04~21~58.84$ & $+28~18~06.6$ & G2 & w & Absorp. & $79\,\pm\,3$ & $14.22\,\pm\,0.08$ & $0.94$ & $1.30$ & $4$\\

T\,Tau\,A\dotfill & $04~21~59.43$ & $+19~32~06.4$ & K1.5 & c & $430\,\pm\,23$ & $23.0\,\pm\,1.2$ & $19.23\,\pm\,0.02$ & $0.30$ & $0.06$ & $3$\\

LkCa\,21\dotfill & $04~22~03.14$ & $+28~25~39.0$ & M3 & c & $277$ & $46\,\pm\,3$ & $14.3\,\pm\,0.3$ & \nodata & \nodata & $1$\\
HD285751\dotfill & $04~23~41.33$ & $+15~37~54.9$ & G5 & w & $125$ & $26.6\,\pm\,1.4$ & $15.22\,\pm\,0.03$ & $0.27$ & $0.27$ & $5$\\
BD\,+26\,718B\dotfill & $04~24~49.04$ & $+26~43~10.4$ & K0 & w & Absorp. & $32.4\,\pm\,1.8$ & $16.23\,\pm\,0.04$ & $1.12$ & $1.17$ & $4$\\
IP\,Tau\dotfill & $04~24~57.08$ & $+27~11~56.5$ & M0 & c & $333\,\pm\,54$ & $12.3\,\pm\,0.8$ & $16.24\,\pm\,0.03$ & $0.62$ & $0.73$ & $5$\\

DG\,Tau\dotfill & $04~27~04.70$ & $+26~06~16.3$ & K6 & c & $484$ & $24.7\,\pm\,0.7$ & $15.4\,\pm\,0.6$ & \nodata & \nodata & $1$\\
BD\,+17\,724B\dotfill & $04~27~05.97$ & $+18~12~37.2$ & G5 & w & Absorp. & $49\,\pm\,3$ & $16.97\,\pm\,0.06$ & $0.14$ & $0.18$ & $4$\\
NTTS\,042417+1744\dotfill & $04~27~10.56$ & $+17~50~42.6$ & K1 & w & Absorp. & $17.6\,\pm\,1.5$ & $15.324\,\pm\,0.015$ & $0.241$ & $0.214$ & $4$\\
DH\,Tau\dotfill & $04~29~41.56$ & $+26~32~58.3$ & M1 & c & $348$ & $10.9\,\pm\,0.6$ & $16.52\,\pm\,0.04$ & \nodata & \nodata & $1$\\
DI\,Tau\dotfill & $04~29~42.48$ & $+26~32~49.3$ & M0 & w & $120\,\pm\,19$ & $12.5\,\pm\,0.6$ & $15.103\,\pm\,0.020$ & $0.266$ & $0.192$ & $4$\\
IQ\,Tau\dotfill & $04~29~51.56$ & $+26~06~44.9$ & M0.5 & c & $411\,\pm\,48$ & $14.4\,\pm\,0.3$ & $15.83\,\pm\,0.03$ & $0.25$ & $0.27$ & $5$\\
UX\,Tau\dotfill & $04~30~04.00$ & $+18~13~49.4$ & K2 & c & $513\,\pm\,44$ & $23.6\,\pm\,1.8$ & $15.45\,\pm\,0.02$ & $2.28$ & $2.90$ & $4$\\

FX\,Tau\,A\dotfill & $04~30~29.61$ & $+24~26~45.0$ & M2 & c & $281\,\pm\,67$ & $9.61\,\pm\,0.19$ & $16.363\,\pm\,0.013$ & $0.179$ & $0.223$ & $4$\\
FX\,Tau\,B\dotfill & $04~30~29.61$ & $+24~26~45.0$ & M1 & c & $413\,\pm\,53$ & $7.9\,\pm\,0.3$ & $17.332\,\pm\,0.015$ & $0.236$ & $0.185$ & $4$\\
DK\,Tau\,A\dotfill & $04~30~44.25$ & $+26~01~24.5$ & K7 & c & $461\,\pm\,54$ & $17.5\,\pm\,1.2$ & $16.29\,\pm\,0.05$ & $0.42$ & $0.29$ & $4$\\
DK\,Tau\,B\dotfill & $04~30~44.25$ & $+26~01~24.5$ & M1 & c & $397\,\pm\,35$ & $14.0\,\pm\,0.8$ & $14.70\,\pm\,0.10$ & $0.85$ & $0.78$ & $5$\\
RX\,J0430.8+2113\dotfill & $04~30~49.18$ & $+21~14~10.6$ & G8 & w & Absorp. & $41\,\pm\,4$ & $15.09\,\pm\,0.03$ & $0.55$ & $0.30$ & $4$\\
HD284496\dotfill & $04~31~16.86$ & $+21~50~25.3$ & G0 & w & Absorp. & $20.0\,\pm\,1.0$ & $14.414\,\pm\,0.019$ & $0.329$ & $0.307$ & $4$\\
NTTS\,042835+1700\dotfill & $04~31~27.17$ & $+17~06~24.9$ & K5 & w & $84\,\pm\,7$ & $14.8\,\pm\,1.3$ & $16.63\,\pm\,0.04$ & $0.38$ & $0.43$ & $4$\\
XZ\,Tau\dotfill & $04~31~40.07$ & $+18~13~57.2$ & M3 & c & $341\,\pm\,13$ & $15.0\,\pm\,1.2$ & $18.30\,\pm\,0.04$ & $0.18$ & $0.20$ & $4$\\
V710\,Tau\,A\dotfill & $04~31~57.79$ & $+18~21~38.1$ & M0.5 & w & $192$ & $21.5\,\pm\,0.4$ & $15.75\,\pm\,0.15$ & \nodata & \nodata & $1$\\
V710\,Tau\,B\dotfill & $04~31~57.79$ & $+18~21~38.1$ & M2 & c & $371$ & $18.31\,\pm\,0.19$ & $17.72\,\pm\,0.09$ & \nodata & \nodata & $1$\\
L1551-51\dotfill & $04~32~09.27$ & $+17~57~22.8$ & K7 & w & $146\,\pm\,26$ & $32.1\,\pm\,1.4$ & $18.39\,\pm\,0.10$ & $0.63$ & $0.59$ & $4$\\
V827\,Tau\dotfill & $04~32~14.57$ & $+18~20~14.7$ & K7 & w & $168\,\pm\,15$ & $20.9\,\pm\,1.3$ & $17.77\,\pm\,0.05$ & $1.67$ & $1.91$ & $4$\\

V928\,Tau\dotfill & $04~32~18.86$ & $+24~22~27.1$ & M0.5 & w & $178\,\pm\,60$ & $31.6\,\pm\,0.7$ & $15.38\,\pm\,0.16$ & $1.67$ & $2.02$ & $4$\\
GG\,Tau\,A\dotfill & $04~32~30.35$ & $+17~31~40.6$ & K7 & c & $512\,\pm\,10$ & $11.5\,\pm\,0.7$ & $18.08\,\pm\,0.03$ & $0.16$ & $0.19$ & $4$\\
RX\,J0432.7+1853\dotfill & $04~32~42.43$ & $+18~55~10.2$ & K1 & w & Absorp. & $25.2\,\pm\,1.6$ & $21.834\,\pm\,0.013$ & $0.128$ & $0.106$ & $4$\\
UZ\,Tau\,A\dotfill & $04~32~43.04$ & $+25~52~31.1$ & M1 & c & $438\,\pm\,41$ & $19.3\,\pm\,0.5$ & $18.03\,\pm\,0.08$ & $0.57$ & $0.58$ & $6$\\
L1551-55\dotfill & $04~32~43.73$ & $+18~02~56.3$ & K7 & w & $94\,\pm\,9$ & $7.7\,\pm\,0.7$ & $18.264\,\pm\,0.013$ & $0.158$ & $0.120$ & $4$\\
RX\,J0432.8+1735\dotfill & $04~32~53.24$ & $+17~35~33.8$ & M2 & w & $105\,\pm\,4$ & $11.18\,\pm\,0.11$ & $18.039\,\pm\,0.015$ & $0.243$ & $0.344$ & $5$\\
GH\,Tau\dotfill & $04~33~06.22$ & $+24~09~34.0$ & M1.5 & c & $472\,\pm\,38$ & $30.3\,\pm\,0.7$ & $16.70\,\pm\,0.10$ & $1.01$ & $1.12$ & $4$\\
V807\,Tau\dotfill & $04~33~06.64$ & $+24~09~55.0$ & K7 & c & $408\,\pm\,18$ & $13.6\,\pm\,0.7$ & $16.85\,\pm\,0.03$ & $0.32$ & $0.30$ & $4$\\
V830\,Tau\dotfill & $04~33~10.03$ & $+24~33~43.4$ & K7 & w & $121$ & $32.0\,\pm\,1.5$ & $18.2\,\pm\,0.2$ & \nodata & \nodata & $1$\\
GI\,Tau\dotfill & $04~33~34.06$ & $+24~21~17.0$ & K7 & c & $302\,\pm\,45$ & $12.7\,\pm\,1.9$ & $17.29\,\pm\,0.04$ & $1.08$ & $1.00$ & $4$\\
RX\,J0433.5+1916\dotfill & $04~33~34.67$ & $+19~16~48.9$ & G6 & w & Absorp. & $58\,\pm\,3$ & $21.9\,\pm\,0.4$ & $4.1$ & $4.1$ & $5$\\
DL\,Tau\dotfill & $04~33~39.06$ & $+25~20~38.2$ & G & c & $581\,\pm\,6$ & $19\,\pm\,4$ & $13.94\,\pm\,0.14$ & $0.86$ & $1.03$ & $5$\\
HN\,Tau\,A\dotfill & $04~33~39.35$ & $+17~51~52.4$ & K5 & c & $595\,\pm\,48$ & $39\,\pm\,10$ & $4.6\,\pm\,0.6\ddagger$ & $11.6$ & $12.6$ & $4$\\

DM\,Tau\dotfill & $04~33~48.72$ & $+18~10~10.0$ & M1 & c & $376\,\pm\,27$ & $4.0\,\pm\,0.7$ & $18.607\,\pm\,0.011$ & $0.100$ & $0.109$ & $4$\\

HBC\,407\dotfill & $04~34~18.04$ & $+18~30~06.7$ & G8 & w & Absorp. & $8.8\,\pm\,1.8$ & $17.75\,\pm\,0.03$ & \nodata & \nodata & $1$\\
AA\,Tau\dotfill & $04~34~55.42$ & $+24~28~53.2$ & K7 & c & $402\,\pm\,89$ & $12.8\,\pm\,1.1$ & $16.98\,\pm\,0.04$ & $0.62$ & $0.26$ & $4$\\
FF\,Tau\dotfill & $04~35~20.90$ & $+22~54~24.2$ & K7 & w & $160\,\pm\,86$ & $5.6\,\pm\,0.8$ & $14.165\,\pm\,0.015$ & $0.141$ & $0.139$ & $5$\\

HBC\,412\,A\dotfill & $04~35~24.51$ & $+17~51~43.0$ & M1.5 & w & $105\,\pm\,4$ & $4.9\,\pm\,0.3$ & $18.587\,\pm\,0.012$ & $0.024$ & $0.024$ & $2$\\
HBC\,412\,B\dotfill & $04~35~24.51$ & $+17~51~43.0$ & M1.5 & w & $104\,\pm\,6$ & $4.1\,\pm\,0.2$ & $18.295\,\pm\,0.012$ & $0.168$ & $0.168$ & $2$\\
DN\,Tau\dotfill & $04~35~27.37$ & $+24~14~58.9$ & M0 & c & $336\,\pm\,16$ & $12.3\,\pm\,0.6$ & $16.30\,\pm\,0.02$ & $0.24$ & $0.32$ & $5$\\
HQ\,Tau\dotfill & $04~35~47.34$ & $+22~50~21.7$ & K0 & c & $442\,\pm\,93$ & $48\,\pm\,2$ & $16.65\,\pm\,0.11$ & $0.85$ & $0.81$ & $4$\\
HP\,Tau/G2\dotfill & $04~35~54.15$ & $+22~54~13.5$ & G0 & w & Absorp. & $127\,\pm\,4$ & $16.6\,\pm\,1.0$ & $1.4$ & $1.3$ & $4$\\
RX\,J0435.9+2352\dotfill & $04~35~56.83$ & $+23~52~05.0$ & M1 & w & $125\,\pm\,21$ & $4.2\,\pm\,0.5$ & $16.605\,\pm\,0.011$ & $0.129$ & $0.139$ & $4$\\
LkCa\,14\dotfill & $04~36~19.09$ & $+25~42~59.0$ & M0 & w & $122\,\pm\,20$ & $22.7\,\pm\,1.0$ & $16.65\,\pm\,0.04$ & $0.32$ & $0.36$ & $5$\\
HD283759\dotfill & $04~36~49.12$ & $+24~12~58.8$ & F2 & w & Absorp. & $57\,\pm\,6$ & $32.1\,\pm\,0.2\ddagger$ & $1.2$ & $1.5$ & $4$\\
RX\,J0437.2+3108\dotfill & $04~37~16.86$ & $+31~08~19.5$ & K4 & w & $82\,\pm\,9$ & $11.3\,\pm\,0.8$ & $15.718\,\pm\,0.016$ & $0.441$ & $0.417$ & $4$\\
RX\,J0438.2+2023\dotfill & $04~38~13.04$ & $+20~22~47.1$ & K2 & w & Absorp. & $16.1\,\pm\,1.7$ & $14.954\,\pm\,0.016$ & $0.519$ & $0.579$ & $4$\\
RX\,J0438.2+2302\dotfill & $04~38~15.62$ & $+23~02~27.6$ & M1 & w & $111\,\pm\,29$ & $4.5\,\pm\,0.4$ & $16.493\,\pm\,0.009$ & $0.201$ & $0.318$ & $4$\\
RX\,J0438.4+1543\dotfill & $04~38~27.66$ & $+15~43~38.0$ & K3 & ? & \nodata & $6\,\pm\,2$ & $15.211\,\pm\,0.009$ & $0.126$ & $0.126$ & $5$\\
DO\,Tau\dotfill & $04~38~28.58$ & $+26~10~49.4$ & M0 & c & $480\,\pm\,56$ & $10.5\,\pm\,1.0$ & $16.04\,\pm\,0.17$ & $1.86$ & $2.00$ & $5$\\
HD285957\dotfill & $04~38~39.07$ & $+15~46~13.7$ & K2 & w & Absorp. & $22.5\,\pm\,1.2$ & $17.991\,\pm\,0.015$ & $0.370$ & $0.385$ & $4$\\
VY\,Tau\dotfill & $04~39~17.41$ & $+22~47~53.4$ & M0 & w & $124\,\pm\,16$ & $5.8\,\pm\,1.0$ & $17.716\,\pm\,0.016$ & $0.320$ & $0.357$ & $5$\\
LkCa\,15\dotfill & $04~39~17.80$ & $+22~21~03.5$ & K5 & c & $451\,\pm\,51$ & $13.9\,\pm\,1.2$ & $17.65\,\pm\,0.03$ & $0.65$ & $0.69$ & $5$\\
IW\,Tau\dotfill & $04~41~04.71$ & $+24~51~06.2$ & K7 & w & $170\,\pm\,32$ & $8.7\,\pm\,0.8$ & $15.880\,\pm\,0.010$ & $0.259$ & $0.264$ & $5$\\
CoKu\,Tau/4\dotfill & $04~41~16.81$ & $+28~40~00.1$ & M1 & w & $185\,\pm\,33$ & $25.8\,\pm\,0.4$ & $15.98\,\pm\,0.04$ & $0.42$ & $0.40$ & $7$\\

HD283798\dotfill & $04~41~55.16$ & $+26~58~49.5$ & G2 & w & Absorp. & $25.2\,\pm\,1.2$ & $13.774\,\pm\,0.014$ & $0.140$ & $0.082$ & $4$\\

RX\,J0444.3+2017\dotfill & $04~44~23.55$ & $+20~17~17.5$ & K1 & ? & \nodata & $60\,\pm\,3$ & $15.09\,\pm\,0.16$ & $0.90$ & $0.94$ & $5$\\

HD30171\dotfill & $04~45~51.29$ & $+15~55~49.7$ & G5 & w & Absorp. & $108\,\pm\,4$ & $21.13\,\pm\,0.17$ & $1.37$ & $1.24$ & $4$\\

V1001\,Tau\,A\dotfill & $04~46~58.98$ & $+17~02~38.2$ & K6 & c & $525\,\pm\,48$ & $12.1\,\pm\,1.2$ & $22.45\,\pm\,0.05\ddagger$ & $0.72$ & \nodata & $2$\\
V1001\,Tau\,B\dotfill & $04~46~58.98$ & $+17~02~38.2$ & K6 & c & $405\,\pm\,24$ & $7.0\,\pm\,0.3$ & $21.74\,\pm\,0.08$ & $0.50$ & $1.41$ & $3$\\
DR\,Tau\dotfill & $04~47~06.21$ & $+16~58~42.8$ & K4 & c & $370\,\pm\,7$ & $6.26\,\pm\,0.12$ & $21.10\,\pm\,0.04$ & $0.19$ & $0.16$ & $5$\\

RX\,J0447.9+2755\,A\dotfill & $04~48~00.44$ & $+27~56~19.6$ & G2.5 & w & Absorp. & $30.5\,\pm\,1.8$ & $16.29\,\pm\,0.10$ & $1.03$ & $1.03$ & $2$\\
RX\,J0447.9+2755\,B\dotfill & $04~48~00.44$ & $+27~56~19.6$ & G2 & w & Absorp. & $27.9\,\pm\,1.4$ & $15.67\,\pm\,0.04$ & $0.97$ & $0.97$ & $3$\\
RX\,J0450.0+2230\dotfill & $04~50~00.20$ & $+22~29~57.5$ & K1 & ? & \nodata & $57\,\pm\,3$ & $15.04\,\pm\,0.08$ & $0.13$ & $0.12$ & $4$\\
UY\,Aur\,A\dotfill & $04~51~47.38$ & $+30~47~13.5$ & K7 & c & $324\,\pm\,19$ & $23.8\,\pm\,1.3$ & $13.92\,\pm\,0.07$ & $0.87$ & $0.97$ & $5$\\
RX\,J0452.5+1730\dotfill & $04~52~30.75$ & $+17~30~25.8$ & K4 & w & $89$ & $8.8\,\pm\,0.6$ & $16.820\,\pm\,0.010$ & $0.196$ & $0.222$ & $4$\\
RX\,J0452.8+1621\dotfill & $04~52~50.15$ & $+16~22~09.2$ & K6 & w & $123\,\pm\,12$ & $24.9\,\pm\,1.2$ & $19.28\,\pm\,0.06$ & $0.40$ & $0.40$ & $4$\\
RX\,J0452.9+1920\dotfill & $04~52~57.08$ & $+19~19~50.4$ & K5 & w & $89\,\pm\,8$ & $4.8\,\pm\,1.3$ & $14.699\,\pm\,0.010$ & $0.117$ & $0.107$ & $4$\\
HD31281\dotfill & $04~55~09.62$ & $+18~26~30.9$ & G0 & w & Absorp. & $79\,\pm\,4$ & $14.57\,\pm\,0.12$ & $0.26$ & $0.20$ & $4$\\
GM\,Aur\dotfill & $04~55~10.98$ & $+30~21~59.5$ & K7 & c & $505\,\pm\,11$ & $14.8\,\pm\,0.9$ & $15.15\,\pm\,0.04$ & $0.50$ & $0.41$ & $4$\\
LkCa\,19\dotfill & $04~55~36.96$ & $+30~17~55.3$ & K0 & w & $154\,\pm\,40$ & $20.1\,\pm\,1.1$ & $13.578\,\pm\,0.013$ & $1.120$ & $0.674$ & $4$\\

SU\,Aur\dotfill & $04~55~59.38$ & $+30~34~01.6$ & G2 & c & $561\,\pm\,58$ & $59\,\pm\,2$ & $14.26\,\pm\,0.05$ & $0.40$ & $0.25$ & $4$\\

RX\,J0456.2+1554\dotfill & $04~56~13.57$ & $+15~54~22.0$ & K7 & w & $106\,\pm\,20$ & $9.7\,\pm\,0.6$ & $18.966\,\pm\,0.016$ & $0.092$ & $0.106$ & $4$\\
HD286179\dotfill & $04~57~00.65$ & $+15~17~53.1$ & G0 & w & Absorp. & $17.1\,\pm\,1.2$ & $10.069\,\pm\,0.009\ddagger$ & $0.211$ & $0.090$ & $4$\\

RX\,J0457.2+1524\dotfill & $04~57~17.67$ & $+15~25~09.4$ & K1 & w & Absorp. & $42\,\pm\,2$ & $19.77\,\pm\,0.03$ & $0.70$ & $0.75$ & $4$\\

RX\,J0458.7+2046\dotfill & $04~58~39.74$ & $+20~46~44.1$ & K7 & w & Absorp. & $7.8\,\pm\,0.5$ & $19.043\,\pm\,0.013$ & $0.108$ & $0.124$ & $5$\\
RX\,J0459.7+1430\dotfill & $04~59~46.17$ & $+14~30~55.4$ & K4 & w & $53$ & $14.5\,\pm\,0.6$ & $19.875\,\pm\,0.012$ & $0.150$ & $0.092$ & $4$\\
V836\,Tau\dotfill & $05~03~06.60$ & $+25~23~19.7$ & K7 & c & $403\,\pm\,58$ & $13.4\,\pm\,1.1$ & $18.15\,\pm\,0.03$ & $0.53$ & $0.44$ & $4$\\
RX\,J0507.2+2437\dotfill & $05~07~12.07$ & $+24~37~16.4$ & K6 & w & $126\,\pm\,14$ & $19.7\,\pm\,1.0$ & $18.74\,\pm\,0.04$ & $1.24$ & $1.45$ & $5$\\

RW\,Aur\,B\dotfill & $05~07~49.54$ & $+30~24~05.1$ & K1 & c & $618\,\pm\,78$ & $14.5\,\pm\,1.8$ & $15.00\,\pm\,0.03$ & $0.66$ & $0.54$ & $4$\\

\enddata
\tablenotetext{a}{c: Classical T~Tauri star, w: Weak-lined T~Tauri star, ?: Unknown.}
\tablenotetext{b}{The H$\alpha$ 10\% widths were adopted from \citet{2009ApJ...695.1648N} where available, or measured using the same method otherwise.}
\tablenotetext{c}{The H$\alpha$ 10\% width uncertainty does not correspond to the measurement uncertainty, but to the scatter in our multi-epoch data.}
\tablenotetext{d}{The $v~\!\sin~\!i$ were adopted from \citet{2009ApJ...695.1648N} where available, or measured using the same method, otherwise.}
\tablenotetext{e}{The $v~\!\sin~\!i$ uncertainty represents the combined measurement scatter between results using different template spectra, and over different epochs.}
\tablenotetext{f}{The symbol $\ddagger$ indicates that the overall radial velocity deviates from that of the associated star-forming region.}
\tablenotetext{g}{This is the weighted standard deviation of the radial velocity as described in \S\ref{sec:SB1}.}
\tablenotetext{h}{This is the systematic noise of the radial velocity as described in \S\ref{sec:SB1}.}
\end{deluxetable}

\clearpage
\end{landscape}

\LongTables
\begin{landscape}
\begin{deluxetable}{cclcccccccccc}
\tablecolumns{12}
\tabletypesize{\scriptsize}
\tablewidth{0pt}
\tablecaption{Measurements of Stars with Close Companions in Cha~I \& Tau-Aur\label{tbl:Binaries}
}
\tablehead{
\colhead{SB\tablenotemark{a}} & \colhead{Region} &  \colhead{Object} & \colhead{R.A.} & \colhead{Dec.} & \colhead{SpT} & \colhead{TTS\tablenotemark{b}} & \colhead{$10\%$\,width\tablenotemark{cd}} & \colhead{\# of} & \colhead{Comp.} & \colhead{$v~\!\sin~\!i$\tablenotemark{ef}} & \colhead{Flux\tablenotemark{g}}\\
\colhead{type} & & & \colhead{(J2000.0)} & \colhead{(J2000.0)} & & & \colhead{(km~s$^{-1}$)} & \colhead{obs.} & & \colhead {(km~s$^{-1}$)} & \colhead{ratio} &
}
\startdata

SB1 & Cha & T39\,B\dotfill & $11~09~11.72$ & $-77~29~12.5$ & M1.5 & w & $106\,\pm\,13$ & $5$ &  & $12.8\,\pm\,0.4$ & \nodata\\
SB1 & Tau & RX\,J0415.8+3100\dotfill & $04~15~51.38$ & $+31~00~35.6$ & G6 & w & Absorp. & $4$ &  & $31.7\,\pm\,1.9$ & \nodata\\
SB1 & Tau & RX\,J0457.5+2014\dotfill & $04~57~30.66$ & $+20~14~29.7$ & K3 & w & Absorp. & $4$ &  & $33\,\pm\,3$ & \nodata\\

SB1? & Cha & T7\dotfill & $10~59~01.09$ & $-77~22~40.7$ & K8 & c & $365\,\pm\,76$ & $6$ &  & $11.3\,\pm\,0.8$ & \nodata\\
SB1? & Cha & CHXR\,28\,B\dotfill & $11~07~55.89$ & $-77~27~25.8$ & G9 & w & $175$ & $2$ &  & $61\,\pm\,4$ & \nodata\\
SB1? & Tau & RY\,Tau\dotfill & $04~21~57.40$ & $+28~26~35.5$ & F8 & c & $600\,\pm\,24$ & $4$ &  & $48\,\pm\,3$ & \nodata\\
SB1? & Tau & CI\,Tau\dotfill & $04~33~52.00$ & $+22~50~30.2$ & G & c & $572\,\pm\,9$ & $4$ &  & $13\,\pm\,2$ & \nodata\\

SB2 & Cha & CHXR\,12\dotfill & $11~03~56.83$ & $-77~21~33.0$ & M3.5 & w & $101\,\pm\,8$ & $6$ & A & $8.99\,\pm\,0.14$ & \nodata\\
 &  &  &  &  &  &  &  &  & B & $8\,\pm\,2$ & $0.20\,\pm\,0.08$\\
SB2 & Cha & T42\dotfill & $11~09~53.41$ & $-76~34~25.5$ & K5 & c & $543\,\pm\,83$ & $7$ & A & $11.4\,\pm\,1.1$ & \nodata\\
 &  &  &  &  &  &  &  &  & B & $11.5\,\pm\,1.5$ & $0.8\,\pm\,0.3$\\
SB2 & Tau & V826\,Tau\dotfill & $04~32~15.84$ & $+18~01~38.7$ & K7 & w & $139\,\pm\,13$ & $5$ & A & $8.5\,\pm\,0.5$ & \nodata\\
 &  &  &  &  &  &  &  &  & B & $9.3\,\pm\,0.7$ & $0.88\,\pm\,0.06$\\
SB2 & Tau & DQ\,Tau\dotfill & $04~46~53.05$ & $+17~00~00.2$ & M0 & c & $340\,\pm\,22$ & $4$ & A & $14.7\,\pm\,1.6$ & \nodata\\
 &  &  &  &  &  &  &  &  & B & $11.3\,\pm\,0.7$ & $0.78\,\pm\,0.18$\\
SB2 & Tau & HBC\,427\dotfill & $04~56~02.02$ & $+30~21~03.8$ & K5 & w & $145\,\pm\,13$ & $4$ & A & $9.9\,\pm\,0.3$ & \nodata\\
 &  &  &  &  &  &  &  &  & B & $14.5\,\pm\,0.5$ & $0.157\,\pm\,0.014$\\

SB2$^{\dagger}$ & Cha & T21\dotfill & $11~06~15.41$ & $-77~21~56.8$ & G5 & w & Absorp. & $1$ & A & $94.1\,\pm\,0.7$ & \nodata\\
 &  &  &  &  &  &  &  &  & B & $14.5\,\pm\,0.3$ & $0.101\,\pm\,0.005$\\
SB2$^{\dagger}$ & Cha & CHXR\,47\dotfill & $11~10~38.02$ & $-77~32~39.9$ & K3 & w & Absorp. & $2$ & A & $59.5\,\pm\,0.4$ & \nodata\\
 &  &  &  &  &  &  &  &  & B & $24\,\pm\,3$ & $0.24\,\pm\,0.06$\\
SB2$^{\dagger}$ & Tau & HD285281\dotfill & $04~00~31.07$ & $+19~35~20.7$ & K0 & w & Absorp. & $4$ & A & $78.0\,\pm\,0.3$ & \nodata\\
 &  &  &  &  &  &  &  &  & B & $17.0\,\pm\,1.9$ & $0.07\,\pm\,0.05$\\
SB2$^{\dagger}$ & Tau & RX\,J0406.8+2541\,A\dotfill & $04~06~51.35$ & $+25~41~28.3$ & K4.5 & c & $277\,\pm\,74$ & $2$ & a & $47\,\pm\,4$ & \nodata\\
 &  &  &  &  &  &  &  &  & b & $9.8\,\pm\,0.3$ & $0.32\,\pm\,0.07$\\
SB2$^{\dagger}$ & Tau & DF\,Tau\dotfill & $04~27~02.80$ & $+25~42~22.3$ & M3 & c & $369\,\pm\,19$ & $4$ & A & $46.6\,\pm\,1.8$ & \nodata\\
 &  &  &  &  &  &  &  &  & B & $9.8\,\pm\,0.6$ & $0.49\,\pm\,0.11$\\
SB2$^{\dagger}$ & Tau & RX\,J0441.4+2715\dotfill & $04~41~24.00$ & $+27~15~12.4$ & G8 & w & Absorp. & $4$ & A & $37.0\,\pm\,0.6$ & \nodata\\
 &  &  &  &  &  &  &  &  & B & $12.6\,\pm\,1.5$ & $0.32\,\pm\,0.07$\\
SB2$^{\dagger}$ & Tau & RX\,J0443.4+1546\dotfill & $04~43~25.97$ & $+15~46~03.9$ & G7 & w & Absorp. & $4$ & A & $86.5\,\pm\,1.6$ & \nodata\\
 &  &  &  &  &  &  &  &  & B & $24.0\,\pm\,0.8$ & $0.83\,\pm\,0.10$\\
SB2$^{\dagger}$ & Tau & RX\,J0455.7+1742\dotfill & $04~55~47.67$ & $+17~42~02.0$ & K3 & w & Absorp. & $4$ & A & $8\,\pm\,2$ & \nodata\\
 &  &  &  &  &  &  &  &  & B & $9.5\,\pm\,0.4$ & $0.337\,\pm\,0.014$\\

SB2? & Cha & T11\dotfill & $11~02~24.91$ & $-77~33~35.7$ & K6 & c & $367\,\pm\,38$ & $4$ & A & $11.2\,\pm\,0.3$ & \nodata\\
 &  &  &  &  &  &  &  &  & B & $13.0\,\pm\,0.3$ & $1.0\,\pm\,0.4$\\
SB2? & Cha & Hn\,4\dotfill & $11~05~14.67$ & $-77~11~29.1$ & M3.25 & w & $176\,\pm\,126$ & $4$ & A & $9.0\,\pm\,0.5$ & \nodata\\
 &  &  &  &  &  &  &  &  & B & $90\,\pm\,15$ & $0.62\,\pm\,0.09$\\
SB2? & Cha & T31\,A\dotfill & $11~08~01.49$ & $-77~42~28.9$ & K3 & c & $471\,\pm\,55$ & $3$ & a & $13\,\pm\,2$ & \nodata\\
 &  &  &  &  &  &  &  &  & b & $13\,\pm\,3$ & $0.7\,\pm\,0.3$\\
SB2? & Tau & Hubble\,4\dotfill & $04~18~47.04$ & $+28~20~07.3$ & K7 & w & $188\,\pm\,16$ & $4$ & A & $12.1\,\pm\,0.3$ & \nodata\\
 &  &  &  &  &  &  &  &  & B & $13.1\,\pm\,1.9$ & $0.20\,\pm\,0.05$\\

SB3 & Cha & T55\dotfill & $11~13~33.57$ & $-76~35~37.4$ & M4.5 & w & $165\,\pm\,59$ & $5$ & A & $23\,\pm\,2$ & \nodata\\
 &  &  &  &  &  &  &  &  & B & $34\,\pm\,4$ & $0.38\,\pm\,0.08$\\
 &  &  &  &  &  &  &  &  & C & $31.0\,\pm\,1.6$ & $0.19\,\pm\,0.08$\\
SB3 & Tau & RX\,J0412.8+2442\dotfill & $04~12~51.22$ & $+24~41~44.3$ & G9 & w & Absorp. & $3$ & A & $28.50\,\pm\,0.12$ & \nodata\\
 &  &  &  &  &  &  &  &  & B & $15.1\,\pm\,1.5$ & $0.88\,\pm\,0.12$\\
 &  &  &  &  &  &  &  &  & C & $7.4\,\pm\,0.3$ & $0.36\,\pm\,0.04$\\
SB3 & Tau & V773\,Tau\dotfill & $04~14~12.92$ & $+28~12~12.4$ & K3 & c & $433\,\pm\,51$ & $4$ & A & $27\,\pm\,4$ & \nodata\\
 &  &  &  &  &  &  &  &  & B & $28\,\pm\,4$ & $1.0\,\pm\,0.3$\\
 &  &  &  &  &  &  &  &  & C & $21\,\pm\,7$ & $0.09\,\pm\,0.06$\\
SB3 & Tau & LkCa\,3\dotfill & $04~14~47.97$ & $+27~52~34.7$ & M1 & w & $187\,\pm\,25$ & $4$ & Aa & $12.0\,\pm\,1.0$ & \nodata\\
 &  &  &  &  &  &  &  &  & Ab1 & $16\,\pm\,2$ & $0.75\,\pm\,0.13$\\
 &  &  &  &  &  &  &  &  & Ab2 & $14\,\pm\,4$ & $0.21\,\pm\,0.05$\\
 
\enddata
\tablenotetext{a}{SB1: Single-lined spectroscopic binary, SB2: Double-lined spectroscopic binary, SB3: Triple-lined spectroscopic binary; ?: suspected, $^{\dagger}$: long-period.}
\tablenotetext{b}{c: Classical T~Tauri star, w: Weak-lined T~Tauri star, ?: Unknown.}
\tablenotetext{c}{The H$\alpha$ 10\% widths were adopted from \citet{2009ApJ...695.1648N} where available, or measured using the same method otherwise.}
\tablenotetext{d}{The H$\alpha$ 10\% width uncertainty does not correspond to the measurement uncertainty, but to the scatter in our multi-epoch data.}
\tablenotetext{e}{The $v~\!\sin~\!i$ were adopted from \citet{2009ApJ...695.1648N} where available, or measured using the same method for SB1s and the broadening function method described in \S\ref{sec:RadialVelocities} for SB2s and SB3s, otherwise.}
\tablenotetext{f}{The $v~\!\sin~\!i$ uncertainty represents the combined measurement scatter between results using different template spectra, and over different epochs.}
\tablenotetext{g}{The flux ratio is between the companion and the primary star, see \S\ref{sec:RadialVelocities}.}
\end{deluxetable}

\clearpage
\end{landscape}

\begin{deluxetable}{llccc}
\tablecolumns{5}
\tabletypesize{\scriptsize}
\tablewidth{0pt}
\tablecaption{Radial Velocity Measurements for Candidate and Suspected SB2s in Cha~I \& Tau-Aur\label{tbl:RVSB2}}
\tablehead{
\colhead{Region} & \colhead{Object} & \colhead{MJD} & \colhead{$RV_1$ (km~s$^{-1}$)} & \colhead{$RV_2$ (km~s$^{-1}$)}
}
\startdata
\cutinhead{Candidate Short-Period SB2s}

Cha & CHXR\,12\dotfill & 53787.13005 & $3.49\,\pm\,0.03$ & $36.45\,\pm\,0.30$\\
 & & 53836.09473 & $7.00\,\pm\,0.01$ & $27.98\,\pm\,0.18$\\
 & & 53836.35925 & $8.00\,\pm\,0.02$ & $28.09\,\pm\,0.36$\\
 & & 53837.05400 & $8.00\,\pm\,0.02$ & $27.78\,\pm\,0.45$\\
 & & 54074.35810 & $17.76\,\pm\,0.07$ & $4.27\,\pm\,0.42$\\
 & & 54075.28527 & $17.64\,\pm\,0.07$ & $3.98\,\pm\,0.17$\\
Cha & T42\dotfill & 53786.16759 & $-5.30\,\pm\,0.28$ & $32.30\,\pm\,0.45$\\
 & & 53786.38541 & $-6.10\,\pm\,0.69$ & $33.74\,\pm\,0.65$\\
 & & 53788.13175 & $-4.51\,\pm\,0.31$ & $31.11\,\pm\,0.69$\\
 & & 53788.33456 & $-4.50\,\pm\,0.05$ & $31.67\,\pm\,0.46$\\
 & & 53836.37795 & $-13.00\,\pm\,0.25$ & $25.82\,\pm\,0.53$\\
 & & 54073.33792 & $0.99\,\pm\,0.18$ & $9.50\,\pm\,0.07$\\
 & & 54075.27381 & $4.76\,\pm\,0.39$ & $11.83\,\pm\,1.26$\\
Tau & V826\,Tau\dotfill & 54045.15674 & $1.49\,\pm\,0.05$ & $33.63\,\pm\,0.15$\\
 & & 54071.07067 & $32.50\,\pm\,0.02$ & $2.50\,\pm\,0.02$\\
 & & 54074.06213 & $31.48\,\pm\,0.05$ & $3.95\,\pm\,0.17$\\
 & & 54074.29539 & $35.50\,\pm\,0.01$ & $1.50\,\pm\,0.03$\\
 & & 54075.12450 & $29.82\,\pm\,0.08$ & $5.43\,\pm\,0.14$\\
Tau & DQ\,Tau\dotfill & 54043.29394 & $29.30\,\pm\,0.19$ & $7.76\,\pm\,0.24$\\
 & & 54071.32764 & $38.23\,\pm\,0.28$ & $2.62\,\pm\,0.20$\\
 & & 54072.10274 & $38.13\,\pm\,0.22$ & $4.48\,\pm\,0.21$\\
 & & 54074.31433 & $32.34\,\pm\,0.28$ & $5.58\,\pm\,0.13$\\
Tau & HBC\,427\dotfill & 54042.27028 & $5.25\,\pm\,0.06$ & $34.50\,\pm\,0.02$\\
 & & 54043.26942 & $3.80\,\pm\,0.05$ & $31.67\,\pm\,0.58$\\
 & & 54071.17361 & $4.42\,\pm\,0.04$ & $33.58\,\pm\,0.49$\\
 & & 54073.23627 & $4.58\,\pm\,0.04$ & $34.10\,\pm\,0.64$\\

\cutinhead{Candidate Long-Period SB2s}

Cha & T21\dotfill & 54073.36536 & $16.49\,\pm\,0.58$ & $8.46\,\pm\,0.32$\\
Cha & CHXR\,47\dotfill & 54072.32956 & $17.50\,\pm\,0.06$ & $7.87\,\pm\,0.49$\\
 & & 54075.30733 & $18.05\,\pm\,0.67$ & $6.50\,\pm\,0.03$\\
Tau & HD285281\dotfill & 54044.33142 & $15.55\,\pm\,0.88$ & $21.00\,\pm\,0.01$\\
 & & 54071.26329 & $16.00\,\pm\,0.13$ & $21.37\,\pm\,0.75$\\
 & & 54072.05828 & $14.79\,\pm\,0.47$ & $17.56\,\pm\,15.16$\\
 & & 54074.21922 & $14.44\,\pm\,0.43$ & $16.44\,\pm\,0.47$\\
Tau & RX\,J0406.8+2541\,A\dotfill & 54071.13966 & $21.18\,\pm\,0.25$ & $16.31\,\pm\,0.08$\\
 & & 54074.12534 & $15.82\,\pm\,0.76$ & $15.98\,\pm\,0.16$\\
Tau & DF\,Tau\dotfill & 54045.21462 & $9.78\,\pm\,0.84$ & $16.96\,\pm\,0.16$\\
 & & 54071.24608 & $9.49\,\pm\,0.89$ & $15.11\,\pm\,0.31$\\
 & & 54072.16221 & $11.78\,\pm\,0.61$ & $15.44\,\pm\,0.08$\\
 & & 54074.20299 & $11.10\,\pm\,0.37$ & $16.73\,\pm\,0.09$\\
Tau & RX\,J0441.4+2715\dotfill & 54042.23259 & $22.00\,\pm\,0.04$ & $22.70\,\pm\,0.16$\\
 & & 54071.19589 & $21.35\,\pm\,0.20$ & $21.85\,\pm\,0.19$\\
 & & 54072.21185 & $21.56\,\pm\,0.23$ & $21.93\,\pm\,0.18$\\
 & & 54073.16892 & $23.47\,\pm\,0.32$ & $22.48\,\pm\,0.21$\\
Tau & RX\,J0443.4+1546\dotfill & 54043.29064 & $7.50\,\pm\,0.16$ & $16.00\,\pm\,0.01$\\
 & & 54071.07395 & $6.45\,\pm\,1.28$ & $14.07\,\pm\,0.14$\\
 & & 54073.31655 & $6.91\,\pm\,0.87$ & $15.67\,\pm\,0.16$\\
 & & 54074.06749 & $7.96\,\pm\,1.67$ & $13.74\,\pm\,0.37$\\
Tau & RX\,J0455.7+1742\dotfill & 54044.37308 & $19.81\,\pm\,0.08$ & $4.01\,\pm\,0.39$\\
 & & 54071.33319 & $20.50\,\pm\,0.02$ & $4.50\,\pm\,0.05$\\
 & & 54072.11740 & $20.17\,\pm\,0.05$ & $4.50\,\pm\,0.07$\\
 & & 54074.31905 & $20.50\,\pm\,0.03$ & $2.46\,\pm\,0.29$\\

\cutinhead{Suspected SB2s}

Cha & T11\dotfill & 53786.14279 & $9.95\,\pm\,0.17$ & $19.00\,\pm\,0.01$\\
 & & 53786.32551 & $9.92\,\pm\,0.17$ & $19.50\,\pm\,0.03$\\
 & & 53788.26188 & $10.12\,\pm\,0.12$ & $20.00\,\pm\,0.01$\\
 & & 53836.41081 & $10.65\,\pm\,0.14$ & $24.13\,\pm\,0.30$\\
Cha & Hn\,4\dotfill & 53788.27834 & $16.55\,\pm\,0.05$ & $2.00\,\pm\,0.51$\\
 & & 53835.04491 & $15.48\,\pm\,0.06$ & $-2.00\,\pm\,6.85$\\
 & & 53835.36297 & $15.98\,\pm\,0.08$ & $-3.48\,\pm\,2.59$\\
 & & 53837.21652 & $15.05\,\pm\,0.08$ & $6.63\,\pm\,3.23$\\
Cha & T31\,A\dotfill & 53788.26999 & $23.72\,\pm\,0.29$ & $5.79\,\pm\,0.41$\\
 & & 53835.16847 & $27.16\,\pm\,0.21$ & $6.02\,\pm\,0.37$\\
 & & 54071.37156 & $26.31\,\pm\,0.26$ & $2.51\,\pm\,0.35$\\
Tau & Hubble\,4\dotfill & 54043.25711 & $17.27\,\pm\,0.10$ & $0.28\,\pm\,0.45$\\
 & & 54071.21366 & $17.68\,\pm\,0.10$ & $4.49\,\pm\,0.10$\\
 & & 54072.23986 & $18.95\,\pm\,0.11$ & $0.74\,\pm\,0.41$\\
 & & 54073.18295 & $18.10\,\pm\,0.10$ & $1.36\,\pm\,0.35$\\
 
\enddata
\end{deluxetable}

\clearpage

\begin{deluxetable}{llcccc}
\tablecolumns{6}
\tabletypesize{\scriptsize}
\tablewidth{0pt}
\tablecaption{Radial Velocity Measurements for Candidate SB3s in Cha~I \& Tau-Aur\label{tbl:RVSB3}}
\tablehead{
\colhead{Region} & \colhead{Object} & \colhead{MJD} & \colhead{$RV_1$ (km~s$^{-1}$)} & \colhead{$RV_2$ (km~s$^{-1}$)} & \colhead{$RV_3$ (km~s$^{-1}$)}
}
\startdata

Cha & T55\dotfill & 53786.19360 & $15.30\,\pm\,0.16$ & $66.46\,\pm\,0.45$ & $-56.74\,\pm\,1.11$\\
 & & 53835.24146 & $16.52\,\pm\,0.26$ & $16.52?$ & $16.52?$\\
 & & 53836.16456 & $13.50\,\pm\,0.05$ & $-31.93\,\pm\,0.87$ & $67.93\,\pm\,2.64$\\
 & & 53837.32253 & $14.04\,\pm\,0.28$ & $56.51\,\pm\,0.40$ & $-45.00\,\pm\,0.07$\\
 & & 54071.34865 & $14.84\,\pm\,0.12$ & $79.51\,\pm\,0.60$ & $-65.27\,\pm\,0.67$\\
Tau & RX\,J0412.8+2442\dotfill & 54071.24851 & $-25.50\,\pm\,0.03$ & $87.92\,\pm\,0.11$ & $32.50\,\pm\,0.02$\\
 & & 54072.12813 & $-22.57\,\pm\,0.20$ & $86.02\,\pm\,0.18$ & $32.00\,\pm\,0.03$\\
 & & 54074.20523 & $77.43\,\pm\,0.32$ & $-16.65\,\pm\,0.12$ & $32.48\,\pm\,0.05$\\
Tau & V773\,Tau\dotfill & 54043.22853 & $10.80\,\pm\,0.27$ & $44.63\,\pm\,0.35$ & $-27.03\,\pm\,0.41$\\
 & & 54071.20801 & $2.51\,\pm\,0.10$ & $40.71\,\pm\,0.15$ & $-19.00\,\pm\,3.13$\\
 & & 54072.23483 & $3.39\,\pm\,0.32$ & $38.63\,\pm\,0.26$ & $-24.49\,\pm\,1.29$\\
 & & 54074.15653 & $3.42\,\pm\,0.46$ & $45.25\,\pm\,0.39$ & $-26.00\,\pm\,0.26$\\
Tau & LkCa\,3\dotfill & 54043.21590 & $12.77\,\pm\,0.07$ & $42.26\,\pm\,0.12$ & $-37.50\,\pm\,0.02$\\
 & & 54071.22094 & $-5.13\,\pm\,0.11$ & $56.40\,\pm\,0.19$ & $-52.50\,\pm\,0.37$\\
 & & 54072.24682 & $-6.12\,\pm\,0.07$ & $16.46\,\pm\,0.13$ & $61.53\,\pm\,0.10$\\
 & & 54074.16123 & $-2.14\,\pm\,0.09$ & $6.67\,\pm\,0.11$ & $40.50\,\pm\,0.10$\\

\enddata
\end{deluxetable}

\clearpage

\begin{deluxetable}{llcc}
\tablecolumns{4}
\tabletypesize{\scriptsize}
\tablewidth{0pt}
\tablecaption{Radial Velocity Measurements for Candidate and Suspected SB1s in Cha~I \& Tau-Aur\label{tbl:RVSB1}}
\tablehead{
\colhead{Region} & \colhead{Object} & \colhead{MJD} & \colhead{$RV$ (km~s$^{-1}$)}
}
\startdata
\cutinhead{Candidate SB1s}

Cha & T39\,B\dotfill & 53788.30821  & $16.71\,\pm\,0.10$\\
 & & 53835.18985  & $15.17\,\pm\,0.20$\\
 & & 53836.07753  & $15.04\,\pm\,0.47$\\
 & & 53837.02627  & $15.78\,\pm\,0.08$\\
 & & 54074.36553  & $14.10\,\pm\,0.13$\\
Tau & RX\,J0415.8+3100\dotfill & 54043.24631  & $56.12\,\pm\,0.35$\\
 & & 54071.18164  & $-13.81\,\pm\,0.13$\\
 & & 54072.20036  & $53.74\,\pm\,0.09$\\
 & & 54073.14788  & $-2.22\,\pm\,0.15$\\
Tau & RX\,J0457.5+2014\dotfill & 54043.34779  & $11.92\,\pm\,0.06$\\
 & & 54071.31745  & $18.39\,\pm\,0.07$\\
 & & 54072.13301  & $18.00\,\pm\,0.05$\\
 & & 54074.28725  & $18.46\,\pm\,0.05$\\

\cutinhead{Suspected SB1s}

Cha & T7\dotfill & 53786.16446  & $16.23\,\pm\,0.10$\\
 & & 53786.34625  & $16.59\,\pm\,0.10$\\
 & & 53788.31133  & $16.12\,\pm\,0.11$\\
 & & 53836.38144  & $15.48\,\pm\,0.32$\\
 & & 53836.41500  & $15.02\,\pm\,0.34$\\
 & & 54071.36880  & $17.18\,\pm\,0.09$\\
Cha & CHXR\,28\,B\dotfill & 54072.36538  & $30.96\,\pm\,2.83$\\
 & & 54075.36595  & $23.36\,\pm\,2.15$\\
Tau & RY\,Tau\dotfill & 54043.25397  & $19.37\,\pm\,0.28$\\
 & & 54071.21608  & $16.72\,\pm\,0.23$\\
 & & 54072.24213  & $16.96\,\pm\,0.21$\\
 & & 54073.18576  & $16.76\,\pm\,0.22$\\
Tau & CI\,Tau\dotfill & 54043.29708  & $17.61\,\pm\,0.09$\\
 & & 54071.12631  & $16.49\,\pm\,0.08$\\
 & & 54073.28333  & $16.43\,\pm\,0.10$\\
 & & 54074.11409  & $16.55\,\pm\,0.10$\\
\enddata
\end{deluxetable}

\clearpage

\begin{deluxetable}{lccccc}
\tablecolumns{6}
\tabletypesize{\scriptsize}
\tablewidth{0pt}
\tablecaption{Wide Companions in Cha~I \& Tau-Aur\label{tbl:WideCompanions}}
\tablehead{
\colhead{Object} & \colhead{Pairing} & \colhead{Separation ('')} & \colhead{$\Delta K$} & \colhead{$R$-band} & \colhead{References\tablenotemark{b}} \\
& & & & \colhead{flux ratio\tablenotemark{a}} &
}
\startdata
\cutinhead{Cha~I}
T5\dotfill & A--B & 0.159 & 0.05 & 0.87 & 6\\
T6\dotfill & A--B & 5.122 & 3.79 & $\lesssim$0.01 & 6\\
CHXR\,9C\,A\dotfill & A--Ba & 0.852 & 0.36 & 0.56 & 6\\
CHXR\,9C\,Ba+Bb\dotfill & Ba--Bb & 0.13 & 0.7 & 0.18 & 6\\
CHXR\,71\dotfill & A--B & 0.572 & 1.63 & 0.02 & 6\\

T26\dotfill & Ba--Bb & 0.066 & 0.06 & 0.86 & 6\\

CHXR\,28\dotfill & Aa--Ab & 0.143 & 0.4 & 0.55 & 6\\

ISO\,126\dotfill & A--B & 0.292 & 0.66 & 0.23 & 6\\
T33\dotfill & A--B & 2.434 & 1.93 & 0.01 & 6\\

T39\,A\dotfill & Aa--Ab & 1.242 & 0.31 & 0.44 & 6\\

CHXR\,37\dotfill & A--B & 0.079 & 1.07 & 0.08 & 6\\
CHXR\,79\dotfill & A--B & 0.885 & 2.45 & 0.01 & 6\\
CHXR\,40\dotfill & A--B & 0.151 & 0.12 & 0.83 & 6\\
T43\dotfill & A--B & 0.796 & 1.41 & 0.02 & 6\\
T45\dotfill & A--B & 0.752 & 2.67 & $\lesssim$0.01 & 6\\
T46\dotfill & A--B & 0.123 & 1.57 & 0.01 & 6\\

T51\dotfill & A--B & 1.977 & 2.35 & $\lesssim$0.01 & 6\\
T54\dotfill & A--B & 0.247 & 1.45 & 0.19 & 6\\
CHXR\,59\dotfill & A--B & 0.148 & 0.02 & 0.95 & 6\\
CHXR\,62\dotfill & A--B & 0.12 & 0.06 & 0.86 & 6\\
Hn\,21W\dotfill & A--B & 5.495 & 0.95 & 0.07 & 6\\
B53\dotfill & A--B & 0.295 & 1.52 & 0.02 & 6\\
CHXR\,68\,A\dotfill & Aa--Ab & 0.101 & 0.22 & 0.6 & 6\\
CHXR\,68\,B\dotfill & Aa--B & 4.367 & 0.95 & 0.08 & 6\\

\cutinhead{Tau-Aur}

NTTS\,034903+2431\dotfill & A--B & 0.61 & 1.64 & 0.01 & 1\\
NTTS\,035120+3154SW\dotfill & A--B & 8.6 & 0.28 & 0.67 & 1\\

HD284135\dotfill & A--B & 0.378 & 0.21 & 0.8 & 5\\

RX\,J0406.8+2541\,B\dotfill & A--B & 0.977 & 0.04 & 0.94 & 5\\

RX\,J0409.1+2901\dotfill & A--B & 6.764--6.786 & 1.53--1.59 & 0.16--0.17 & 5\\
RX\,J0412.8+1937\dotfill & A--B & 2.568 & 1.05 & 0.1 & 5\\
RX\,J0413.4+3352\dotfill & A--B & 1.008 & 3.13 & $\lesssim$0.01 & 5\\

RX\,J0415.3+2044\dotfill & A--B & 0.589 & 1.92 & 0.06 & 5\\

V410\,Tau\dotfill & A--B & 0.123 & 1.94--2.5 & 0--0.01 & 2,4\\
 & AB--C & 0.2871 & $\cdots$ & $\cdots$ & 2,4\\

DD\,Tau\,A\dotfill & A--B & 0.56--0.57 & 0.48--0.84 & 0.05--0.28 & 1,2\\
V819\,Tau\dotfill & A--B & 10.5 & 3.64 & $\lesssim$0.01 & 1\\
LkCa\,7\dotfill & A--B & 1.05 & 0.63 & 0.26 & 1\\
T\,Tau\,A\dotfill & A--B & 0.71--0.73 & 1.99--2.6 & 0--0.04 & 1,2\\

BD\,+26\,718B\dotfill & B--b & 0.155--0.166 & 0.53--0.73 & 0.43--0.54 & 5\\

BD\,+17\,724B\dotfill & A--B & 0.083--0.1 & 1.55--2.28 & 0.07--0.16 & 5\\
DI\,Tau\dotfill & A--B & 0.12 & 2.26--2.3 & 0.01 & 2,3\\
UX\,Tau\dotfill & A--B & 5.9 & 1.42 & 0.09 & 1\\
 & A--C & 2.7 & 2.9 & 0.09 & 1\\
FX\,Tau\,A\dotfill & A--B & 0.9--0.91 & 0.4--0.73 & 0.06--0.35 & 1,2,3\\

DK\,Tau\dotfill & A--B & 2.53--2.8 & 1.3--1.51 & 0.01--0.03 & 1,2,3\\

RX\,J0430.8+2113\dotfill & A--B & 0.389 & 3.58 & $\lesssim$0.01 & 5\\
HD284496\dotfill & A--B & 4.598 & 4.25 & $\lesssim$0.01 & 5\\
XZ\,Tau\dotfill & A--B & 0.3--0.311 & 0.73--1.14 & 0.04--0.07 & 1,2\\
V710\,Tau\dotfill & A--B & 3.24 & 0.2 & 0.71 & 1\\

V827\,Tau\dotfill & A--B & 0.0909 & 0.51 & 0.31 & Priv.Comm.\\

V928\,Tau\dotfill & A--B & 0.165--0.18 & 0.14--0.6 & 0.31--0.79 & 1,2,3\\
GG\,Tau\,A\dotfill & A--a & 0.2502--0.288 & 0.48--1.51 & 0.01--0.33 & 1,2,3,4\\
UZ\,Tau\,A\dotfill & A--B & 0.34--0.3678 & 0.5--1.11 & 0.03--0.37 & 1,2,3\\
GH\,Tau\dotfill & A--B & 0.314--0.35 & 0.1--0.64 & 0.22--0.86 & 1,2\\
V807\,Tau\dotfill & A--B & 0.375--0.41 & 0.84--1.07 & 0.08--0.15 & 1,2\\
GI\,Tau\dotfill & GK--GI & 12.2 & 0.23 & 0.5 & 1\\
HN\,Tau\,A\dotfill & A--B & 3.1 & 3.44 & $\lesssim$0.01 & 1\\
FF\,Tau\dotfill & A--B & 0.026 & 1 & 0.1 & 3\\

HBC\,412\dotfill & A--B & 0.7 & 0 & 1 & 1\\
HP\,Tau/G2\dotfill & G2--G3 & 9.9--10 & 1.55 & 0.15 & 1,3\\
RX\,J0435.9+2352\dotfill & A--B & 0.069 & 1.28 & 0.02 & 5\\
 & AB--C & 11.315 & 1.92 & 0.02 & 5\\
RX\,J0437.2+3108\dotfill & A--B & 0.109 & 1.03 & 0.15 & 5\\
RX\,J0438.2+2023\dotfill & A--B & 0.464 & 0.1 & 0.87 & 5\\
RX\,J0438.2+2302\dotfill & A--B & 9.19 & 2.49 & 0.01 & 5\\
HD285957\dotfill & A--B & 9.463--9.504 & 1.74--1.84 & 0.04--0.05 & 5\\
 & A--C & 10.345--10.396 & 4.63--4.72 & 0.04--0.05 & 5\\
VY\,Tau\dotfill & A--B & 0.66 & 1.46--1.5 & 0.01--0.02 & 1,3\\
IW\,Tau\dotfill & A--B & 0.27 & 0--0.1 & 0.66--1 & 1,3\\
CoKu\,Tau/4\dotfill & A--B & 0.0536 & 0.2 & 0.73 & 7\\

HD283798\dotfill & A--B & 1.631 & 3.42 & 0.01 & 5\\
 & AB--C & 7.147 & 5.75 & 0.01 & 5\\
RX\,J0444.3+2017\dotfill & A--B & 9.868 & 2.45 & 0.01 & 5\\

HD30171\dotfill & A--B & 12.926 & 1.71 & 0.14 & 5\\
V1001\,Tau\,A\dotfill & A--B & 2.7 & 0.87 & 0.17 & 1\\

RX\,J0447.9+2755\,A\dotfill & A--B & 0.639 & 0.12 & 0.79 & 5\\
RX\,J0450.0+2230\dotfill & A--B & 2.072 & 3.74 & $\lesssim$0.01 & 5\\
 & AB--C & 8.361 & 3.7 & $\lesssim$0.01 & 5\\
UY\,Aur\,A\dotfill & A--B & 0.88--0.89 & 1.14--1.38 & 0.02--0.06 & 1,2\\
RX\,J0452.8+1621\dotfill & A--B & 0.478 & 0.19 & 0.67 & 5\\
RX\,J0452.9+1920\dotfill & A--B & 0.425 & 1.85 & 0.01 & 5\\

HD286179\dotfill & A--B & 0.112 & 1.26 & 0.22 & 5\\
RX\,J0457.2+1524\dotfill & A--B & 0.57 & 0.05 & 0.94 & 5\\

RX\,J0458.7+2046\dotfill & A--B & 6.113 & 6.75 & $\lesssim$0.01 & 5\\

RW\,Aur\,B\dotfill & A--B & 1.5 & 1.6 & 0.08 & 1,2,4\\
 & B--C & 0.12 & 4.03 & 0.08 & 1,2,4\\
\enddata
\tablenotetext{a}{$R$-band flux ratios are given as the fainter object to the brighter object, and are derived from $\Delta K$ as outlined in \S\ref{sec:SB2}.}
\tablenotetext{b}{(1) \citet{1993AA...278..129L}; (2) \citet{1993AJ....106.2005G}; (3) \citet{1995ApJ...443..625S}; (4) \citet{1997ApJ...490..353G}; (5) \citet{1998AA...331..977K}; (6) \citet{2008ApJ...683..844L}; (7) \citet{2008ApJ...678L..59I}}
\end{deluxetable}

\end{document}